\documentclass[final,5p,times,twocolumn,authoryear]{elsarticle}

\usepackage{amssymb}
\usepackage{amsmath}

\usepackage{url}
\usepackage{adjustbox}
\usepackage{setspace}
\usepackage{array}
\usepackage{booktabs}
\usepackage{subcaption} 
\usepackage{calc}
\usepackage{graphicx}
\usepackage{multirow}

\journal{Computers in Biology and Medicine}

\begin{document}

\begin{frontmatter}



\title{AXIAL: Attention-based eXplainability for Interpretable Alzheimer's Localized Diagnosis using 2D CNNs on 3D MRI brain scans} 


\author[1,2]{Gabriele Lozupone\corref{cor1}} 
\ead{gabriele.lozupone@unicas.it}
\cortext[cor1]{Corresponding author:}
\author[1]{Alessandro Bria} 
\author[1]{Francesco Fontanella} 
\author[3]{Frederick J.A. Meijer} 
\author[1]{Claudio De Stefano} 

\affiliation[1]{organization={Department of Electrical and Information Engineering (DIEI), University of Cassino and Southern Lazio}, 
            addressline={Via G. Di Biasio 43}, 
            city={Cassino},
            postcode={03043}, 
            state={FR},
            country={Italy}}

\affiliation[2]{organization={Diagnostic Image Analysis Group, Radboud University Medical Center},
addressline={Geert Grooteplein 10},
city={Nijmegen}, 
postcode={6500HB},
country={Netherlands}
}

\affiliation[3]{organization={Department of Medical Imaging, Radboud University Medical Center},
addressline={Geert Grooteplein 10},
city={Nijmegen}, 
postcode={6500HB},
country={Netherlands}
}

\begin{abstract}
Accurate early diagnosis of Alzheimer's disease (AD) is a critical challenge in clinical practice. Emerging deep learning-aided diagnostic systems utilizing 3D MRI show promise but often fail to highlight meaningful and well-localized brain areas. This study presents an innovative method for 3D MRI classification via 2D CNNs, designed to enhance the explainability of model decisions. Our approach adopts a soft attention mechanism, enabling 2D CNNs to extract volumetric representations. At the same time, the importance of each slice in decision-making is learned, allowing the generation of a voxel-level attention map to produces an explainable MRI. To test our method and ensure the reproducibility of our results, we chose a standardized collection of MRI data from the Alzheimer's Disease Neuroimaging Initiative (ADNI). On this dataset, our method significantly outperforms state-of-the-art methods in (i) distinguishing AD from cognitive normal (CN) with an accuracy of 0.856 and Matthew's correlation coefficient (MCC) of 0.712, representing improvements of 2.4\% and 5.3\% respectively over the second-best, and (ii) in the prognostic task of discerning stable from progressive mild cognitive impairment (MCI) with an accuracy of 0.725 and MCC of 0.443, showing improvements of 10.2\% and 20.5\% respectively over the second-best. We achieved this prognostic result by adopting a double transfer learning strategy, which enhanced sensitivity to morphological changes and facilitated early-stage AD detection. Our approach identified predominant AD-related brain regions: \emph{hippocampus}, \emph{parahippocampus}, \emph{amygdala}, and \emph{inferior lateral ventricles}. Differently from the current literature, the areas are ordered by the algorithm in the same order of importance that radiologists give to diagnose AD. In addition, our model highlighted 17 well-localized regions (considering the left and right sides separately), with a strong focus on the first 14 regions. In contrast, the state-of-the-art methods highlighted 88 regions, demonstrating the more targeted and precise identification of our model. Furthermore, our approach consistently found the same AD-related areas across multiple cross-validation folds, demonstrating its robustness and accuracy in highlighting areas that closely align with known pathological markers of the disease.

\end{abstract}




\end{frontmatter}

\section{Introduction}
\label{sec:intro}
Alzheimer's disease (AD) is a chronic neurodegenerative disorder characterized by the irreversible progression of cognitive impairment and gradual death of nerve cells throughout the brain. AD diagnosis and progression prediction present a significant challenge in clinical practice. Initially characterized by symptoms such as Mild Cognitive Impairment (MCI), AD progressively leads to more severe cognitive decline, behavioral alterations, and loss of functional independence, eventually leading to death \citep{lam2013clinical, loewenstein2006cognitive}. AD is a disorder with a rapidly increasing prevalence trend, predicted to affect 1 in 85 people globally by 2050 \citep{brookmeyer2007forecasting}. 
An early diagnosis of AD is fundamental for both present and future patients once disease-modifying pharmacological treatments are available. In both cases, it will significantly improve the effectiveness of the available treatments, improving quality of life and reducing care costs \citep{nicoll2019persistent,winblad2016defeating, wong2020economic}. 

Magnetic resonance imaging (MRI) is a key diagnostic and prognostic tool for AD. It provides a non-invasive means to observe and analyze in vivo pathological changes in the brain related to AD, facilitating the study of disease evolution \citep{ewers2011neuroimaging}. MRI analysis is significant as AD is identified by structural and functional changes that occur in dynamically changing morphological patterns, which are appropriately captured with high-resolution MRI \citep{duchesne2008mri, jack2003mri, kloppel2008automatic, vemuri2009mri}. It is also noteworthy that brain atrophy, a distinctive AD symptom, can be identified through MRI. This form of atrophy serves as a reliable marker of the disease. It is indicative of its progression, as well as being associated with tau deposition and neuropsychological impairments, essential factors in the clinical manifestation of AD \citep{frisoni2010clinical}.

In recent years, deep learning (DL), particularly with convolutional neural networks (CNNs), has transformed neuroimaging data analysis for AD, moving beyond the traditional machine learning (ML) that focuses on approaches that rely on handcrafted features and classifiers \citep{ML_falahati2014multivariate, ML_haller2011principles, ML_rathore2017review}. DL's ability to autonomously extract features at different abstraction levels minimizes the need for extensive image pre-processing and feature selection, offering a more objective and less biased approach in medical imaging. Several researchers have recently proposed various DL-based approaches to diagnose and predict AD using MRI. These methods can be divided into three main categories: (i) analysis of the entire three-dimensional (3D) volume with 3D CNNs \citep{3d_basaia2019automated, 3d_feng2022deep, 3d_wu2022attention, 3d_venugopalan2021multimodal}, (ii) methods based on extracting 3D patches from the volume \citep{3d_patch_goenka2022patch, 3d_patch_liu2023patch, 3d_patch_park2023deep, 3d_patch_qiu2020development}, and (iii) classification of two-dimensional (2D) slices selected from a specific plane \citep{2d_slice_ebrahimi2021deep, 2d_slice_hon2017towards, 2d_slice_kang2021multi, 2d_slice_pan2020early, 2d_slice_tanveer2021classification, 2d_slice_zhang2022diagnosis}.
Each of these methods has several advantages and limitations. 3D volume analysis with 3D CNN offers a comprehensive approach. However, it is computationally expensive and suffers from the scarcity of pre-trained models. This limits the opportunities for using transfer learning strategies, which is essential when dealing with small datasets \citep{yosinski2014transferable}. The 3D patch-based approach, while reducing computational complexity, shares similar limitations due to the lack of pre-trained models. Furthermore, this method does not directly allow the extraction of global 3D volume features crucial for accurate analysis.
On the other hand, methods utilizing 2D slices can benefit from the abundance of 2D models pre-trained on extensive image datasets such as ImageNet \citep{russakovsky2015imagenet}. This enables transfer learning, which can significantly increase performance in limited data regimes. Nevertheless, this method also faces a significant limitation: it fails to retain the comprehensive spatial information intrinsic to 3D volumes, which is crucial for discerning neurodegenerative patterns in AD.

The interpretability of DL models remains a significant challenge that hinders the deployment of deep learning-aided diagnostic systems (DLADS) in real-world scenarios \citep{EXP_singh2020explainable, EXP_van2022explainable}. 
Most DLADS often base their explainability on methods that provide post hoc explanations, such as visual inspection by saliency maps. A popular approach in this category is the Gradient-weighted Class Activation Mapping (GradCAM) \citep{selvaraju2017grad}. This class of explainable artificial intelligence (XAI) methods generally struggles to offer the information needed to identify specific brain regions affected by AD, thus limiting their effectiveness in providing meaningful explanations for model decisions \citep{EXP_no_gradcam_viswan2024explainable}. These challenges highlight the need for methodologies that not only achieve high diagnostic accuracy but also improve the explainability of the model by detailing the involvement of specific brain areas. 

\begin{figure*}[!h]
\centering
\includegraphics[width=\textwidth]{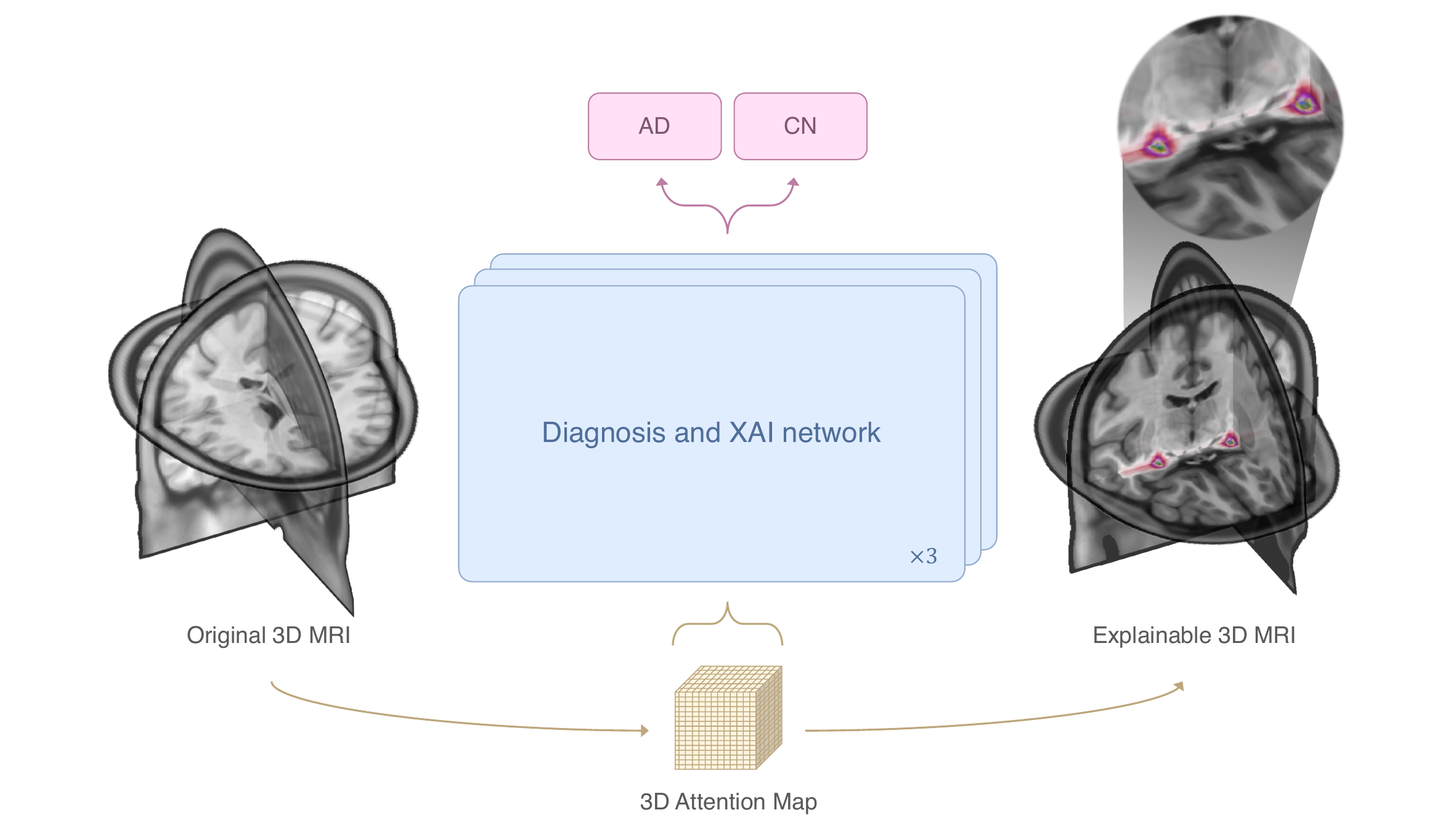}
\caption{Schematic representation of the proposed diagnostic framework. The Diagnosis and XAI framework processes a 3D sMRI brain image to generate two key outputs: three diagnosis networks identifying the condition as either AD or CN from the three possible slicing axes and a corresponding 3D attention map that can be overlapped to the input image to visually highlight the brain regions the network focuses on to derive its diagnosis.}
\label{fig:approach_overview}
\end{figure*}

Our research introduces an explainable method for AD detection using 3D MRI. This advancement leverages an attention fusion mechanism of feature maps, enhancing interpretability and accuracy even with limited data. As illustrated in Fig. \ref{fig:approach_overview}, the framework generates two key outputs: diagnosis and a 3D attention map using three networks based on different MRI slicing axes. The 3D attention map overlay on the original scan produces an explainable MRI highlighting brain regions for visual diagnosis. The significant contributions of this work are summarized as follows:
\begin{enumerate}
  \item we introduce a novel classification and XAI approach capable of highlighting brain areas highly correlated with AD without sacrificing performance, even in the case of limited datasets.
  \item we tested a double transfer learning strategy for distinguishing between stable Mild Cognitive Impairment (sMCI) and progressive Mild Cognitive Impairment (pMCI), enhancing model sensitivity to morphological changes indicative of early-stage disease progression.
  \item we evaluated our approach using a small standardized subset of the Alzheimer's Disease Neuroimaging Initiative (ADNI), proposed in \citep{adni_wyman2013standardization}, and compared its effectiveness to recent transformer-based approaches and related XAI techniques. Using this standardized dataset in combination with standardized pipelines makes our results easily accessible and reproducible, aiding comparison with future studies. The framework implementation to reproduce all XAI, diagnostic and prognostic results can be found here \url{https://github.com/GabrieleLozupone/AXIAL.git}.
  \item we propose an approach to quantify how important specific brain regions are in a model's decision-making process, facilitating the alignment of XAI with medical knowledge about AD, thus promoting valuable insights in clinical practice.
\end{enumerate}

The remainder of the paper is organized as follows: Section \ref{sec:rel} discusses the related work that supports our study. In Section \ref{sec:materials}, we describe the materials, including the dataset and pipeline utilized. Section \ref{sec:methods} outlines our methods, focusing on the diagnosis network and the proposed XAI approach. Section \ref{sec:experiments} presents the experimental settings, along with diagnostic and XAI results. Discussions of the findings are provided in Section \ref{sec:discussions}, and Section \ref{sec:conclusions} concludes the paper with a summary of our contributions.

\section{Related work}
\label{sec:rel} As anticipated in Section \ref{sec:intro}, DL-based analysis of MRI images to diagnose and predict AD can be divided into three broad categories: (i) approaches that use 3D CNNs to analyze the entire 3D volume, (ii) approaches that analyze 3D patches from the whole 3D volume, and (iii) approaches that classify 2D slices. The following subsections report recent research activities belonging to these categories. Furthermore, we discuss the current challenges and issues in DL-based AD analysis, focusing on reliability and interpretability. 
\subsection{3D CNNs}
\label{subsec:rel_3dcnn}
In \cite{3d_basaia2019automated}, the authors propose a fully convolutional 3D CNN model replacing typical max-pooling operations with standard convolution layers and test the model to distinguish AD, pMCI and sMCI from the entire subject's brain MRI scan. The results presented in  \cite{3d_feng2022deep} show that an MRI-based 3D CNN outperforms other neuroimaging biomarkers of neurodegeneration in prodromal AD and also outperforms amyloid or tau pathology biomarkers. More sophisticated approaches introduce attention mechanisms into 3D CNNs, e.g., the study presented in \cite{3d_jin2019attention} investigates a novel attention-based 3D ResNet architecture to diagnose AD and explore potential biological markers. Also, a novel Attention-based 3D Multi-scale CNN model was proposed in \cite{3d_wu2022attention} to capture better and integrate multiple spatial-scale features of AD. These approaches are very promising and comprehensive since analyzing the entire volume allows the extraction of global information on a spatial scale, unlike techniques based on conventional 2D CNNs. However, they are computationally expensive and require a large dataset during the training phase to generalize well. Furthermore, given the scarcity of pre-trained models, this problem cannot be easily addressed with a transfer learning strategy \citep{3d_review_klaiber2021systematic}.
\subsection{3D Patch-based}
\label{subsec:rel_3dpatch}
One way to reduce the computational cost and the overfitting risk of 3D CNN-based approaches consists of reducing the model inputs. For this reason, several approaches have been proposed to analyze 3D patches extracted from MRI scans. In \cite{3d_patch_qiu2020development}, the authors present a novel computationally efficient patch-level training strategy to train a 3D CNN. They developed a 3D CNN model for the AD classification task, using random subvolumes of MRI scans as training data. This model generates patient-specific disease probability maps that are used to train a Multilayer Perceptron (MLP) to distinguish between AD and CN subjects. On the other hand, in \cite{3d_patch_park2023deep}, the authors introduce a framework that utilizes a 3D CNN to extract local features from 3D patches in brain MRI scans, forming the basis for patch-level responses. These responses are then processed through a dual-branch approach, combining patch classification and location identification. These methods enhance computational efficiency by focusing on patches of the 3D volume rather than processing the entire volume. However, the capacity for recognizing and learning global-level patterns within the data is limited. Moreover, the encoding process for 3D patches relies on 3D CNNs and restricts the application of transfer learning strategies.
\subsection{2D Slice-level}
\label{subsec:rel_2dslice}
An alternative to 3D CNNs are the more widespread 2D CNNs. Given the 3D nature of the input, to use 2D CNNs, 2D images must be extracted from the MRI image. This process is commonly known as slicing. Slicing can be performed according to one of the three planes of the MRI brain scan: (i) axial, (ii) coronal, and (iii) sagittal \citep{review_zhou2023survey}. 

In \cite{2d_slice_wang2018classification}, the authors propose an eight-layer custom CNN with leaky rectified learning units to classify single-slice MRI images. Recently, the work presented in \cite{2d_slice_carcagni2023convolution} investigates the ability of different CNN and Transformer-based models to classify single slices selected from MRI with a mechanism based on Shannon entropy. Although using 2D CNNs for straightforward classification of 2D images is a standard practice, this method is insufficient for making direct predictions about an individual patient. To achieve accurate subject-level predictions, it is necessary to integrate results from multiple-slice analyses. 

The study presented in \cite{2d_slice_kang2021multi} introduces an ensemble learning architecture based on 2D CNNs, using a multi-model and multi-slice ensemble, in which the majority voting scheme is used to merge the multi-slice decisions of each model. In  \cite{2d_slice_kushol2022addformer}, the authors use a fusion transformer block to merge the outputs of a ViT \citep{vit_dosovitskiy2020image} and a GFNet \citep{gfnet_rao2021global}, both pre-trained on ImageNet. The proposed approach allows them to correlate features extracted from spatial and frequency domains, but predictions are made at the slice level. Thus, they are combined using a majority voting approach to make a decision at the subject level.
Classifying each slice independently and then using majority voting eliminates the opportunity to learn patterns and features that span across slices. 

To overcome this limitation, \cite{2d_slice_altay2021preclinical} introduced the Attention Transformer model. This model comprises a VGG-16 as a Base Network and a Transformer Unit. The Base Network extracts feature representations from each brain MRI image slice. Unlike the traditional Transformer architecture that generates a new representation of the input sequence through self-attention, the Transformer Unit performs a cross-attention operation. This operation consists of considering as Query the feature map relating to the central slice of the 3D volume and as Key and Value the feature maps of the others. This method allows the fusion of information from the entire volume, capturing the interrelationships between the central and surrounding slices.
Similarly, \cite{2d_slice_hu2023conv} proposed the Conv-Swinformer architecture consisting of a CNN and a Transformer encoder module. The CNN module summarizes the planar features of the MRI slices, and the Transformer module establishes semantic connections in 3D space for these planar features. In this case, the Transformer encoder comprises four Swin Transformer blocks that perform self-attention operations on multiple windows extracted from the feature maps sequence.
However, the reliance on cross-attention or self-attention in Transformer-based models, as suggested in \citep{2d_slice_altay2021preclinical, 2d_slice_hu2023conv}, poses challenges in medical scenarios with limited datasets. The extensive parameterization inherent in these mechanisms increases the risk of overfitting when training on smaller and often imbalanced medical datasets. This undermines the model's generalizability to new data, crucial in medical diagnostics.

The overparameterization problem can be addressed using less complex self-attention variants to build lightweight architectures. In the study \cite{2d_slice_wang2024joint}, the authors proposed a diagnosis network composed of two innovative modules in combination with basic ResNet blocks. The first is the slice-aware module designed to interpret significant slices and regions. This module performs self-attention between coded sections in an optimized way to be more parameter efficient. The second is the slice-shift module, which allows joint inter- and intra-slice modelling to exchange information with neighbouring slices.
\subsection{Challenges in DL-Based AD Analysis}
\label{subsec:challenges}
Despite the potential of DL in diagnosing and prognosis AD using MRI images, several critical issues undermine the reliability and reproducibility of the findings. Furthermore, the “black box” nature of DL impacts clinical confidence and transparency of decision-making.

This subsection provides an analysis of the challenges in the AD analysis in terms of reliability, reproducibility and XAI.

\subsubsection{Reliability and Reproducibility}
Recent reviews \citep{review_wen2020convolutional, review_zhou2023survey} highlight several reliability and reproducibility issues in DL-aided AD MRI analysis:
\begin{itemize}
    \item \textbf{Wrong Data Split}: Data leakage due to incorrect data splitting is recurring. When datasets are not split at the subject level, data from the same subject may appear in both training and test sets, leading to overestimated model performance. This problem is particularly pronounced in patch or slice-based approaches \citep{review_wen2020convolutional}.
    \item \textbf{Late Split in Data Processing}: Data processing steps such as data augmentation and feature selection must be conducted post-data splitting to avoid contamination of the test set. If these steps involve test data, it biases the model, rendering the evaluation unreliable \citep{review_wen2020convolutional}.
    \item \textbf{Biased Transfer Learning}: Transfer learning, while beneficial, can introduce bias, particularly when the source and target domains overlap. For instance, using a network pre-trained on an AD vs CN task for an MCI vs CN task can cause bias if the CN subjects overlap in both tasks' training or validation sets \citep{review_wen2020convolutional}.
    \item \textbf{Absence of an Independent Test Set}: The integrity of model evaluation is compromised if the test set is used for any purpose other than final performance evaluation. Hyperparameter optimization must rely on a separate validation set. The absence of an independent test set in some studies leads to inflated performance metrics \citep{review_wen2020convolutional}.
    \item \textbf{Differences in Diagnostic Criteria}: Variability in diagnostic criteria for AD across different studies creates inconsistencies in ground truth labeling. This variability complicates the comparison of results across studies and hampers the development of universally applicable models \citep{review_zhou2023survey}.
    \item \textbf{Lack of Reproducibility}: A significant challenge in the field is the non-availability of frameworks and models for public use. Without access to open-source code and detailed implementation procedures, reproducibility is hampered. This lack of transparency affects the ability to validate and build upon existing work \citep{review_zhou2023survey}.
\end{itemize}
Furthermore, \cite{review_wen2020convolutional} highlights that the prevalent use of the 2D slice approach in AD analysis is often associated with a series of methodological problems. In particular, the authors highlight that among the numerous studies using this approach, only a few have successfully addressed the abovementioned challenges. This fact underlines the need for rigorous methodological standards in the field.

In response to these challenges, our work was carefully developed to avoid common weaknesses identified in previous studies. Section \ref{sec:experiments} details the data splitting for model selection and data augmentation used for this study. Section \ref{sec:double_trans} outlines the double transfer learning strategy investigated. To ensure reproducibility, we provide an open-source repository with all necessary code and implementation procedures at \url{https://github.com/GabrieleLozupone/AXIAL.git}.

\subsubsection{Challenges in current XAI approaches}
Several studies have highlighted the difficulties that prevent using DLADS on medical images due to the current challenges of XAI methods \citep{EXP_van2022explainable, EXP_singh2020explainable, EXP_no_gradcam_viswan2024explainable}. XAI researchers often use self-intuition to determine a good explanation without validating with a medical professional. This missing validation is partially due to the inability to provide quantitative but only qualitative explanations. Qualitative explanations provided for small and specific subsets of subjects make a global evaluation and comparison of XAI techniques difficult. Furthermore, there is no specific correlation between the prediction and the associated brain region. Model-agnostic XAI techniques, such as GradCAM, often produce saliency maps that highlight extensive regions of the brain without sufficient specificity or consistency. Another critical limitation of GradCAM-like methods, mainly when applied to the analysis of 2D slices from MRI scans, is their focus on intra-slice features while neglecting the inter-slice relationships crucial for underlying the 3D brain regions related to AD. Aggregating these saliency maps to form a 3D visualization produces an overly generalized activation map that often highlights different and sparse areas. 

As an alternative to traditional saliency-based methods, recent advancements have explored the use of attention-based models to enhance the interpretability of DL models. In this context, \cite{attention_exp_jetley2018learn} proposed a trainable attention mechanism to highlight where and in what proportion the network paid attention to input images for classification. Attention amplifies relevant areas and suppresses irrelevant ones, improving interpretability. In \cite{attention_exp_schlemper2019attention}, grid attention captures anatomical information in medical images, and attention coefficients were used to explain which areas of the image the network focused on. Recently \cite{2d_slice_wang2024joint}, described in Section \ref{subsec:rel_2dslice}, introduced a slice-aware module to advance XAI in their methodology. Their variant of the attention mechanism enables the extraction of attention weights to determine the importance of each slice in the decision-making process, offering a more precise understanding of model focus. 

\section{Materials}

\bgroup
\def\arraystretch{1.5}
\begin{table}[!t]
    \caption{Summary of participant demographics, mini-mental state examination (MMSE) and global clinical dementia rating (CDR) scores}
    \label{tab:demographic}
    \centering
    \begin{adjustbox}{width=\columnwidth}
    \begin{tabular}{cccccc}
    \hline
    & Subjects & Samples & Age & MMSE & CDR \\  \hline
    CN & 204 & 586 & \(76.31\pm5.22\) & \(29.11\pm1.05\) & \(0.01\pm0.16\) \\ 
    AD & 191 & 474 & \(75.23\pm7.34\) & \(22.46\pm3.38\) & \(0.86\pm0.43\) \\ 
    pMCI & 121 & 162 & \(75.01\pm6.68\) & \(26.47\pm1.83\) & \(0.481\pm0.09\) \\ 
    sMCI & 110 & 154 & \(75.09\pm7.02\) & \(27.80\pm1.74\) & \(0.45\pm0.17\) \\ \hline
    
    \end{tabular}
\end{adjustbox}
\end{table}
\egroup

\label{sec:materials}
The ADNI MRI Core in \cite{adni_wyman2013standardization} has created standardized dataset collections comprising scans that met minimum quality control requirements to promote greater rigour in analysis and meaningful comparison of different algorithms. These standardized datasets within the ADNI archive allow researchers to download a complete and consistent set of images efficiently, thus facilitating comparative research into neurodegenerative diseases. These collections represent a small portion of the ADNI dataset and generally are not used in DL literature. 
However, we choose the ``ADNI1: Complete 1Yr 1.5T'' collection as it directly addresses the problem of lack of reproducibility facilitating model performance comparisons. This collection of 1.5 Tesla MRI scans includes only 639 subjects of the 2.720 available in the ADNI study. All subjects have screening, along with 6 and 12-month follow-up scans. Furthermore, this small data subset is an excellent candidate for testing our method in the recurrent medical imaging data scarcity condition. 

We converted the dataset into the Brain Imaging Data Structure (BIDS) format \citep{bids_gorgolewski2016brain}. A significant advantage of converting the ADNI dataset to a BIDS structure is the availability of specialized Python libraries for handling BIDS-structured data. These libraries, such as PyBIDS \citep{pybids_yarkoni_2023_8253830, pybids_Yarkoni2019}, provide powerful tools for querying, organizing, and processing neuroimaging data. To perform this conversion, we used Clinica \citep{clinica_routier2021, clinica_samper2018reproducible}, an open-source software platform explicitly designed to make clinical neuroscience studies more accessible and reproducible.

During the conversion phase, the ADNI-to-BIDS converter provided in Clinica checks whether the data meets specific criteria, e.g. images that fail quality control are filtered out or, if there are multiple scans for a visit, the `preferred scan' is selected according to specific decision rules \citep{clinica_routier2021}. Consequently, some of the MRIs in the original dataset may be discarded at this stage. The ``ADNI1: Complete 1Yr 1.5T'' dataset contains 2,042 acquisitions from 639 subjects. During the conversion, 151 scans did not meet the criteria, so the resulting dataset was 1,891 samples without varying the number of subjects.
Each subject had a maximum of 3 scans from baseline, 6-month and 12-month follow-up, and consequently, samples from the same patient with a different diagnosis are unlikely. Therefore, to define sMCI patients and pMCI patients, we used the complete clinical data from ADNI, which contains the diagnoses after several years of follow-up.

From the resulting dataset, we have extracted four classes:
\begin{itemize}
    \item CN: 3D brain images of patients diagnosed as CN at the time of image acquisition.
    \item AD: 3D brain images of patients diagnosed with AD at the time of image acquisition.
    \item sMCI: 3D brain images of patients diagnosed with MCI at the time of image acquisition and who remain MCI over time.
    \item pMCI: 3D brain images of patients initially diagnosed with MCI at the time of scan but later diagnosed with AD 36 months post-acquisition.
\end{itemize}
The demographic information of the subjects in the ``ADNI1: Complete 1Yr 1.5T'' dataset is presented in Table \ref{tab:demographic}, which provides essential context for our analysis and findings.

\section{Methods}
\label{sec:methods}
\begin{figure}[!t]
\centering
\includegraphics[scale=.222]{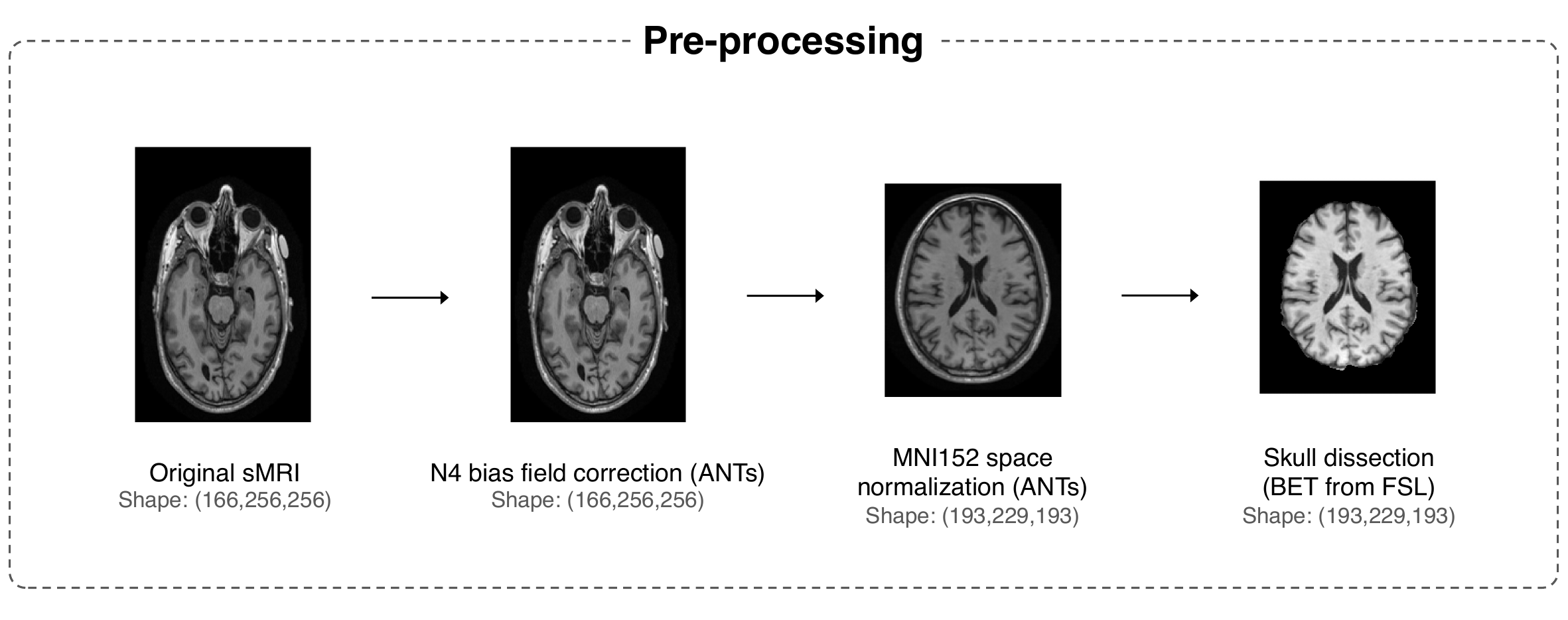}
\caption{MRI data pre-processing pipeline: (i) Original sMRI, (ii) N4 bias field correction, (iii) MNI152 space normalization, and (iv) skull dissection.}
\label{fig:preprocessing}
\end{figure}

This section presents the explainable diagnostic pipeline developed in this study. The process starts with a pre-processing phase described in Fig. \ref{fig:preprocessing}. The XAI diagnosis network illustrated in Fig. \ref{fig:diagnosis} processes the images as a series of 2D slices, producing diagnosis and attention weights to synthesize the slice's importance in decision-making. Finally, the attention weights of the slices across the three distinct views are combined to create a 3D attention map, as shown in Fig. \ref{fig:XAI_net}. 

In the following sections, we provide full details of each step.

\subsection{MRI images pre-processing}
\label{sec:preproc}
As mentioned in Section \ref{subsec:challenges}, classification results are difficult to compare between studies due to their limited reproducibility. In the study \cite{review_wen2020convolutional}, the authors highlight that variations in components such as participant selection and image pre-processing are critical aspects that directly influence this limitation. It is, therefore, sufficient to avoid ambiguity in both cases to facilitate the comparison between models. In Section \ref{sec:materials}, we described the procedures required to avoid ambiguity in participant selection.
The pre-processing pipeline used comprises three steps:
\begin{enumerate}
    \item Bias field correction using the N4ITK method \citep{n4itk_tustinson_2010};
    \item Affine registration using the SyN algorithm \citep{avants2008symmetric} from ANTs \citep{avants2014insight} to align each image to the Montreal Neurological Institute (MNI) 152 space with the ICBM 2009c nonlinear symmetric template \citep{fonov2009unbiased, fonov2011unbiased};
    \item Skull dissection to remove non brain-tissue from the 3D image using Brain Extraction Tool \citep{bet_smith2002fast} from FSL \citep{jenkinson2012fsl}.
\end{enumerate}
We used the t1-linear pipeline of Clinica \citep{clinica_routier2021, review_wen2020convolutional} to perform the first two steps.

\subsection{Diagnosis network}

\begin{figure*}[!h]
\centering
\includegraphics[width=\textwidth]{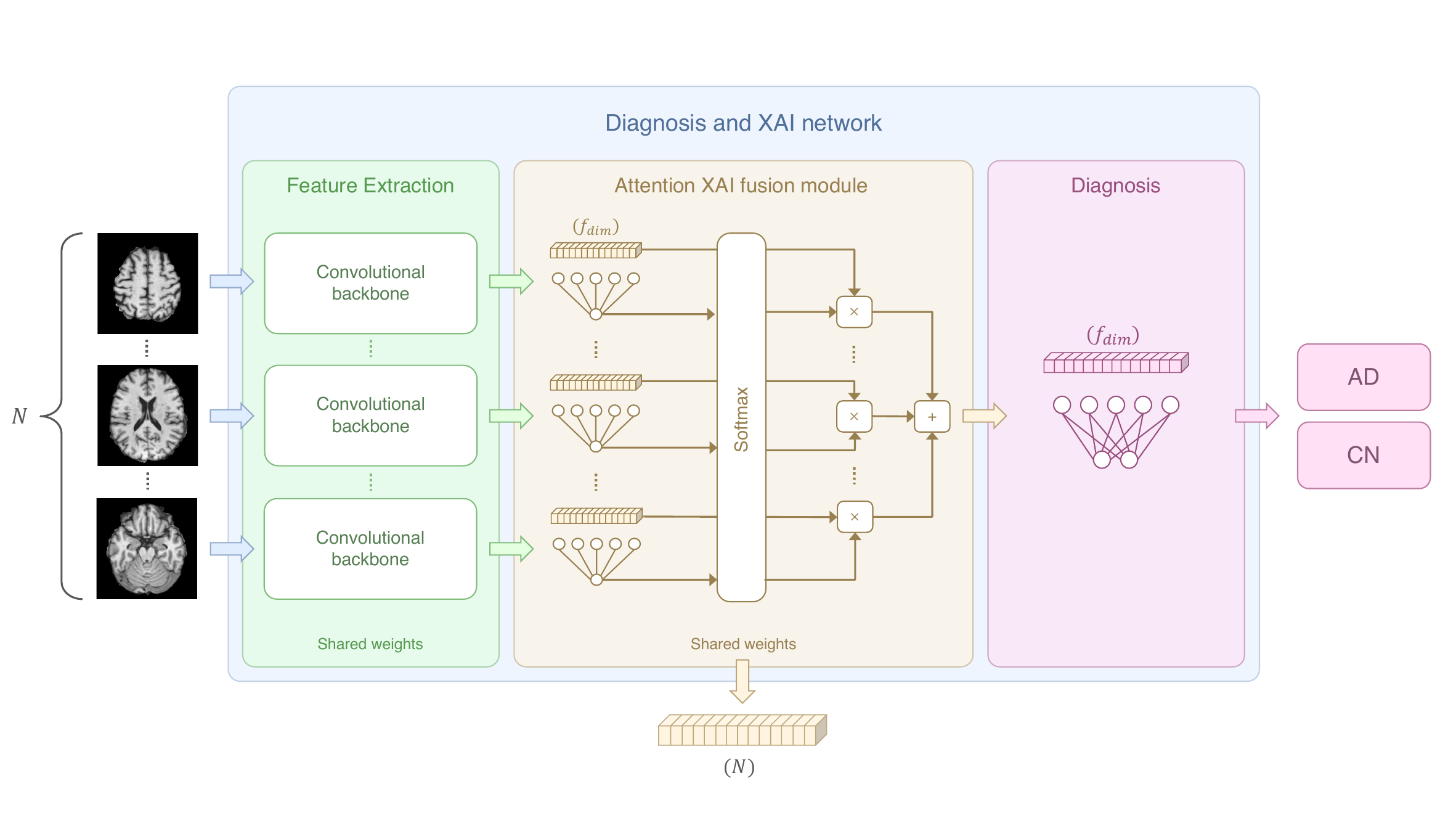}
\caption{
The proposed Diagnosis and XAI network: (i) Feature Extraction module, (ii)  Attention XAI Fusion module, and (iii) Diagnosis module.
}
\label{fig:diagnosis} 
\end{figure*}

Radiologists typically review a series of 2D slice images when analyzing medical imaging data, even when examining 3D images. With this in mind, we have developed an efficient and explainable network for diagnosing AD that learns inter-slice relationships. This network, as illustrated in Fig. \ref{fig:diagnosis}, comprises three key modules described in detail in this section.

\subsubsection{Feature Extraction}

This module employs a pre-trained 2D convolutional backbone to process a series of $N$ 2D slices. The convolutional backbone, which can be any architecture such as VGG or ResNet, is pre-trained on the ImageNet dataset to handle cases of small dataset sizes in AD diagnosis tasks.

The original data are 3D images with a value for each voxel, so the resulting sliced 2D images are 1-channel images. Since pre-trained weights are for 3-channel images, we summed the pre-trained convolutional filters of the backbone's first layer. Since these filters operate linearly and sum their results across channels, we can sum all the filters over the input channel dimension in the first layer. This operation is equivalent to replicating the single-channel image three times, but it is more computationally efficient. The extracted slices were resized to $224 \times 224$ and normalized. 

Each convolutional backbone ends with a max-pooling operation that converts the feature map of each slice into a feature vector of dimension $f_{\text{dim}}$. This sequential input processing is achieved through sharing convolutional weights across the sequence, similar to a Recurrent Neural Network approach:
\[f_{i} = \text{MaxPool}(\text{Conv}(...(x_{i})))\]
where $x_{i}$ represents the $i$-th slice in the sequence, $f_{i}$ is the resulting feature vector for that slice, and $\text{Conv}$ denotes the convolutional operations performed by a convolutional architecture.

\subsubsection{Double Transfer Learning}
\label{sec:double_trans}
Models pre-trained on large datasets can help overcome challenges posed by limited data, but they may not be enough for complex tasks. As shown in Table \ref{tab:demographic}, the data available for training a network to differentiate between AD and CN individuals is usually more than for distinguishing between sMCI and pMCI. Although both tasks are complex, the latter is considered more difficult. Therefore, it is more likely that transfer learning can distinguish AD from CN more effectively than recognizing the difference between sMCI and pMCI.
Since doctors rely on their expertise to assess a patient's cognitive decline, leading to an AD diagnosis, we suggest fine-tuning a model pre-trained on the AD vs. CN task to the sMCI vs. pMCI task. 
Following our definition of classes in Section \ref{sec:materials}, the $AD \cup CN$ set is disjoint from $(sMCI \cup pMCI) \subset MCI$. As a result, adopting this strategy under these assumptions does not imply data leakage issues.

\subsubsection{Attention XAI Fusion Module}

This module enables the network to learn inter-slice dependencies and global 3D patterns through the Feature Extraction module. During backpropagation, the shared weights of the convolutional module are updated based on the entire image, thus enabling the learning of 2D patterns in relation to the importance of that slice at a global level. The Attention XAI Fusion module leverages a fully connected (FC) layer to assign importance to each section based on its feature vector. This introduces a parameter-efficient approach, with only $f_{\text{dim}} + 1$ learnable parameters:

\[w_{i} = \text{FC}_{attention}(f_{i})\]
where $w_{i}$ is the computed weight for slice $i$, and $\text{FC}$ denotes the fully connected layer operation.
Normalization of these weights is performed using a softmax function, ensuring they sum to one and represent the slice-importance distribution:

\[ \alpha_{i} = \textit{softmax}(w_{i}) =\frac{e^{w_{i}}}{\sum_{j=1}^{N} e^{w_{j}}} \]
where $\alpha_{i}$ is the normalized importance weight for the $i$-th slice.
The module performs fusion by calculating a weighted sum of the feature vectors using their corresponding weights, generating a composite feature vector representing the entire brain image:

\[F = \sum_{i=1}^{N} \alpha_{i} f_{i}\]
This module yields a feature vector $F$ that synthesizes the brain image and a set of attention weights $\{\alpha_{i}\}$ highlighting the significance of each slice.

\begin{figure*}[!h]
\centering
\includegraphics[width=\textwidth]{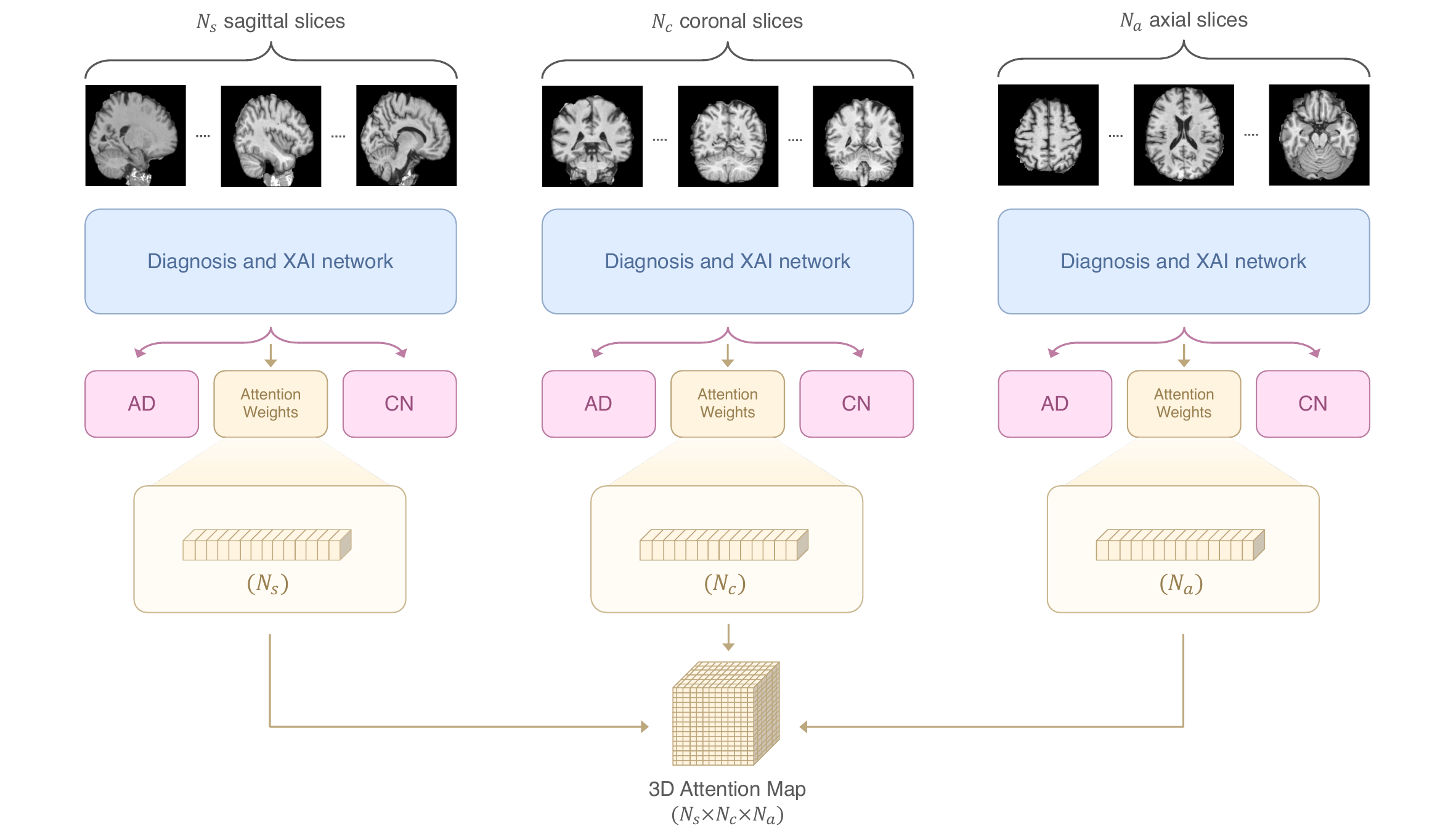}
\caption{Representation of the XAI attention-based approach proposed. Three distinct network are trained for each slicing plane and the slice attention weights for each plane are combined to produce a 3D attention map.}
\label{fig:XAI_net}
\end{figure*}

\subsubsection{Diagnosis}

In the final \textit{Diagnosis} module, the synthesized feature vector $F$ is fed into a head network comprising a fully connected layer with two output neurons for the binary classification task (e.g., AD vs. CN):

\[D = \textit{softmax}(\text{FC}_{\text{head}}(F))\]
$D$ represents the softmax output, providing a probability distribution over the diagnosis classes, and $\text{FC}_{\text{head}}$ denotes the fully connected layer specific to the diagnosis task.

\subsection{XAI Approach}
\label{sec:xai_approach}

Within our diagnostic framework for AD, we introduce an advanced XAI strategy to generate a 3D attention map, facilitating a deeper understanding of the neural network's diagnostic focus. As shown in Fig. \ref{fig:XAI_net}, our approach harnesses attention weights across sagittal, coronal, and axial slicing planes, merging these into a comprehensive representation that highlights key brain regions indicative of AD.

\subsubsection{Attention Weight Generation}

The implemented XAI method yields attention weights for each of the three principal slicing planes of the brain: sagittal ($s$), coronal ($c$), and axial ($a$). We trained a dedicated Diagnosis and XAI Network for each plane to assess slice importance distribution from different views. The MRI brain image is sliced in the three planes, resulting in three sequences of 2D slices: $N_s$, $N_c$, and $N_a$ slices for the sagittal, coronal, and axial planes. 

Each network's Attention XAI module computes a set of attention weights $\mathbf{\alpha}_s$, $\mathbf{\alpha}_c$, and $\mathbf{\alpha}_a$ for its corresponding plane. The computation is formalized as follows:
\[ \mathbf{\alpha}_p = \textit{softmax}\left(\text{FC}_{attention}(f_p)\right) \]
where $p \in \{s, c, a\}$ represents each plane, $f_p$ denotes the feature vectors derived from slices within that plane.

\subsubsection{3D Attention Map Synthesis}

We integrate the attention weights $\mathbf{\alpha}_s$, $\mathbf{\alpha}_c$, and $\mathbf{\alpha}_a$ into a unified 3D attention map. The 3D attention map \(A\) is constructed by applying the following operation to each voxel within the brain's imaging volume:
\[ A[i, j, k] = \mathbf{\alpha}_s[i] \cdot \mathbf{\alpha}_c[j] \cdot \mathbf{\alpha}_a[k] \]
For every voxel at coordinates \([i, j, k]\), its attention value is derived from the product of the corresponding sagittal, coronal, and axial attention weights. This process retains and amplifies the diagnostic importance perceived across all three anatomical planes, yielding a comprehensive 3D depiction of attention distribution.

To facilitate interpretation and comparison, we normalize the entire 3D attention map in the range [0, 1], ensuring that the highest values correspond to areas of high diagnostic relevance. We used min-max normalization:
\[ A = \frac{A - \min(A)}{\max(A) - \min(A)} \]
The resulting map highlights the brain regions most significantly associated with the diagnostic output of the network, offering insights into the pathological hallmarks of AD as learned by the model through its training process.

\subsubsection{Brain regions importance quantification}
\label{sec:xai_metrics}

To achieve comparability across subjects, each MRI image, denoted by \( I \), was normalized to the MNI 152 standard space using the transformation function \( f_{MNI152} \), thus \( I_{norm} = f_{MNI152}(I) \). This step is crucial for aligning brain structures across different individuals to facilitate the identification of AD-specific biomarkers.
\newline \newline \noindent
\textbf{Generation of Binary Heatmap} A binary heatmap, \( H_{binary} \), was generated to isolate regions of significant structural patterns associated with AD, utilizing a threshold \( \theta \) set at the 99.9th percentile. The binary heatmap is defined as:

\[
H_{\text{binary}}[i,j,k] = 
\begin{cases} 
1, & \text{if } A[i,j,k] > \theta \\
0, & \text{otherwise}
\end{cases}
\]
\newline \newline \noindent
\textbf{Overlay of Heatmap on MRI Data} For visualization purposes, the MRI data \( I_{norm} \) was augmented by overlaying \( H_{binary} \) to enhance the saliency of the regions implicated in AD:

\[
I_{XAI} = I_{norm} + H_{\text{binary}} \times \delta
\]
where \( \delta \) is an amplification factor set to 10 to increase the prominence of the heatmap overlay.
\newline \newline \noindent
\textbf{Identification and Analysis of Regions of Interest} Employing a template atlas, we determined regions of interest (ROI) within the brain. For each ROI identified with a unique label \( r \), we computed the overlap \( O_r \) as follows:

\[
O_r = \sum_{i,j,k} (H_{\text{binary}}[i,j,k] > 0) \land (T_{atlas}[i,j,k] = r)
\]
where \(T_{atlas}\) represents the template atlas data. 
The quantitative evaluation of each region's overlap involved several statistical measures calculated from the MRI data. The overlap volume of the heatmap in a given region is:
\[V_r = |O_r|\]
which can be useful to indicate predominant regions in decision process. The mean (\( \mu_r \)) intensity of the heatmap within each identified region is given by

\[
\mu_r = \frac{1}{V_r} \sum_{(i,j,k) \in O_r} A[i,j,k],
\]
which represents the average activity level across the voxels of interest, allowing us to pinpoint regions with consistently high AD-related changes. The standard deviation (\( \sigma_r \)) of the heatmap intensities,
\[
\sigma_r = \sqrt{\frac{1}{V_r-1} \sum_{(i,j,k) \in O_r} (A[i,j,k] - \mu_r)^2},
\]
provides insight into the variability of attention within the region, indicating the heterogeneity of AD impact across different brain areas. The maximum (\( A_{max,r} \)) and minimum (\( A_{min,r} \)) values of the heatmap,
\[
A_{\text{max},r} = \max_{(i,j,k) \in O_r} A[i,j,k]
\]
and
\[
A_{\text{min},r} = \min_{(i,j,k) \in O_r} A[i,j,k],
\]
respectively, highlighting the extremes in activation within the regions, shedding light on the most and least affected areas within the brain's AD-implicated regions. Lastly, the percentage of overlap (\( P_r \)) is calculated as
\[
P_r = \frac{\sum_{i,j,k} H_{\text{binary}}[i,j,k]}{V_r}
\]
to provide a single value that quantifies the portion of a specific brain region involved in the AD-related patterns identified by the network. This measure is a critical indicator of the importance of each region in distinguishing AD patients from CN individuals.

\section{Experiment results}
In this section, we first present the experimental setup developed to validate the method. Then, we present the results of the model in comparison to recent state-of-the-art attentional models on 2D slices. Finally, we present qualitative and quantitative results of our XAI approach compared to other existing XAI methods.

\subsection{Experiment Setup}
\label{sec:experiments}
We implemented all considered methods in PyTorch \citep{paszke2019pytorch} and trained them on an NVIDIA A100 80GB GPU. The pre-processed input 3D images (see Section \ref{sec:preproc}) were converted to a sequence of \(N\) slices at runtime. The \(N\) slices were selected from the center by a slicing operation from the chosen plane. Specifically, given D, the dimension corresponding to the number of slices along the slicing direction, the selected slices belonged to the interval \([D/2 - N/2, D/2 + N/2]\). This slicing operation resulted in \(N < D\) slices, thus reducing computational complexity with \(N\) considered as an optimization hyperparameter. All tested networks were trained with AdamW Optimizer with base learning rate \(1 \times 10^{-4}\) and weight decay \(1 \times 10^{-2}\). An early stopping strategy selected the best model during the training phase on the validation set with patience 15. The training data augmentation strategy consisted of randomly flipping a 2D slice in the series with 0.3 probability. We implemented a 5-fold cross-validation technique to ensure robustness and generalizability of the results. We divided the dataset into five segments, ensuring each segment acted as a test set in one iteration and as part of the training/validation sets in the others. To prevent data leakage and preserve the integrity of our evaluation, we carefully assigned all 3D images from the same subjects exclusively to one of the training, validation, or test sets. In each fold, the 80\% portion allocated for training/validation was further divided using an 80-20 ratio.

\subsubsection{Performance Evaluation}
We used Accuracy (ACC), Sensitivity (SEN), Specificity (SPE), and Matthews Correlation Coefficient (MCC) as performance metrics. ACC is the most common performance measure for classification, and it consists of the percentage of correctly classified samples over the total. MCC \citep{mcc_chicco2020advantages} is a correlation coefficient between predictions and true labels; it provides a more informative and truthful score than accuracy when evaluating binary classifications, allowing a more realistic interpretation of classifier performance, especially in the case of unbalanced datasets. SEN and SPEC serve as critical metrics for evaluating the diagnostic accuracy of our model, highlighting its ability to correctly identify positive and negative cases, respectively. In AD vs. CN task, the AD samples are in the positive class, and CN samples are in the negative one. In sMCI vs. pMCI task, the sMCI samples are in the positive class, and pMCI samples are in the negative one.

All the results obtained are averaged on the test sets from the 5-fold cross-validation.

\subsection{Diagnostic results}
\label{sec:diagnostic_results}
This subsection focuses on our diagnosis network’s results, summarizing the evaluations of various 2D convolutional backbones, the impact of different slicing planes, optimization of network parameters, the effectiveness of double transfer learning, and a comparison with state-of-the-art methods. Almost all experiments use the axial slicing plane due to its common use in clinical applications and because it generally yields better results.
\subsubsection{2D backbone}
\label{sec:2D backbone}
As the first step, we analyzed the effectiveness of different convolutional backbones in the AD vs. CN task. To this aim, two sets of experiments were performed. First, we evaluated the performance of different backbones by considering them in the Feature Extraction Module of our network. Second, to allow an evaluation independent of our method, the best networks for each family were evaluated with a 2D Majority Voting approach, which is also considered one of the baseline methods. In the Majority Voting approach, we train 2D convolutional networks by individually labelling the slices. Consequently, the weights are updated from the slice-level error during training. Then, the predictions of the slices of a single 3D image are aggregated by choosing the most recurring one in the sequence as the final label.

\begin{table}[ht]
\centering
\caption{Performances varying backbone in Feature Extraction module in AD vs. CN task averaged over 5-fold cross-validation}
\label{tab:our_backbone_selection}
\vspace{5pt}
\footnotesize 
\begin{spacing}{1.5}
\begin{tabular}{p{0.2\columnwidth}>{\centering}p{0.14\columnwidth}>{\centering}p{0.14\columnwidth}>{\centering}p{0.14\columnwidth}>{\centering\arraybackslash}p{0.14\columnwidth}}
\toprule
Backbones      & ACC & SPE & SEN & MCC \\ \midrule
VGG16                & \textbf{0.829}            & 0.839               & \textbf{0.816}               & \textbf{0.655}       \\ 
VGG19                & \underline{0.819}            & \textbf{0.875}               & \underline{0.749}               & \underline{0.633}       \\
ResNet34             & 0.770            & 0.843               & 0.681               & 0.534       \\
ResNet50             & 0.804            & \underline{0.860}               & 0.734               & 0.602 \\
ResNet101            & 0.785            & 0.822               & 0.739               & 0.563 \\
EfficientNetV2S      & 0.764            & 0.805               & 0.713               & 0.521       \\
EfficientNetV2M      & 0.764            & 0.833               & 0.679               & 0.520       \\
DenseNet121          & 0.798            & 0.841               & 0.745               & 0.590       \\
\bottomrule
\end{tabular}
\end{spacing}
\normalsize
\end{table}

The first pool of experiments was conducted using a mini-batch of 16 and freezing the first 50\% of the backbone's layers. The convolutional architectures chosen are: (i) VGG16 and VGG19 \citep{vgg_SimonyanZ14a}, (ii) ResNet34, ResNet50 and ResNet101 \citep{resnet_He2016}, (iii) EfficientNetV2 Small and EfficientNetV2 Medium \citep{effnet_tan2021efficientnetv2} and (iv) DenseNet121 \citep{huang2017densely}.
The results of this pool are shown in Table \ref{tab:our_backbone_selection}. In this first analysis, VGG16 had the highest accuracy (ACC: 0.829) and sensitivity (SEN: 0.816), suggesting it is the better-suited backbone for AD patient identification. VGG19, while slightly less accurate, showed superior specificity (SPE: 0.875), indicating it is particularly adept at identifying CN cases. The ResNet series demonstrated that additional depth (as seen in ResNet101) does not equate to improved performance, with ResNet50 showing better accuracy and sensitivity. This could suggest diminishing returns with increased complexity for this task. EfficientNet models trailed behind their counterparts with identical accuracies (ACC: 0.764). DenseNet121, while providing good accuracy (ACC: 0.798), also did not meet the higher results obtained by other architectures. The MCC aligns with these findings, confirming VGG16 as the top-performing model in this case.

Based on the results obtained in Table \ref{tab:our_backbone_selection}, we selected the best backbone from each architecture family as candidates for the second pool of experiments. Since this pool comprises 2D rather than 3D images, we chose 32 as the mini-batch size. To achieve comparable results, the same percentage of backbone layers was frozen. As shown in Table \ref{tab:majority_voting}, the performance trends of the architectures remain consistent with previous observations. VGG16 continued to lead in accuracy (ACC: 0.804) and MCC (MCC: 0.605). This result aligns with other works that design their approaches relying on a VGG as base network \citep{2d_slice_altay2021preclinical, 2d_slice_hu2023conv}. Remarkably, our method outperforms the Majority Voting one across several metrics. The superiority can be seen from the average increase across all tested backbones of ACC by 1.6\% and MCC by 3.08\%. Notably, a good increase was observed with VGG16, with an ACC improvement of 2.5\% and an MCC improvement of 5.0\%.

\begin{table}[ht]
\centering
\caption{Performances varying backbone in Majority Voting approach in AD vs. CN task over 5-fold cross-validation.}
\label{tab:majority_voting}
\vspace{5pt}
\footnotesize 
\begin{spacing}{1.5}
\begin{tabular}{p{0.2\columnwidth}>{\centering}p{0.14\columnwidth}>{\centering}p{0.14\columnwidth}>{\centering}p{0.14\columnwidth}>{\centering\arraybackslash}p{0.14\columnwidth}}
\toprule
Backbones      & ACC & SPE & SEN & MCC \\ \midrule
VGG16          & \textbf{0.804} & \textbf{0.897} & \underline{0.688} & \textbf{0.605}       \\
ResNet50       & \underline{0.798} & \underline{0.873}  & \textbf{0.707} & \underline{0.592}       \\
EfficientNetV2S & 0.759           & 0.839               & 0.664               & 0.516       \\
DenseNet121     & 0.770           & 0.848               & 0.673               & 0.532       \\
\bottomrule
\end{tabular}
\end{spacing}
\normalsize
\end{table}

\subsubsection{Slicing plane}

Although the axial plane is the most commonly used in clinical applications, there could be better choices than this one in this classification scenario. We performed a slicing plane analysis after finding that VGG16 is the best backbone in the diagnostic task. For this set of experiments, we used 100 slices for each view, a batch size of 8, a learning rate 1e-4, and froze the first half of the backbone in training. Table \ref{tab:slicing_plane} shows the performances averaged over the five cross-validation test sets varying the slicing plane. The results show that the axial plane achieved the best performance. This finding aligns with the clinical choice.

\begin{table}[ht]
\centering
\caption{Performances varying slicing plane with our approach in AD vs. CN task over 5-fold cross-validation.}
\label{tab:slicing_plane}
\vspace{5pt}
\footnotesize 
\begin{spacing}{1.5}
\begin{tabular}{p{0.2\columnwidth}>{\centering}p{0.14\columnwidth}>{\centering}p{0.14\columnwidth}>{\centering}p{0.14\columnwidth}>{\centering\arraybackslash}p{0.14\columnwidth}}
\toprule
Slicing plane      & ACC & SPE & SEN & MCC \\ \midrule
Axial              & \textbf{0.839} & \underline{0.872} & \textbf{0.799} & \textbf{0.674} \\
Coronal            & 0.816 & \textbf{0.897} & 0.715 & 0.628  \\
Sagittal           & \underline{0.820} & 0.860 & \underline{0.772} & \underline{0.636}  \\
\bottomrule
\end{tabular}
\end{spacing}
\normalsize
\end{table}

\subsubsection{Parameters Optimization}
Since VGG16 was confirmed as the best backbone among those under investigation and the axial plane provided the best results, we selected this configuration for our Feature Extraction module. The second step consisted of optimizing other parameters: the number of slices extracted \(N\), the batch size, and the percentage of backbone to freeze.

\begin{table*}[t]
\centering
\caption{Network Results using VGG16 as the backbone for Feature Extraction module in AD vs. CN task}
\label{tab:optimization_our_net}
\vspace{5pt}
\footnotesize
\begin{spacing}{1.5}
\begin{tabular}{>{\centering}p{0.1\textwidth}>{\centering}p{0.1\textwidth}>{\centering}p{0.1\textwidth}>{\centering}p{0.1\textwidth}>{\centering}p{0.1\textwidth}>{\centering}p{0.1\textwidth}>{\centering\arraybackslash}p{0.1\textwidth}}
\toprule
Num Slices & Batch Size & Freezing & ACC & SPE & SEN & MCC \\
\midrule
70                  & 8                  & 50\%   & 0.821 & 0.851 & 0.782 & 0.636 \\
\midrule
\multirow{6}{*}{80} & 4                  & 50\%   & 0.825 & 0.867 & 0.774 & 0.646 \\
\cline{2-7}
                    & \multirow{4}{*}{8} & 0\%   & 0.818 & 0.861 & 0.764 & 0.630 \\
                    &                    & 25\%  & \underline{0.840} & \underline{0.892} & 0.776 & \underline{0.677} \\
                    &                    & 50\%  & \textbf{0.856} & \textbf{0.910} & \underline{0.792} & \textbf{0.712} \\
                    &                    & 75\%  & 0.799 & 0.877 & 0.707 & 0.603 \\
\cline{2-7}
                    & 16                 & 50\%   & 0.829 & 0.839 & \textbf{0.816} & 0.655 \\
\midrule
90                  & 8                  & 50\%   & 0.818 & 0.867 & 0.757 & 0.630 \\
100                 & 8                  & 50\%   & 0.839 & 0.872 & 0.799 & 0.674 \\
120                 & 8                  & 50\%   & 0.815 & 0.858 & 0.761 & 0.624 \\
\bottomrule
\end{tabular}
\end{spacing}
\normalsize
\end{table*}

The results in Table \ref{tab:optimization_our_net} indicated that varying the number of slices and the percentage of first backbone layers frozen in the network impacted performances significantly. The configuration with 70 slices, batch size of 8, and freezing 50\% of the layers resulted in an ACC of 0.821 and MCC of 0.636. When the batch size was reduced to 4 under the same conditions, a slight improvement was observed with an ACC of 0.825 and MCC of 0.646. After increasing the number of slices to 80 and employing a batch size of 8, experiments with 0\%, 25\%, 50\%, and 75\% freezing were conducted. The highest accuracy was achieved at 50\% freezing, yielding an ACC of 0.856 and MCC of 0.712, highlighting the best overall performance across all configurations. With a further increase to 90 slices, the performance did not improve, maintaining an ACC of 0.818 with a 50\% freezing and batch size of 8. Finally, the performance slightly fluctuated with 100 and 120 slices but generally did not surpass the optimal results obtained with 80 slices end 50\% freezing.

\begin{table}[ht]
\centering
\caption{Double transfer learning for sMCI vs. pMCI task at 36 months}
\label{tab:double_trans}
\vspace{5pt}
\footnotesize 
\begin{spacing}{1.5}
\begin{tabular}{p{0.3\columnwidth}>{\centering}p{0.1\columnwidth}>{\centering}p{0.07\columnwidth}>{\centering}p{0.07\columnwidth}>{\centering}p{0.07\columnwidth}>{\centering\arraybackslash}p{0.07\columnwidth}}
\toprule
Transfer Learning & Freezing & ACC & SPE & SEN & MCC \\
\midrule
ImageNet                & 50\% & 0.4822 & 0.427 & \underline{0.636} & 0.065 \\
ImageNet + AD vs. CN    & 50\% & \underline{0.703} & \textbf{0.809} & 0.4873 & \underline{0.396} \\
ImageNet + AD vs. CN    & 75\% & \textbf{0.725} & 0.763 & \textbf{0.678} & \textbf{0.443} \\

\bottomrule
\end{tabular}
\end{spacing}
\normalsize
\end{table}
\subsubsection{Double Transfer Learning}
The third step involved investigating the utility of transfer learning for predicting AD progression with the double transfer learning strategy proposed in Section \ref{sec:double_trans}. This approach leverages a backbone pre-trained on the ImageNet dataset, further fine-tuned through a subsequent task distinguishing AD from CN individuals. As mentioned in Section \ref{sec:double_trans}, it is crucial to note that the patient cohorts classified as AD or CN differ from those labeled as sMCI or pMCI, given that the latter are diagnosed as MCI. Consequently, this strategy does not cause data leakage problems.

First, we fine-tuned our model with VGG16 pre-trained on ImageNet in the Feature Extraction module on the entire AD vs. CN dataset. We froze the first 50\% of the layers as this configuration provided the best results in Table \ref{tab:optimization_our_net}. The final model pre-trained on the AD vs. CN task was selected using 10\% of the dataset as a validation set. This step provides a model in which the first convolutional part is highly specialized in extracting fine-grained low-level feature representations and the rest in analyzing AD-related high-level patterns. Second, we ulteriorly fine-tuned the AD pre-trained model on the sMCI vs. pMCI classification task at a 36-month forecast interval. The learning rate for this pool of experiments was reduced to \(1 \times 10^{-5}\) to fine-tune the pre-trained model weights, ensuring that the disease knowledge is preserved during the training process. Table \ref{tab:double_trans} demonstrates the efficacy of this approach in this task. With baseline ImageNet transfer learning and 50\% of the network layers frozen, the model achieved an ACC of 0.4822 and an MCC of 0.065. The MCC value of 0.065 indicates that the model could not find useful correlations to distinguish the two cases. Incorporating AD vs. CN knowledge with a 50\% freezing resulted in a marked enhancement across metrics: an ACC of 0.703 and an MCC of 0.396. Advancing to a 75\% freezing of the network layers under the double transfer learning paradigm, we observed a peak performance of an ACC of 0.725, SPE of 0.763, SEN of 0.678, and an MCC of 0.443.

\subsubsection{Comparison to state-of-the-art methods}

\begin{table*}[!h]
\centering
\setlength{\tabcolsep}{5pt}
\caption{Comparison of performance against other approaches in AD vs. CN task and sMCI vs pMCI}
\label{tab:comparison}
\vspace{5pt}
\footnotesize
\begin{spacing}{1.5}
\begin{tabular}{p{0.26\textwidth}@{\extracolsep{5pt}}>{\centering}p{0.064\textwidth}>{\centering}p{0.064\textwidth}>{\centering}p{0.064\textwidth}>{\centering}p{0.064\textwidth}@{\extracolsep{5pt}}>{\centering}p{0.064\textwidth}>{\centering}p{0.064\textwidth}>{\centering}p{0.064\textwidth}>{\centering\arraybackslash}p{0.064\textwidth}@{\extracolsep{5pt}}}
\toprule
\multirow{2}{*}{Networks} & \multicolumn{4}{c}{AD vs. CN} & \multicolumn{4}{c}{sMCI vs. pMCI} \\
\cline{2-5} \cline{6-9}
                        & ACC   & SPE   & SEN   & MCC   & ACC   & SPE   & SEN   & MCC   \\
\midrule

Attention Transformer \citep{2d_slice_altay2021preclinical}   & 0.826 & \textbf{0.914} & 0.717 & 0.651 & 0.623 & 0.665 & 0.4873 & 0.238 \\
AwareNet Diagnosis \citep{2d_slice_wang2024joint}      & 0.832 & 0.875 & 0.778 & 0.659 & 0.4841     & \textbf{0.774}     & 0.258     & 0.039     \\
Ours                    & \textbf{0.856} & \underline{0.910} & \underline{0.792} & \textbf{0.712} & \textbf{0.725} & \underline{0.763} & \textbf{0.678} & \textbf{0.443}\\
Majority Voting     & 0.804 & 0.897 & 0.688 & 0.605 & 0.614 & 0.601 & 0.629 & 0.229 \\
Attention-Guided Majority Voting & \underline{0.843} & 0.894 & 0.780 & \underline{0.683} & \underline{0.633} & 0.624 & \underline{0.643} & \underline{0.266} \\
Majority Voting 3D & 0.836 & 0.867 & \textbf{0.797} & 0.667 & 0.629 & 0.653 & 0.601 & 0.254 \\
\bottomrule
\end{tabular}
\end{spacing}
\normalsize
\end{table*}

We performed a comparison with current 2D slice attention-based state-of-the-art models. These methods include Attention Transformer \citep{2d_slice_altay2021preclinical} and AwareNet \citep{2d_slice_wang2024joint}. Attention Transformer uses a multi-head self-attention to perform a cross-attention operation resulting in a slice feature maps fusion. AwareNet represents the diagnosis network of a joint learning framework that uses a slice-aware module and a slice-shift module. The slice-aware module leverages the attention mechanism to interpret significant slices and regions. While we have utilized the official implementation of AwareNet, as it is available and provided by the authors, we have implemented the architecture for the Attention Transformer ourselves based on the paper details, as the authors did not provide the official implementation. For both methods, the hyperparameters were optimized to maximize performance. As in our case, Attention Transformer obtained the best results by freezing 50\% of the base network pre-trained on ImageNet and a batch size of 8. The AwareNet network was trained from scratch with a batch size of 2. The network is initialized by the Kaiming method \citep{kaiming_he2015delving} and trained using the Adam optimization algorithm with \(\beta_{1} = 0.48\) and \(\beta_{2} = 0.999\). In both cases, the learning rate was set to \(1 \times 10^{-5}\) to prevent overfitting. 

In the sMCI vs. pMCI task, we adopted the double transfer learning strategy proposed in this work for the Attention Transformer method. The first 75\% of the Attention Transformer base network layers were frozen. Since the AwareNet is trained from scratch on the AD vs. CN task, we froze the first 50\% of the backbone. 

The results of our comparative analysis are summarized in Table \ref{tab:comparison}. To assess the effectiveness of our attentional module in enhancing diagnostic performance, we introduced three additional approaches: Majority Voting, Attention-Guided Majority Voting, and Majority Voting 3D.
We detailed the Majority Voting approach implemented in Section \ref{sec:2D backbone}.
The Attention-Guided Majority Voting approach utilizes the network trained in the majority voting case. However, predictions are made during testing on a subset of slices identified as most informative by our attentional fusion module. Specifically, we identified a contiguous range where the attention values exceeded the 75th percentile. This range was calculated for the axial plane by averaging the attention weight distribution across the five validation sets of the fold cross-validation, resulting in the range [0, 25]. The results of this method validate the module's ability to select feature maps with high information content.
Finally, the 3D Majority Voting approach replaces the attentional fusion module with an averaging module, where feature maps from individual slices are averaged to form a single feature map for making subject-level decisions. This approach further helps to evaluate the impact of slices’ importance weighting compared to considering all slices as equally important. 

Our model demonstrates effectiveness in disease diagnosis and prediction tasks, achieving (i) an accuracy (ACC) of 0.856 and a Matthews correlation coefficient (MCC) of 0.712 in the AD vs. CN comparison, and (ii) an ACC of 0.725 and MCC of 0.443 in the sMCI vs. pMCI task.
While AwareNet shows promise in the diagnosis task with an ACC of 0.832 and an MCC of 0.659, it struggles in the disease prediction task, yielding an MCC of 0.039, likely due to insufficient data for training. On the other hand, Attention Transformer, leveraging double transfer learning, provides a more balanced alternative, with an MCC of 0.651 in the diagnosis task and an MCC of 0.238 in disease prediction. However, its increased parameterization from self-attention prevents it from bridging the gap with lightweight models like ours and AwareNet in the diagnosis task.
Majority voting, lacking a mechanism to learn correlations between sections, underperforms compared to other methods. Incorporating subject-level error evaluation through the Majority Voting 3D approach the MCC improves from 0.605 to 0.667, even surpassing overfitting-prone approaches such as AwareNet and Attention Transformer.
Furthermore, leveraging our attentional mechanism to identify informative slices while using networks pre-trained without considering slice relationships yields an MCC of 0.683. This result underscores our model's ability to select crucial slices for decision-making.

\subsection{XAI results}

\begin{figure*}[!h]
    \centering
    \captionsetup[subfigure]{labelformat=empty}
    \begin{subfigure}{0.33\textwidth}
        \caption{Axial}
        \includegraphics[width=\linewidth]{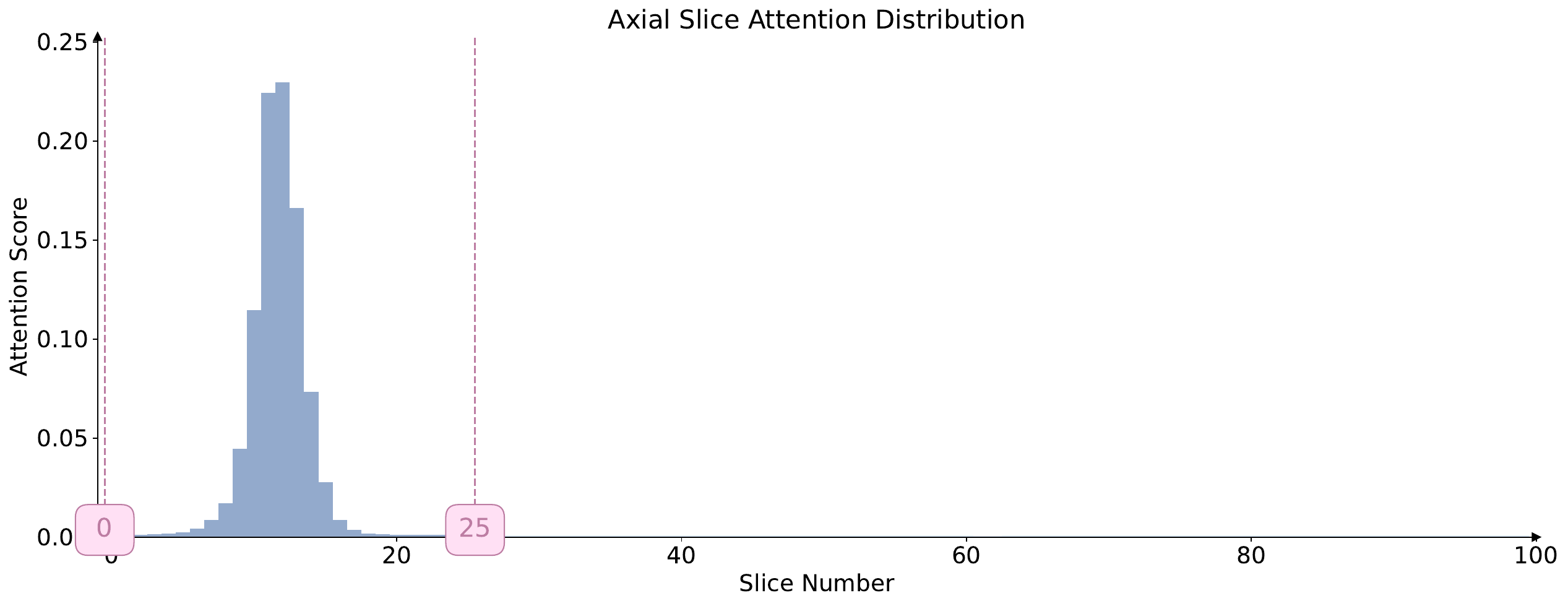}
        \caption{}
    \end{subfigure}
    \hfill
    \begin{subfigure}{0.33\textwidth}
        \caption{Coronal}
        \includegraphics[width=\linewidth]{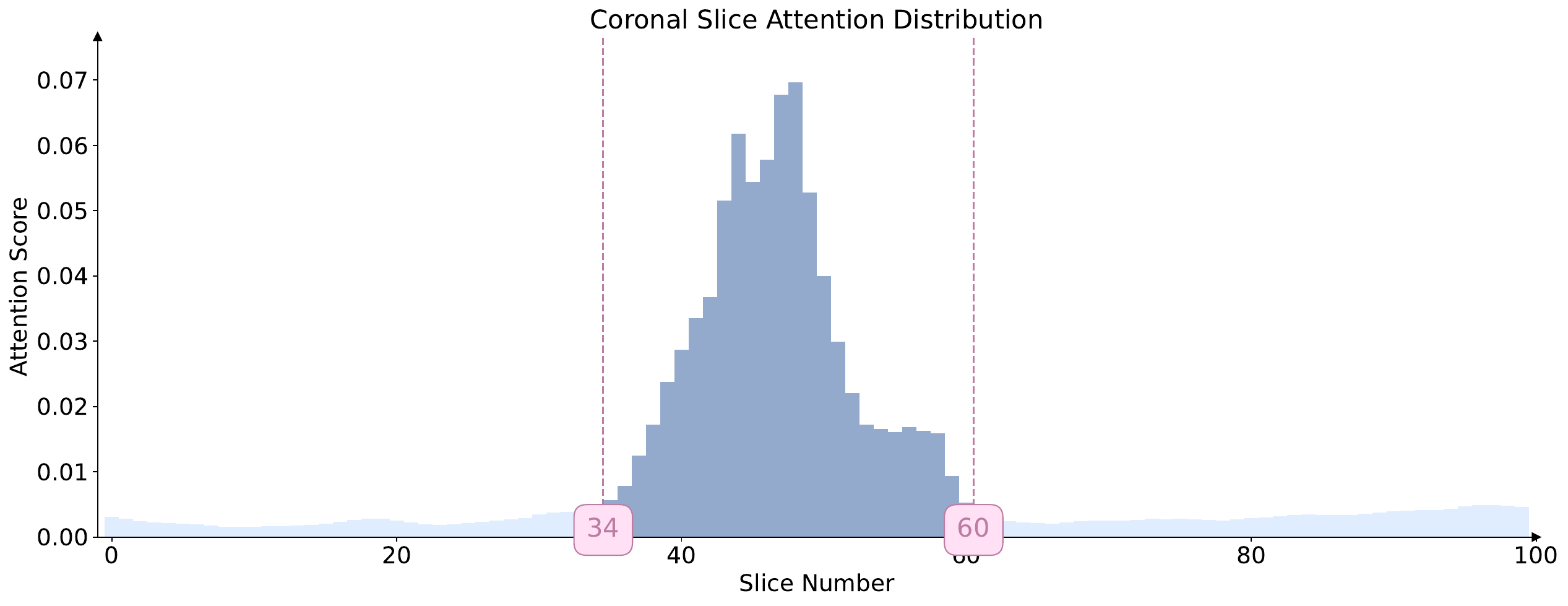}
        \caption{Fold 1}
    \end{subfigure}
    \hfill
    \begin{subfigure}{0.33\textwidth}
        \caption{Sagittal}
        \includegraphics[width=\linewidth]{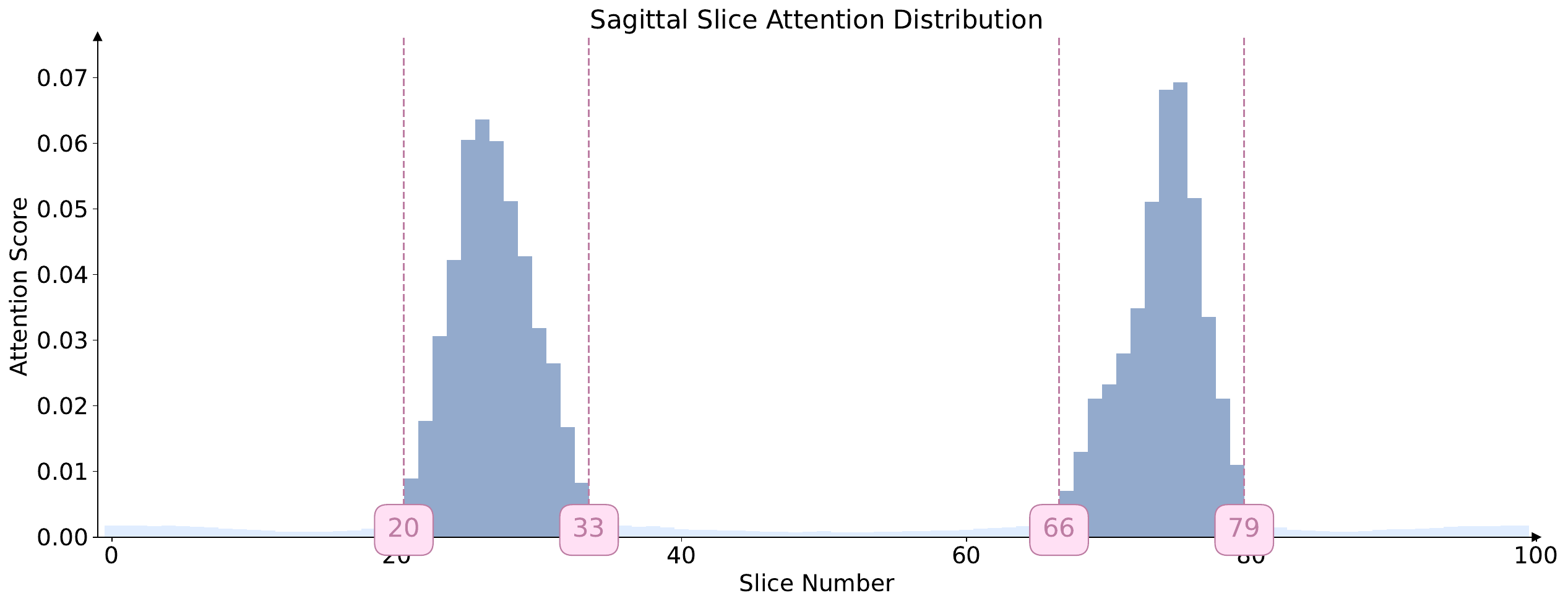}
        \caption{}
    \end{subfigure}
    \hrulefill

    \begin{subfigure}{0.33\textwidth}
        \includegraphics[width=\linewidth]{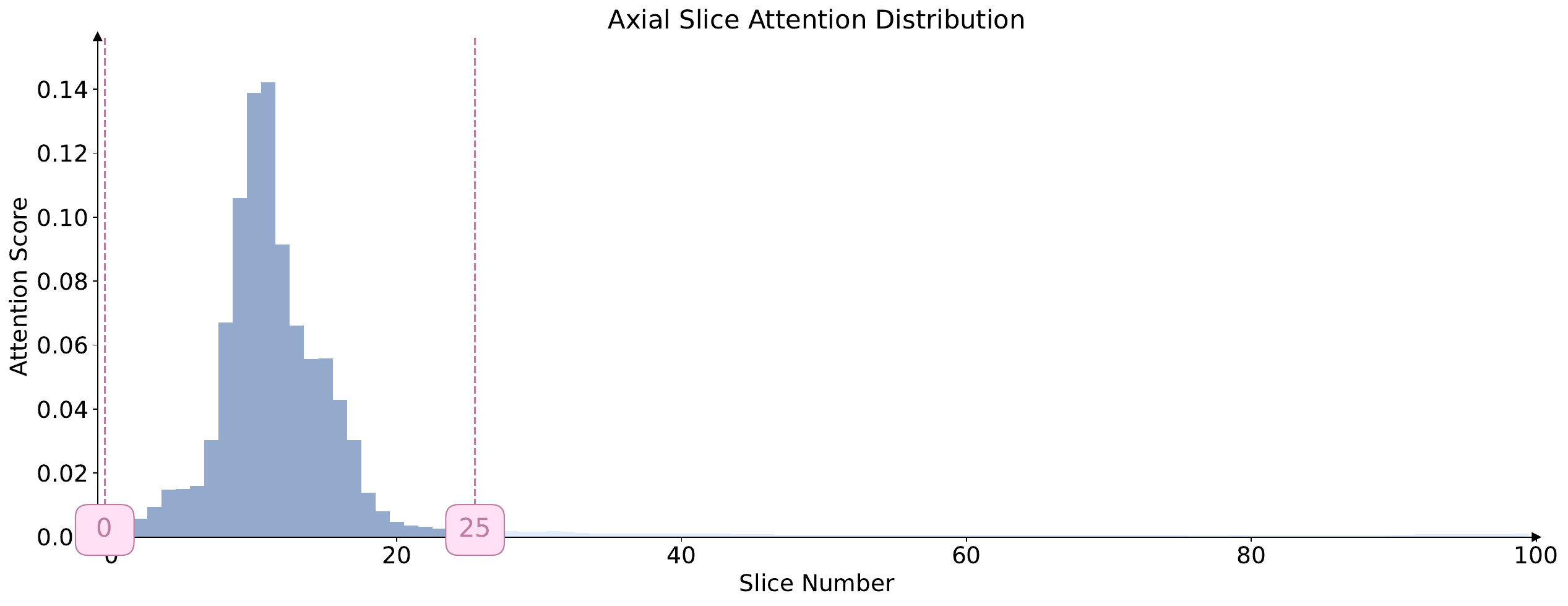}
        \caption{}
    \end{subfigure}
    \hfill
    \begin{subfigure}{0.33\textwidth}
        \includegraphics[width=\linewidth]{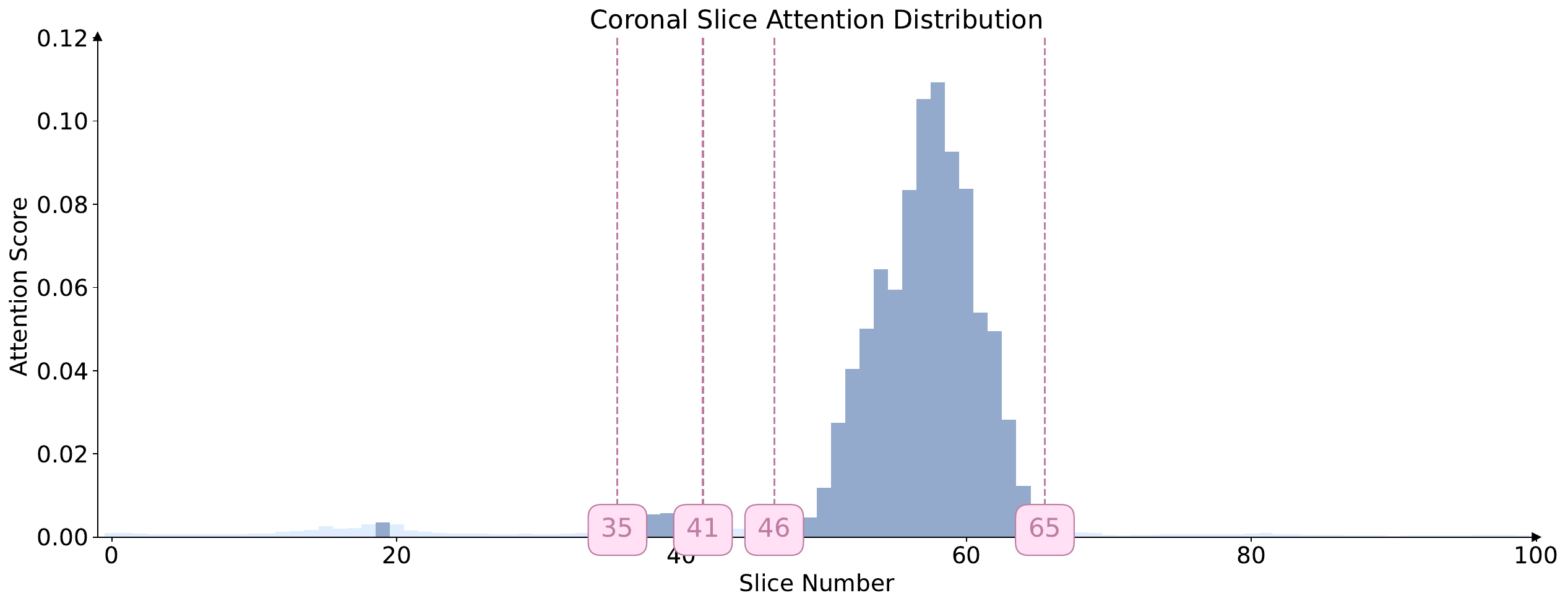}
        \caption{Fold 2}
    \end{subfigure}
    \hfill
    \begin{subfigure}{0.33\textwidth}
        \includegraphics[width=\linewidth]{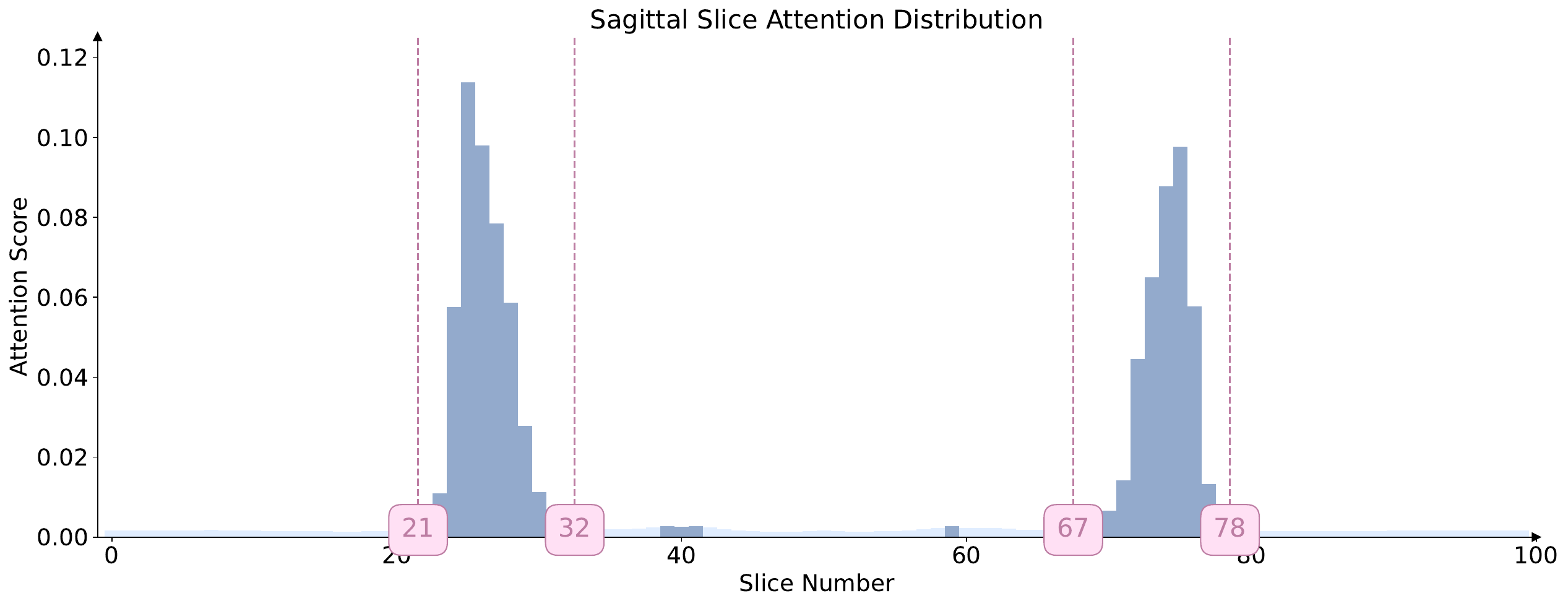}
        \caption{}
    \end{subfigure}
    \hrulefill
    
    \begin{subfigure}{0.33\textwidth}
        \includegraphics[width=\linewidth]{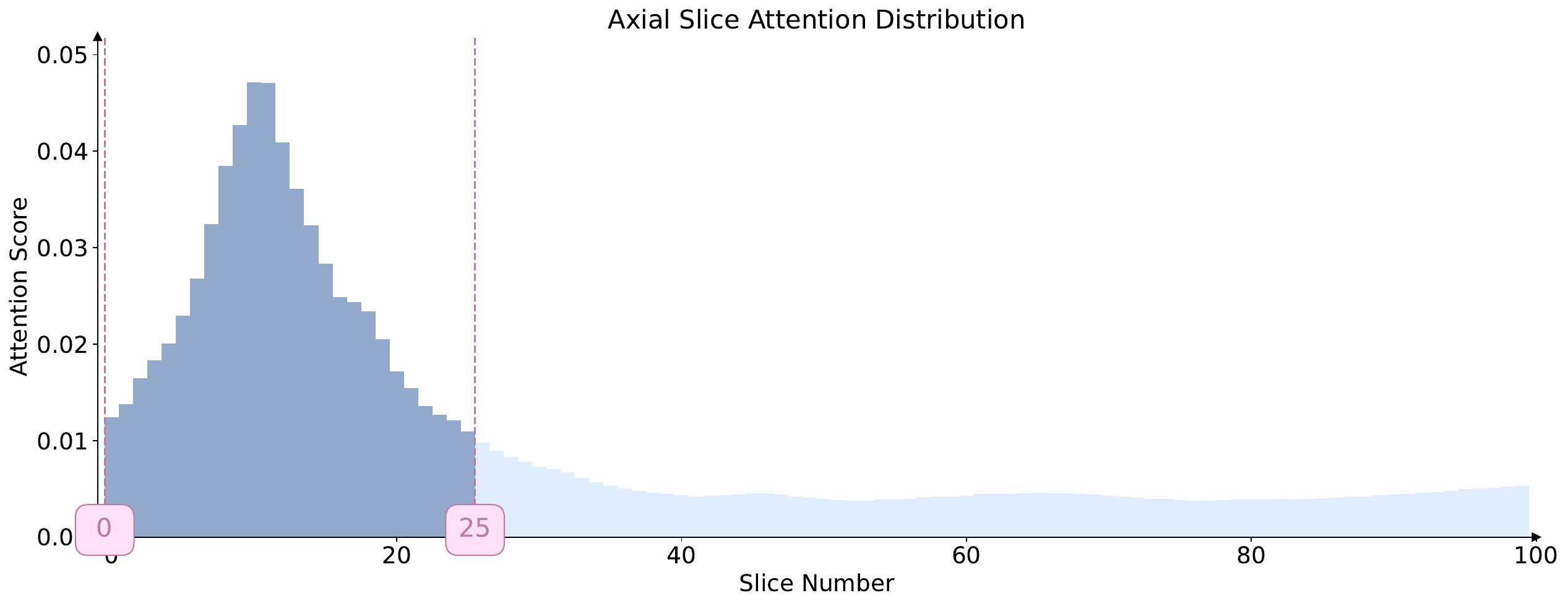}
        \caption{}
    \end{subfigure}
    \hfill
    \begin{subfigure}{0.33\textwidth}
        \includegraphics[width=\linewidth]{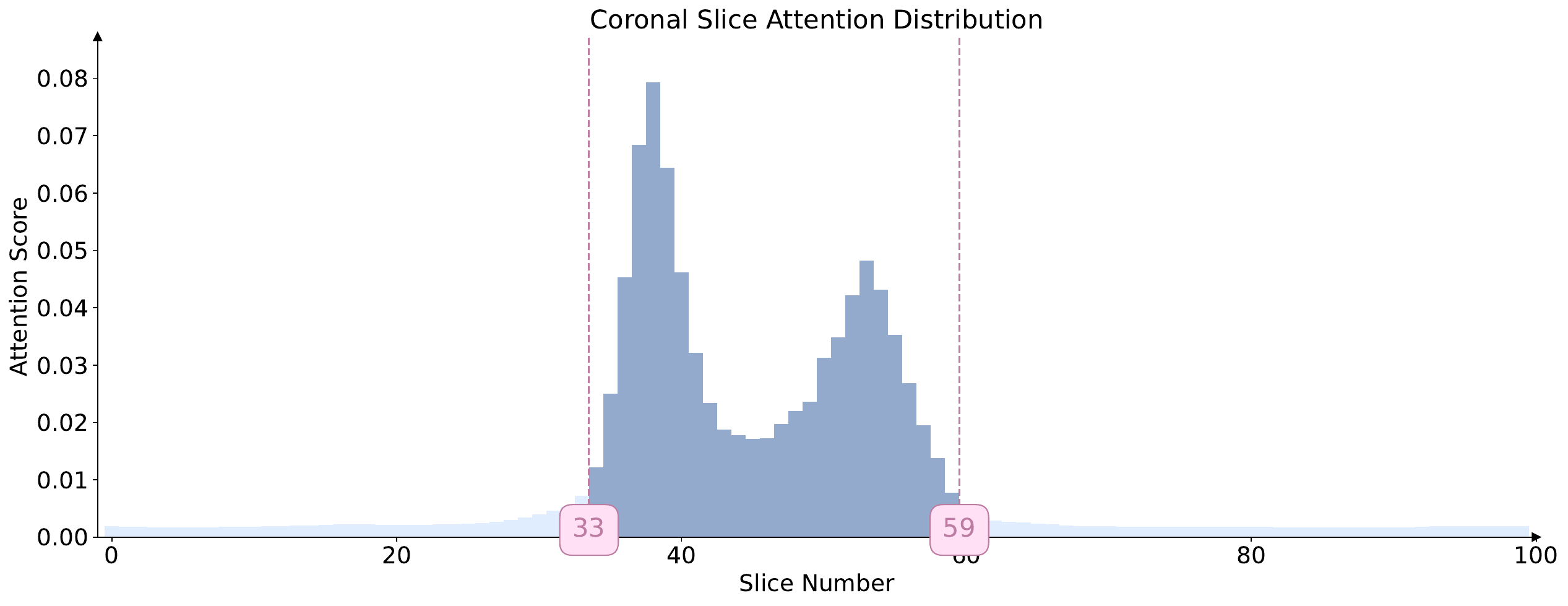}
        \caption{Fold 3}
    \end{subfigure}
    \hfill
    \begin{subfigure}{0.33\textwidth}
        \includegraphics[width=\linewidth]{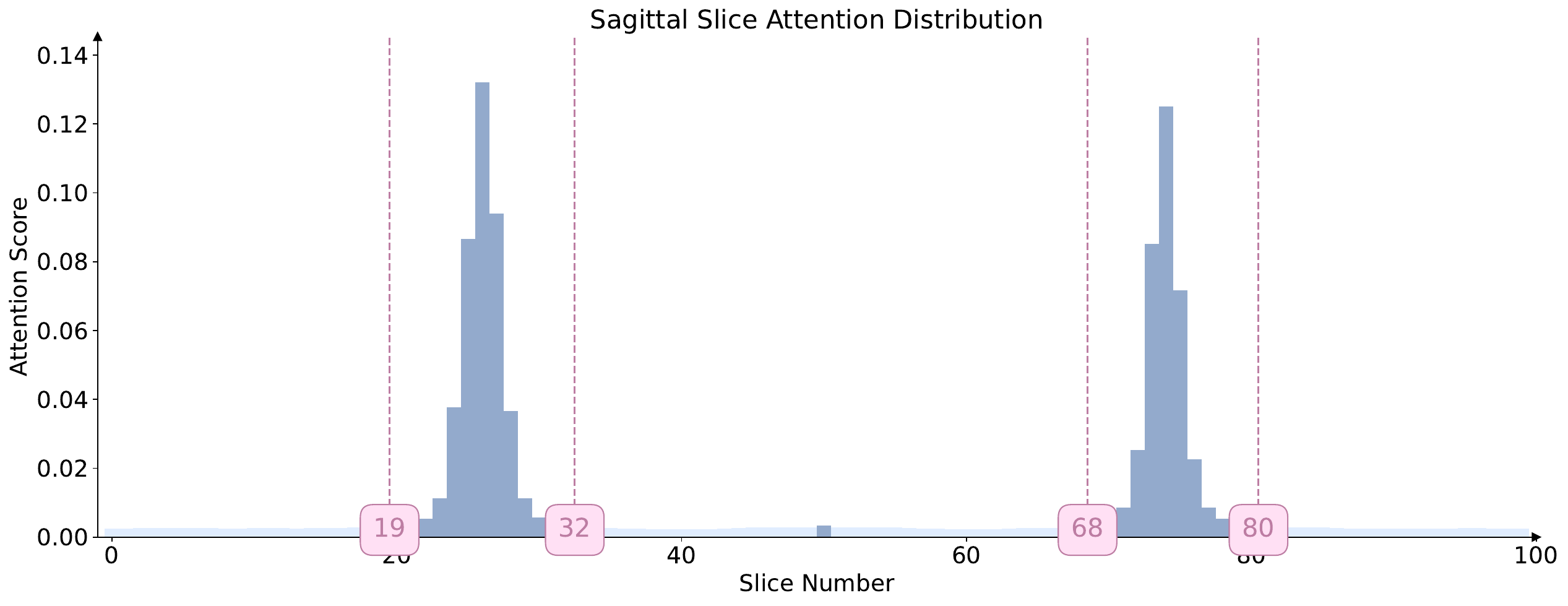}
        \caption{}
    \end{subfigure}
    \hrulefill

    \begin{subfigure}{0.33\textwidth}
        \includegraphics[width=\linewidth]{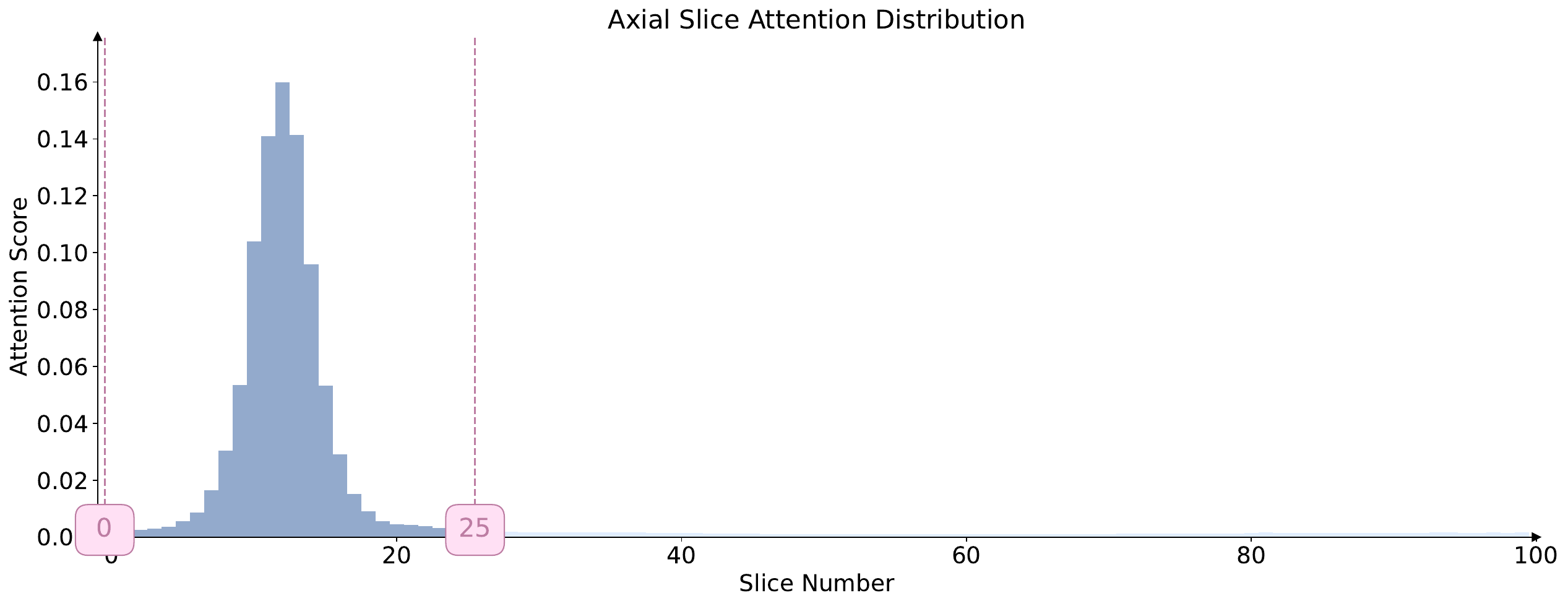}
        \caption{}
    \end{subfigure}
    \hfill
    \begin{subfigure}{0.33\textwidth}
        \includegraphics[width=\linewidth]{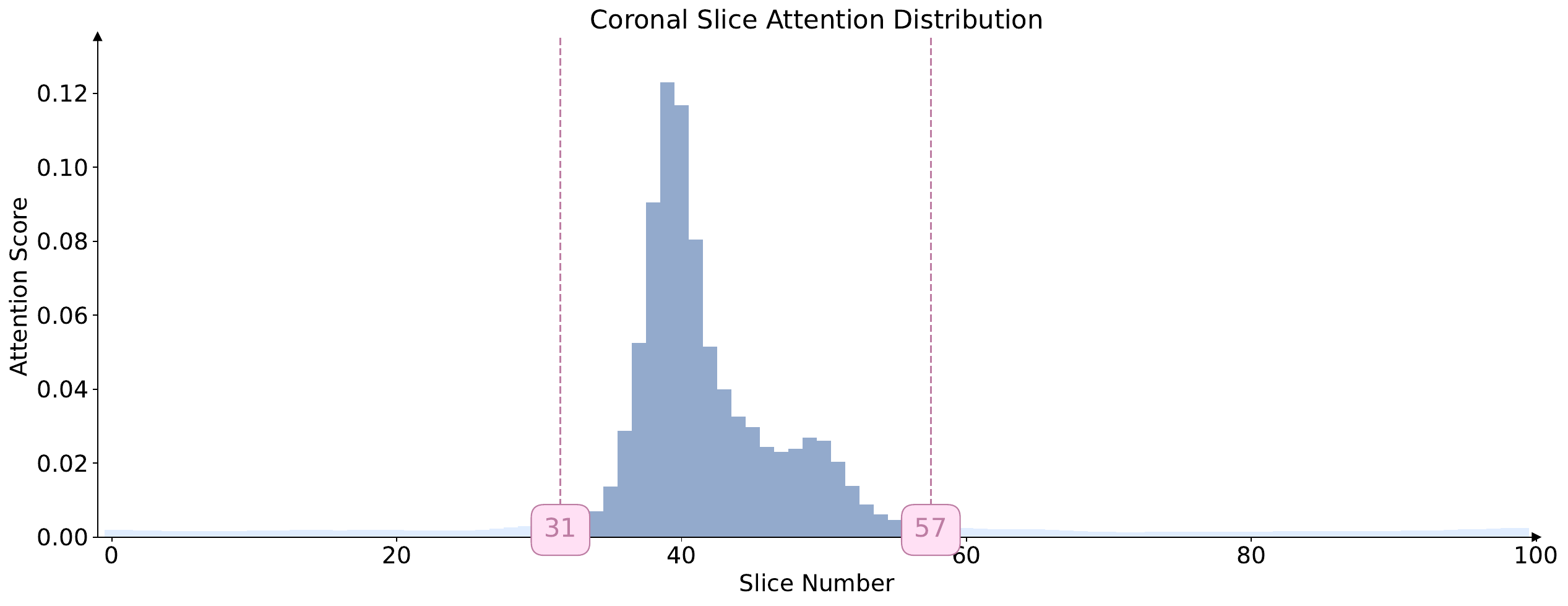}
        \caption{Fold 4}
    \end{subfigure}
    \hfill
    \begin{subfigure}{0.33\textwidth}
        \includegraphics[width=\linewidth]{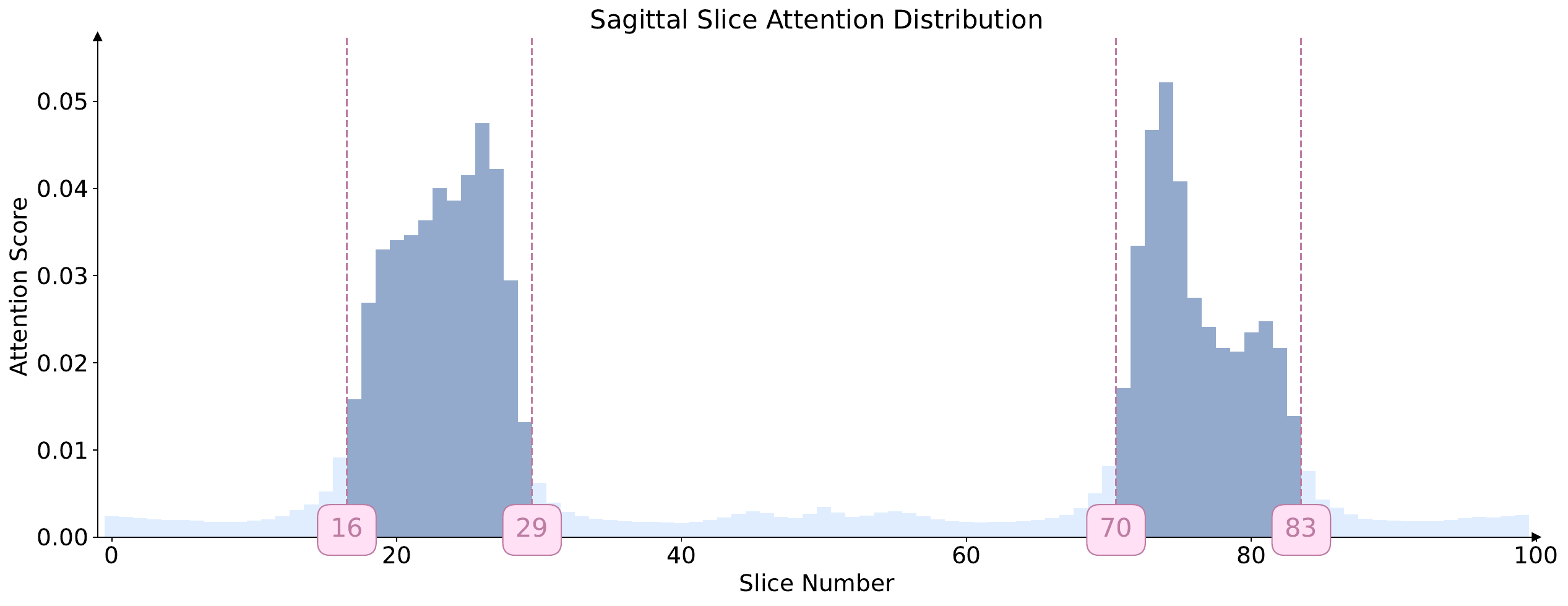}
        \caption{}
    \end{subfigure}
    \hrulefill

    \begin{subfigure}{0.33\textwidth}
        \includegraphics[width=\linewidth]{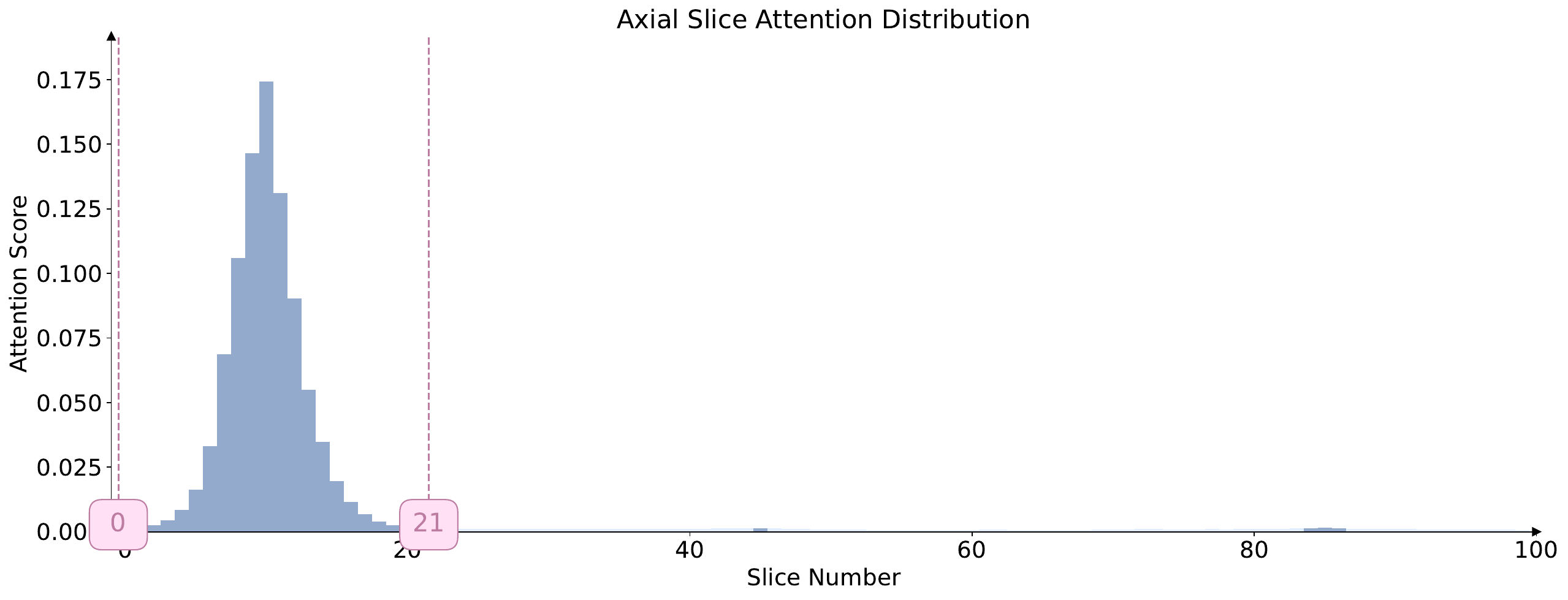}
        \caption{}
    \end{subfigure}
    \hfill
    \begin{subfigure}{0.33\textwidth}
        \includegraphics[width=\linewidth]{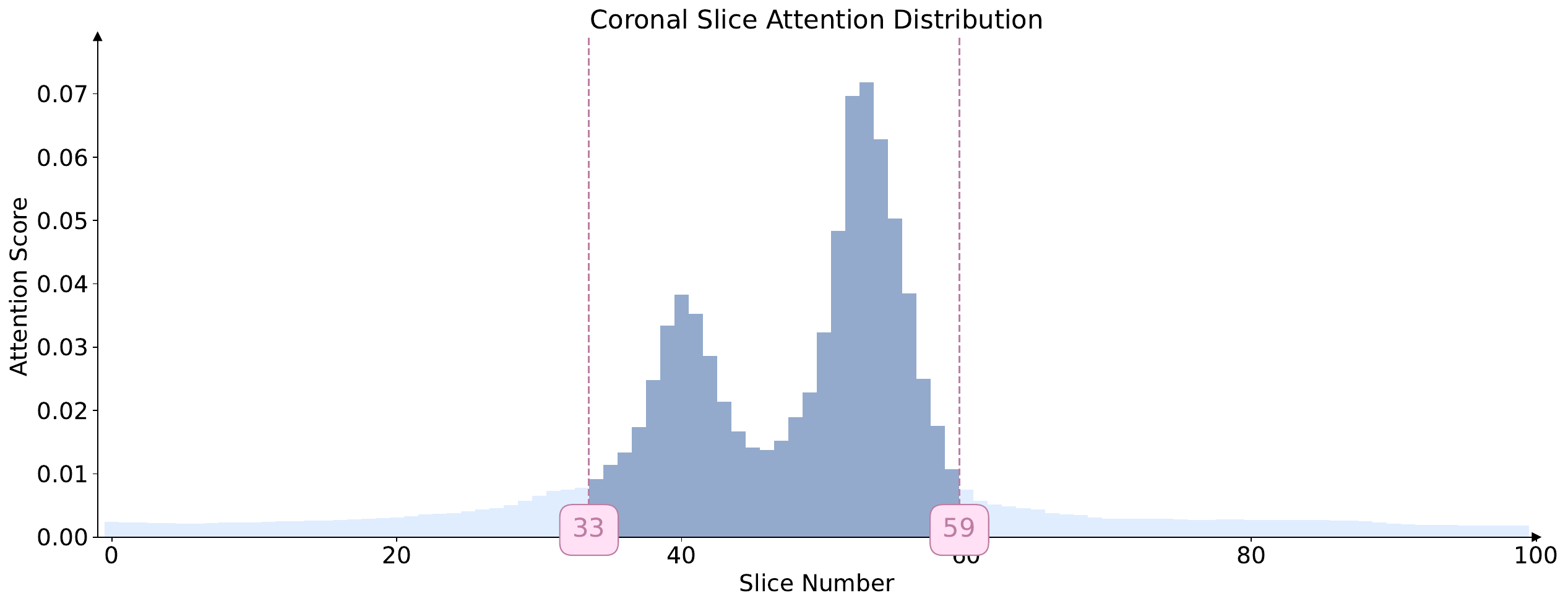}
        \caption{Fold 5}
    \end{subfigure}
    \hfill
    \begin{subfigure}{0.33\textwidth}
        \includegraphics[width=\linewidth]{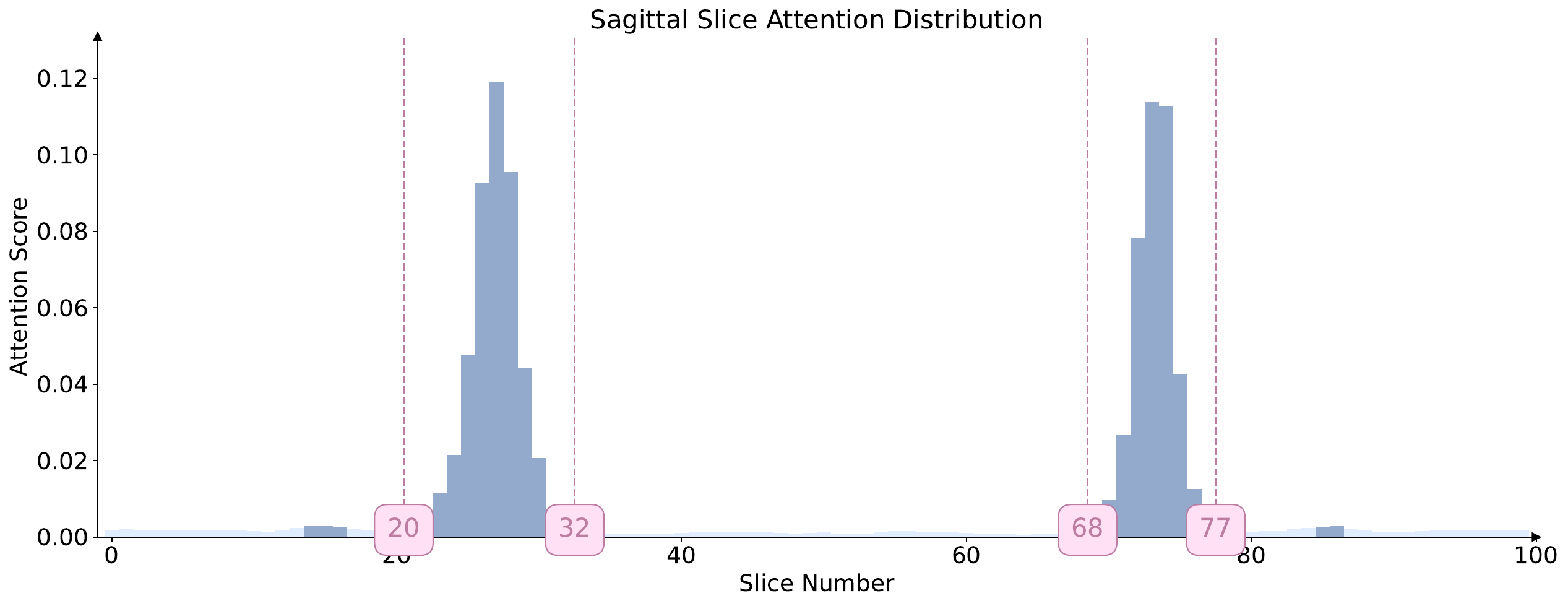}
        \caption{}
    \end{subfigure}
    \hrulefill

    \caption{Average attention weight distributions generate by \textbf{our model} for each fold and each plane}
    \label{fig:consistency_all_views}
\end{figure*}

\begin{figure*}[!h]
    \centering
    \captionsetup[subfigure]{labelformat=empty}
    \begin{subfigure}{0.33\textwidth}
        \caption{Axial}
        \includegraphics[width=\linewidth]{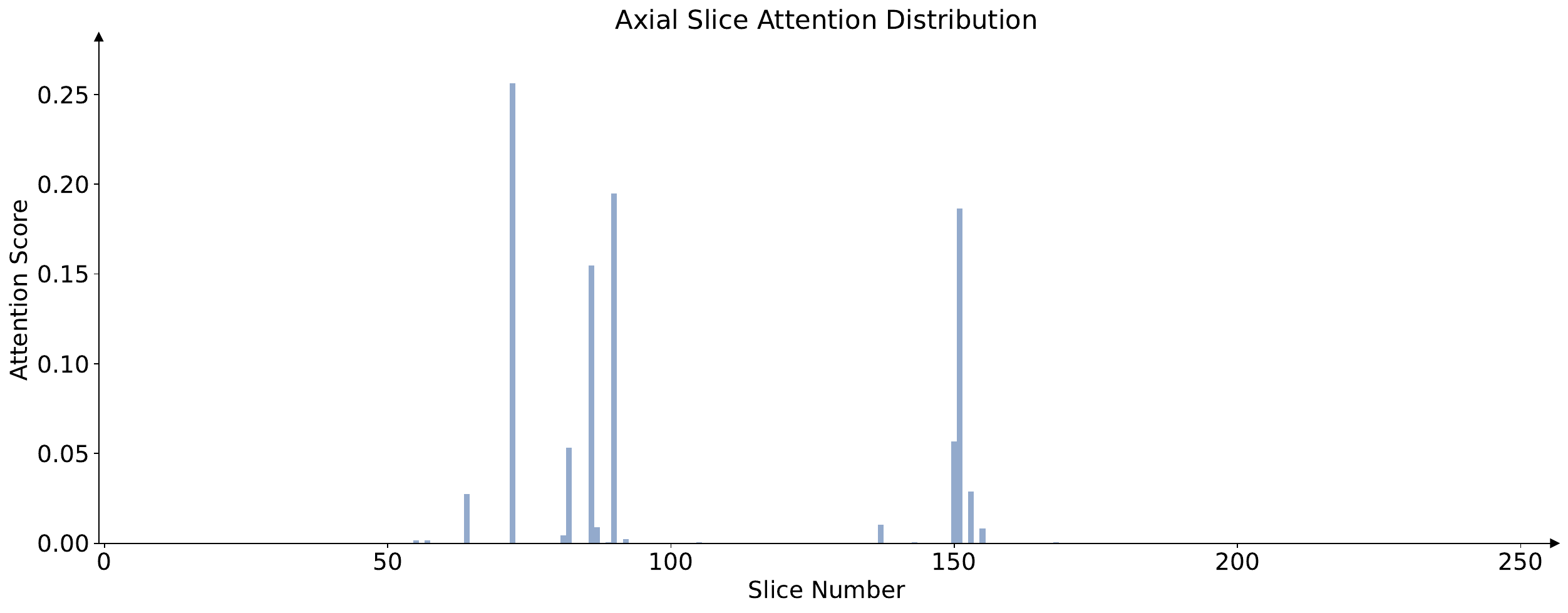}
        \caption{}
    \end{subfigure}
    \hfill
    \begin{subfigure}{0.33\textwidth}
        \caption{Coronal}
        \includegraphics[width=\linewidth]{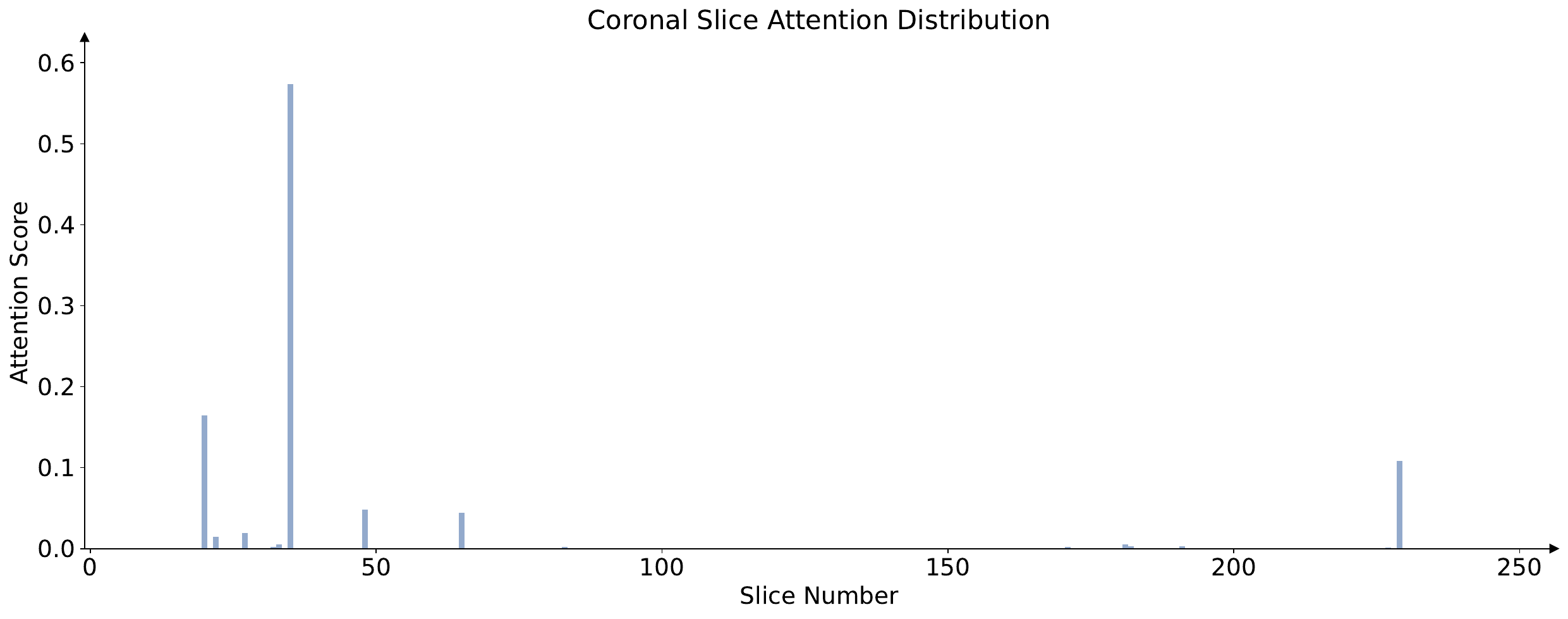}
        \caption{Fold 1}
    \end{subfigure}
    \hfill
    \begin{subfigure}{0.33\textwidth}
        \caption{Sagittal}
        \includegraphics[width=\linewidth]{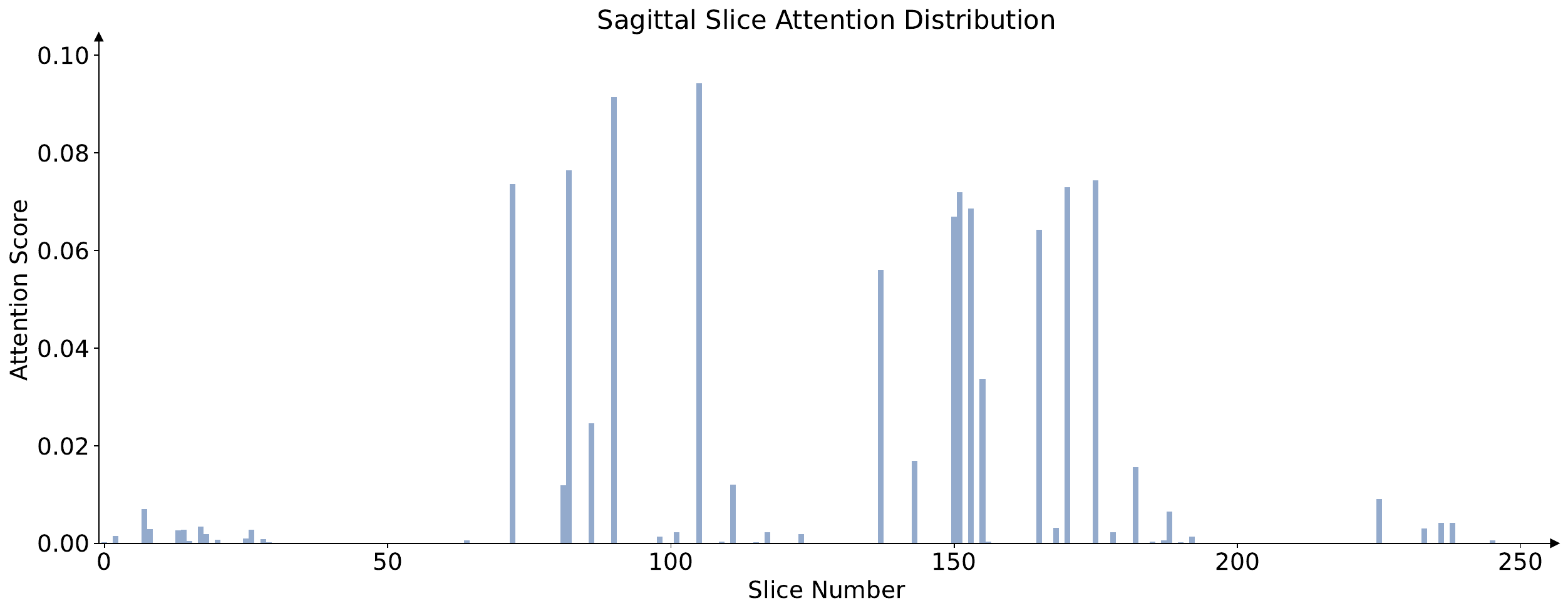}
        \caption{}
    \end{subfigure}
    \hrulefill

    \begin{subfigure}{0.33\textwidth}
        \includegraphics[width=\linewidth]{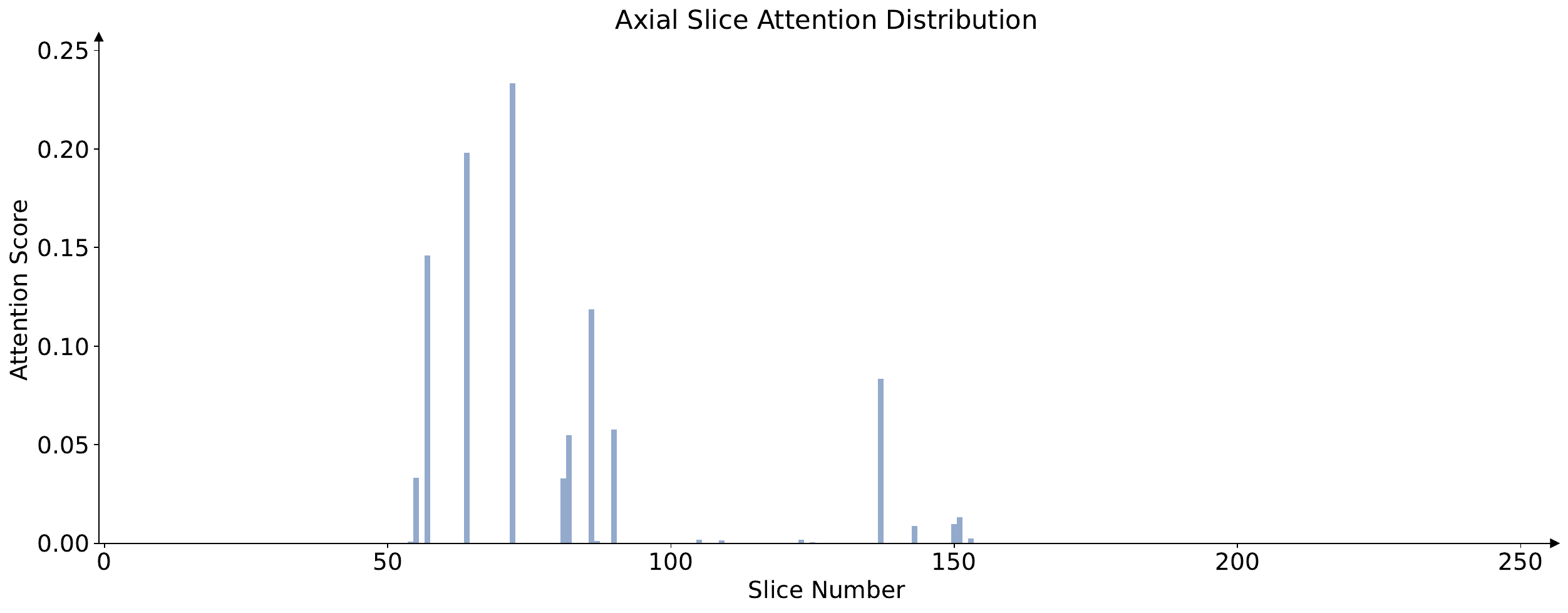}
        \caption{}
    \end{subfigure}
    \hfill
    \begin{subfigure}{0.33\textwidth}
        \includegraphics[width=\linewidth]{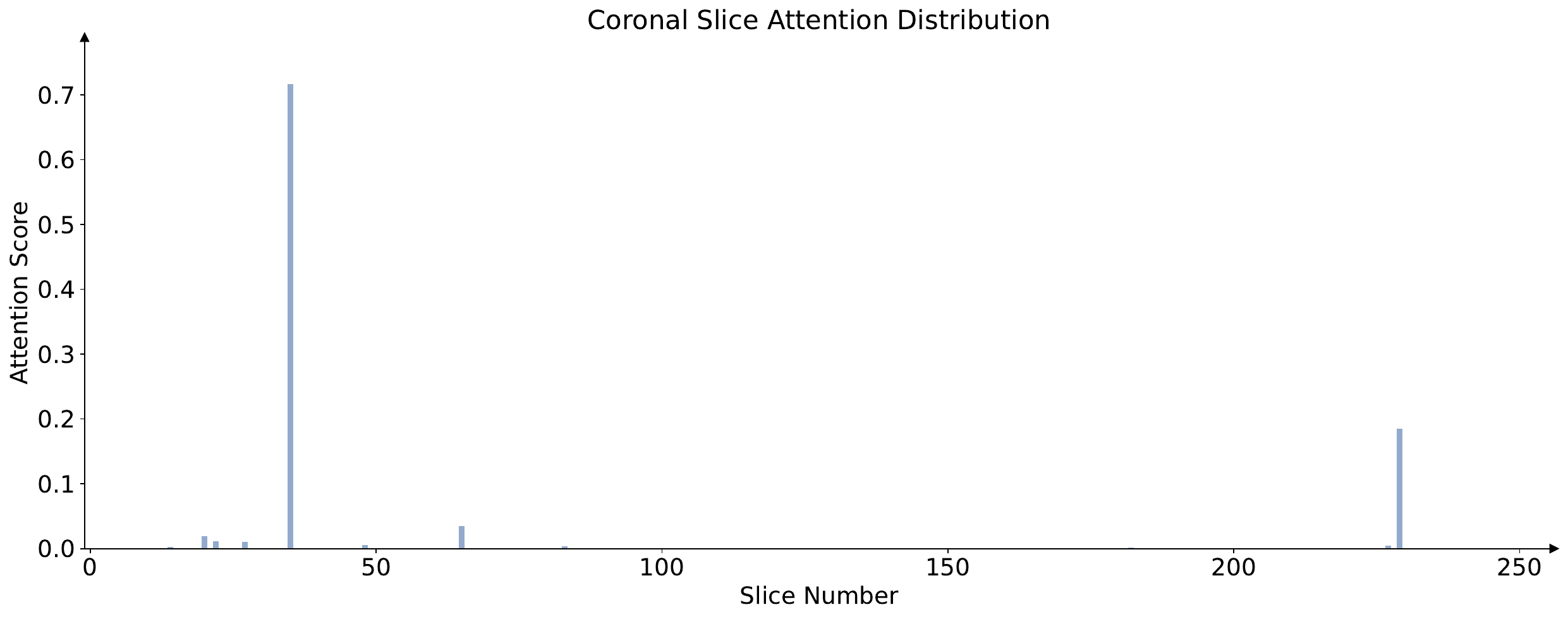}
        \caption{Fold 2}
    \end{subfigure}
    \hfill
    \begin{subfigure}{0.33\textwidth}
        \includegraphics[width=\linewidth]{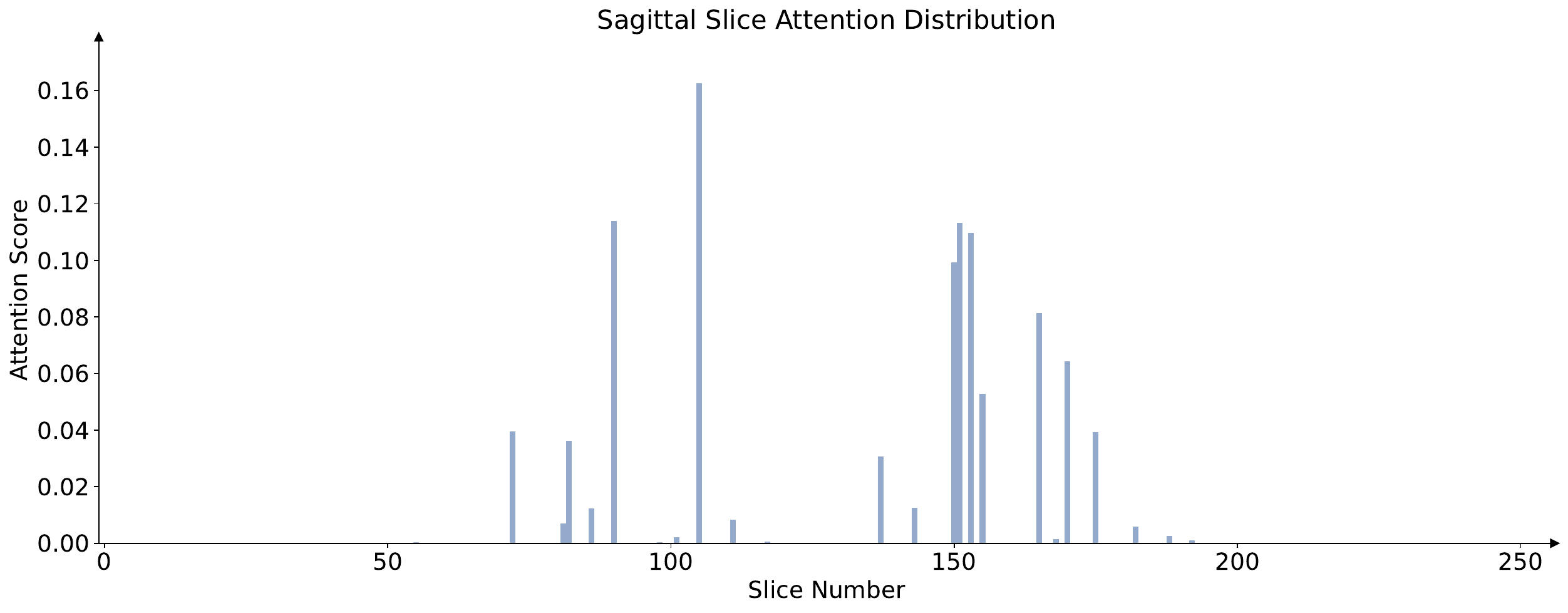}
        \caption{}
    \end{subfigure}
    \hrulefill
    
    \begin{subfigure}{0.33\textwidth}
        \includegraphics[width=\linewidth]{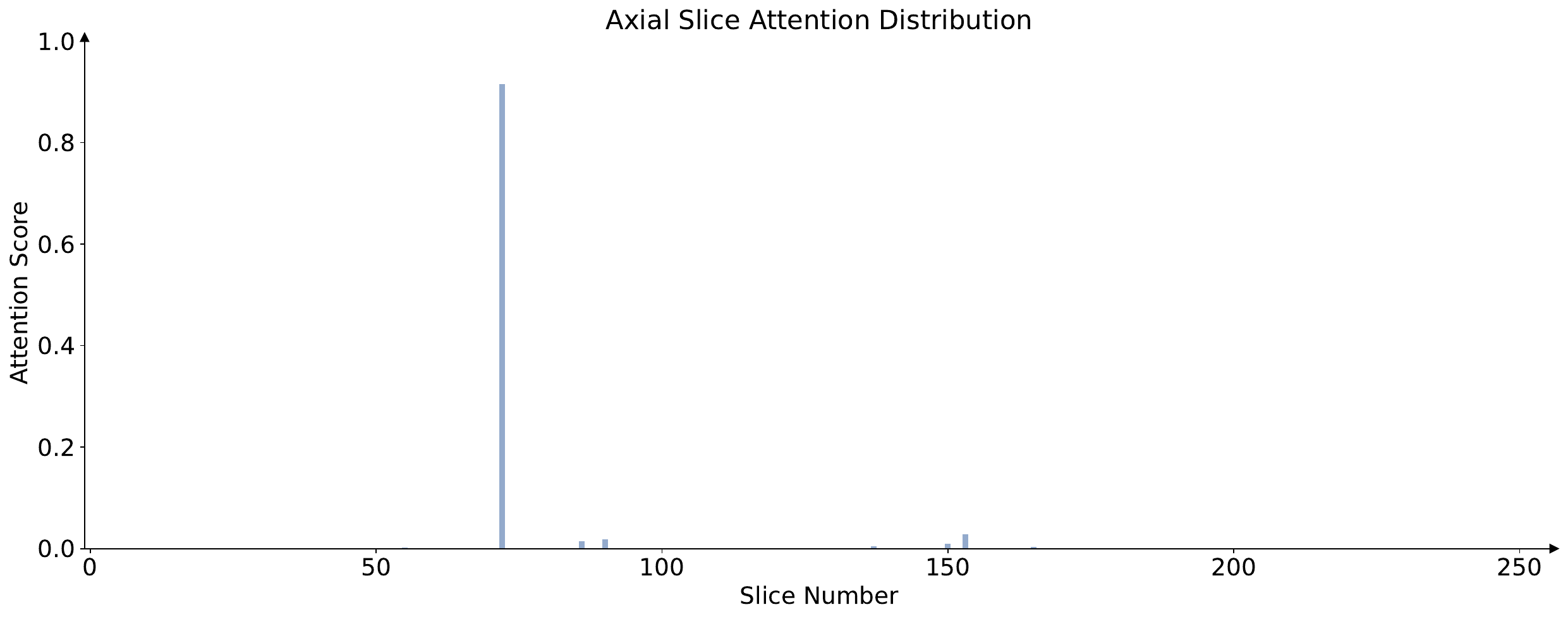}
        \caption{}
    \end{subfigure}
    \hfill
    \begin{subfigure}{0.33\textwidth}
        \includegraphics[width=\linewidth]{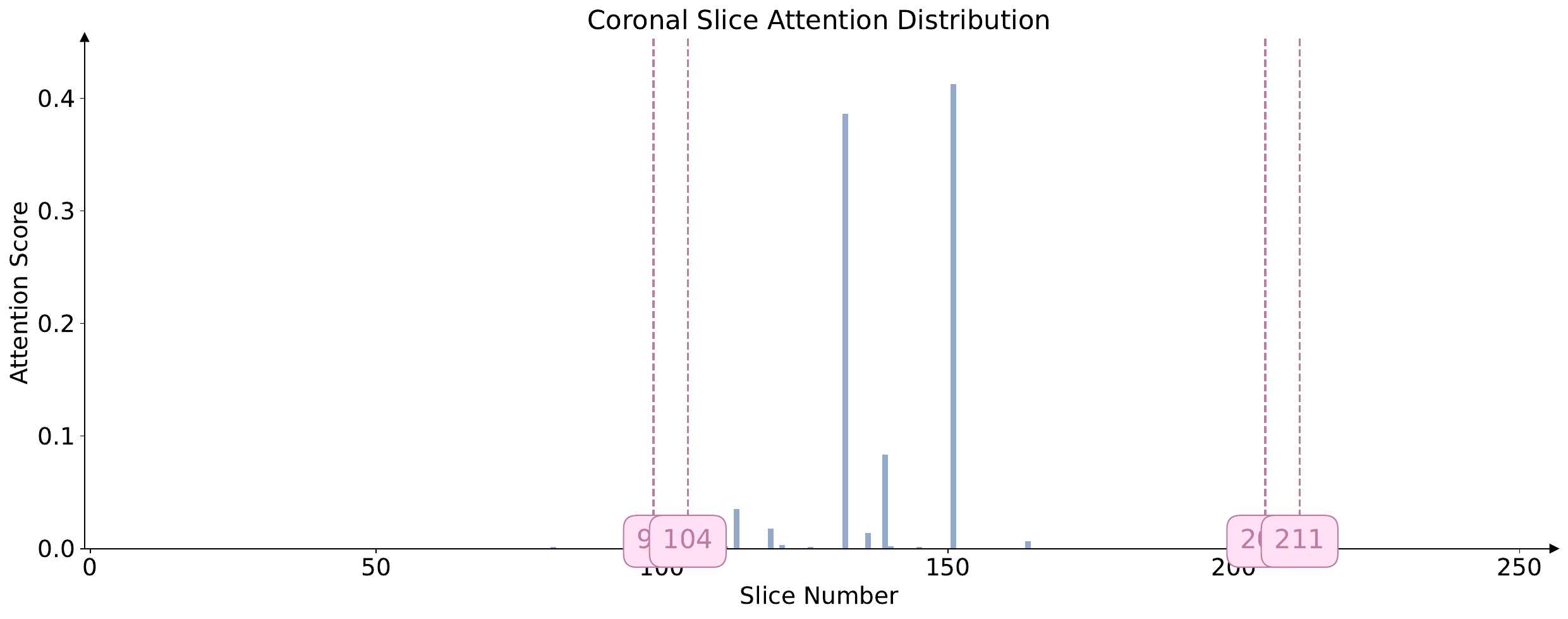}
        \caption{Fold 3}
    \end{subfigure}
    \hfill
    \begin{subfigure}{0.33\textwidth}
        \includegraphics[width=\linewidth]{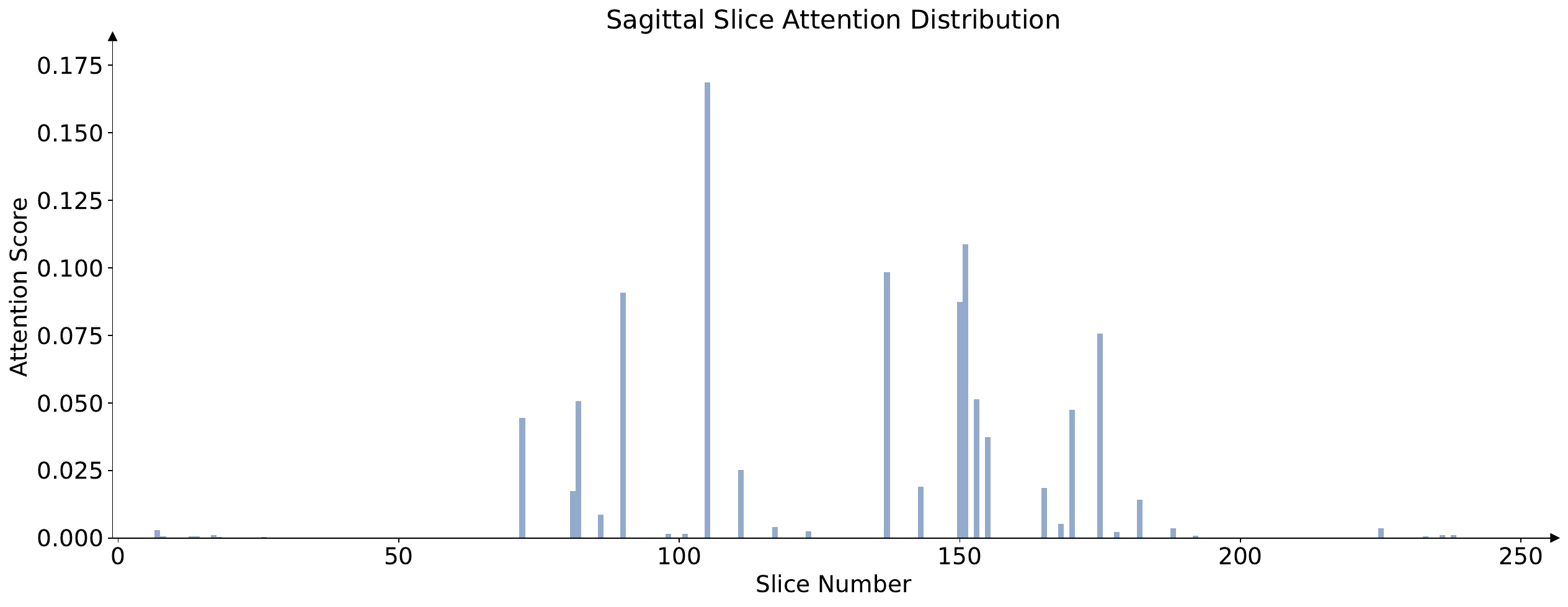}
        \caption{}
    \end{subfigure}
    \hrulefill

    \begin{subfigure}{0.33\textwidth}
        \includegraphics[width=\linewidth]{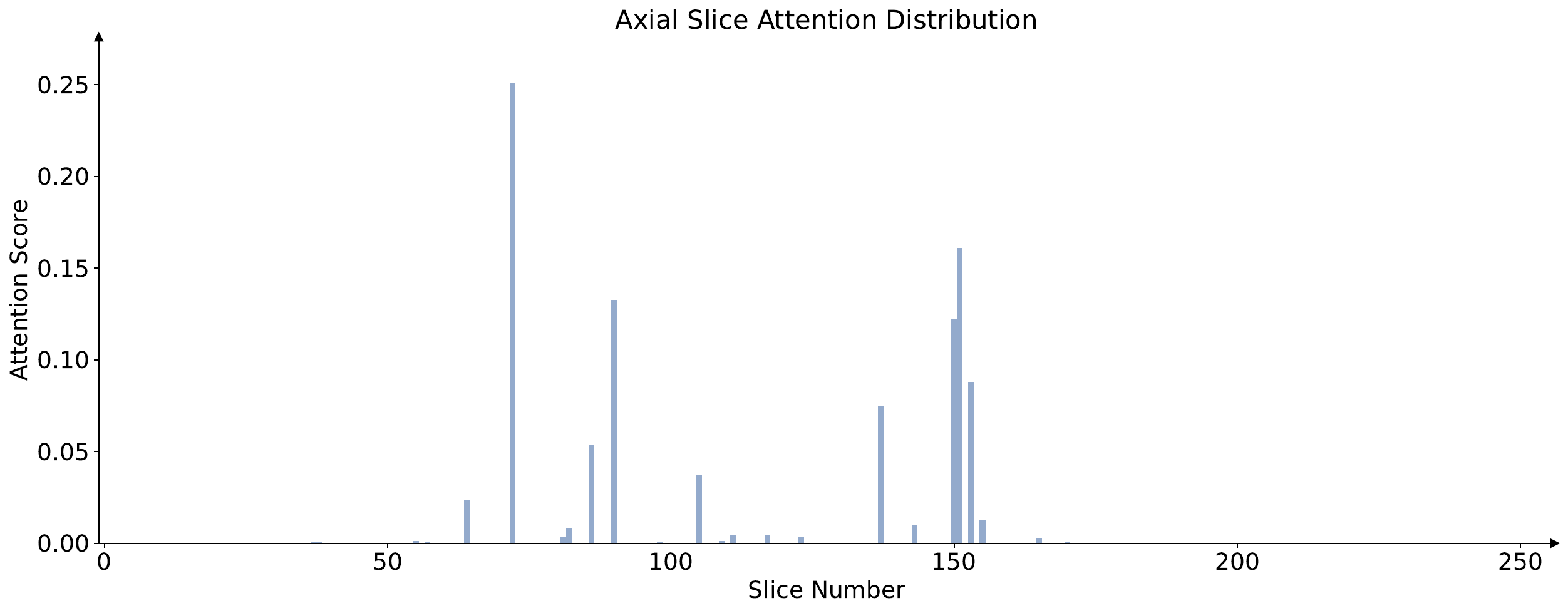}
        \caption{}
    \end{subfigure}
    \hfill
    \begin{subfigure}{0.33\textwidth}
        \includegraphics[width=\linewidth]{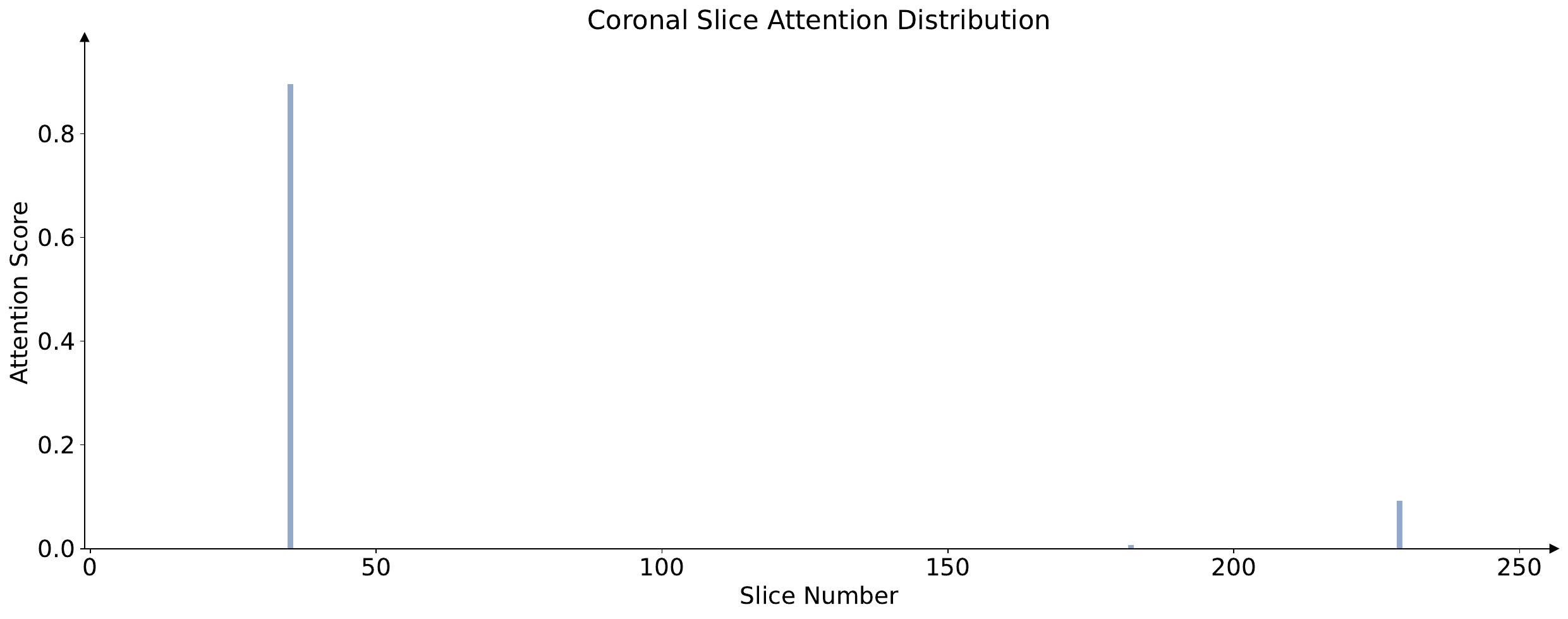}
        \caption{Fold 4}
    \end{subfigure}
    \hfill
    \begin{subfigure}{0.33\textwidth}
        \includegraphics[width=\linewidth]{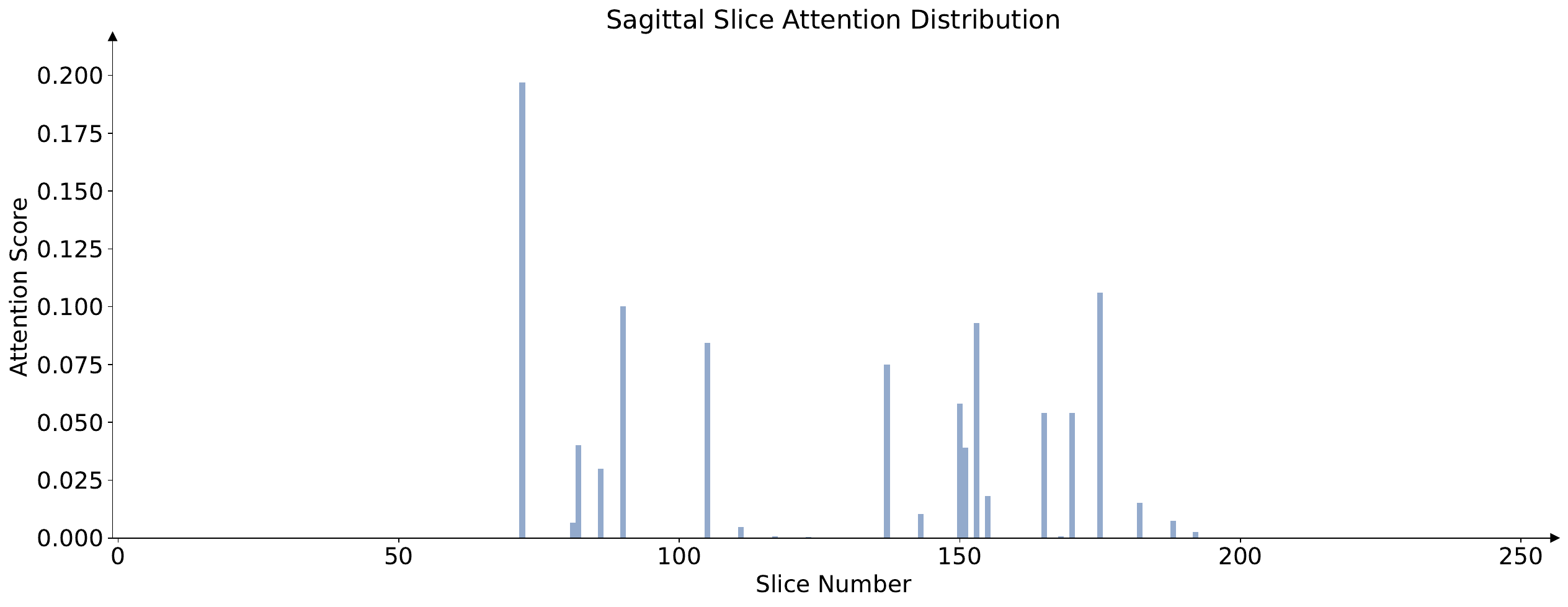}
        \caption{}
    \end{subfigure}
    \hrulefill

    \begin{subfigure}{0.33\textwidth}
        \includegraphics[width=\linewidth]{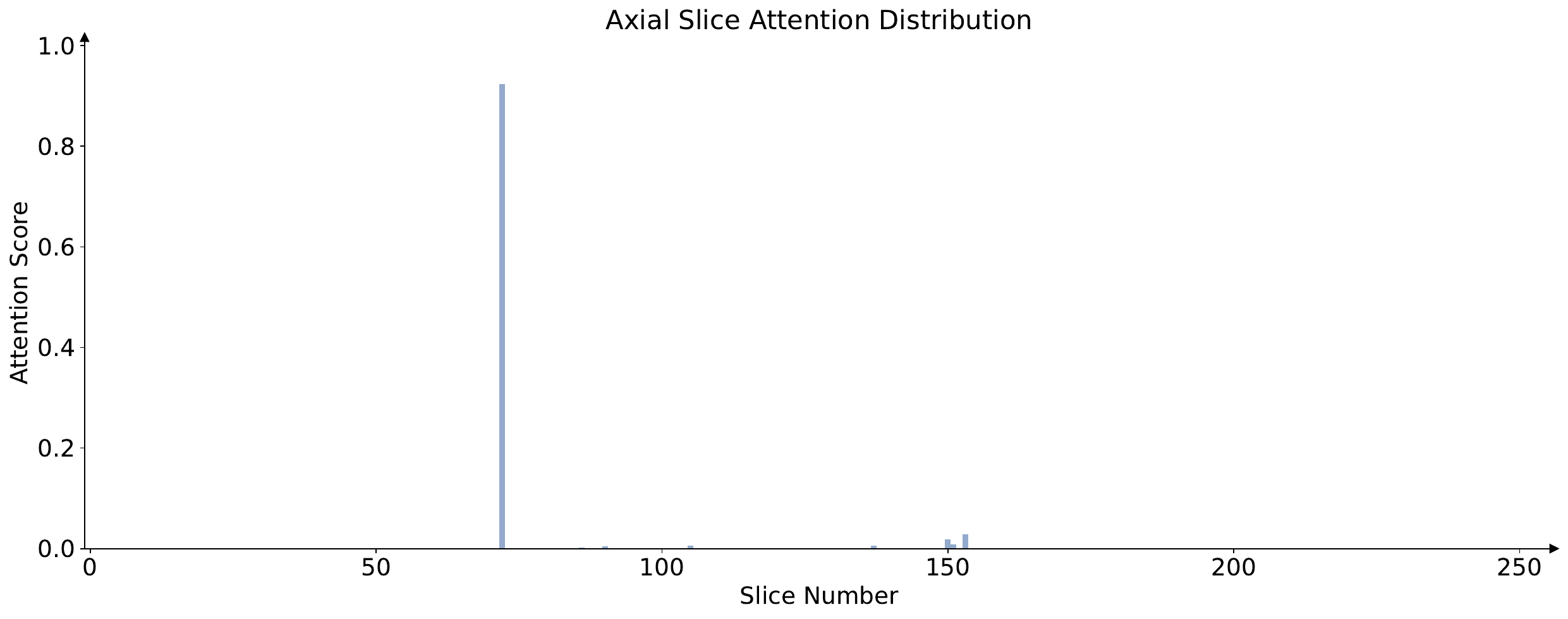}
        \caption{}
    \end{subfigure}
    \hfill
    \begin{subfigure}{0.33\textwidth}
        \includegraphics[width=\linewidth]{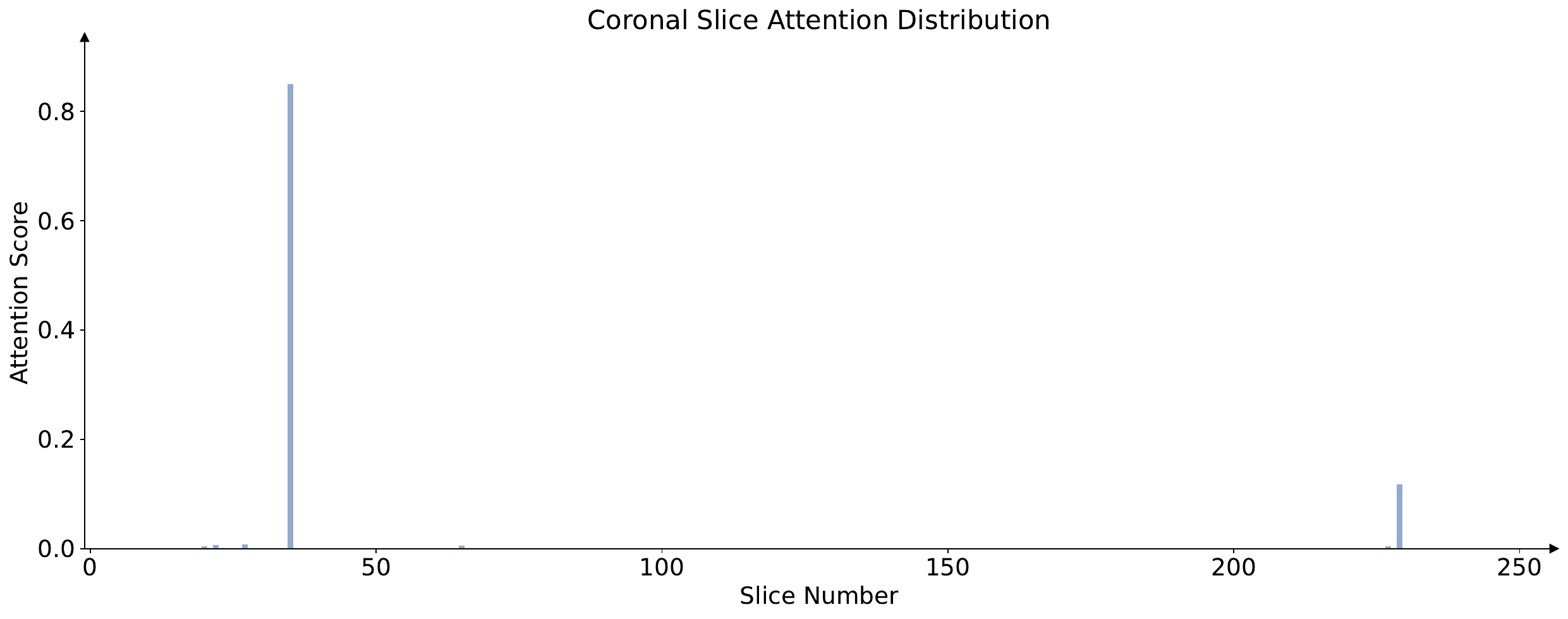}
        \caption{Fold 5}
    \end{subfigure}
    \hfill
    \begin{subfigure}{0.33\textwidth}
        \includegraphics[width=\linewidth]{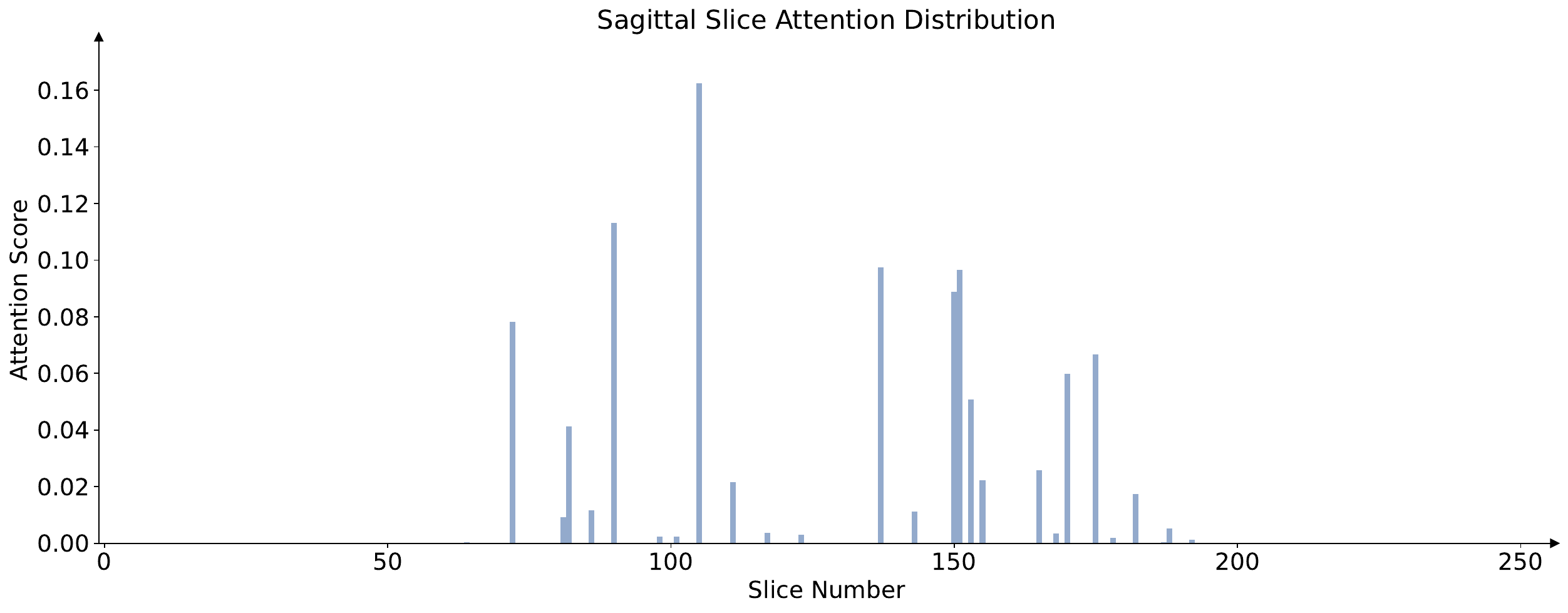}
        \caption{}
    \end{subfigure}
    
    \caption{Average attention weight distributions generate by \textbf{AwareNet} model for each fold and each plane}
    \label{fig:consistency_all_views_awarenet}
\end{figure*}

\begin{figure*}[hbt!]
    \centering
    \begin{minipage}{0.35\textwidth}
        \begin{subfigure}{\linewidth}
            \includegraphics[width=\linewidth]{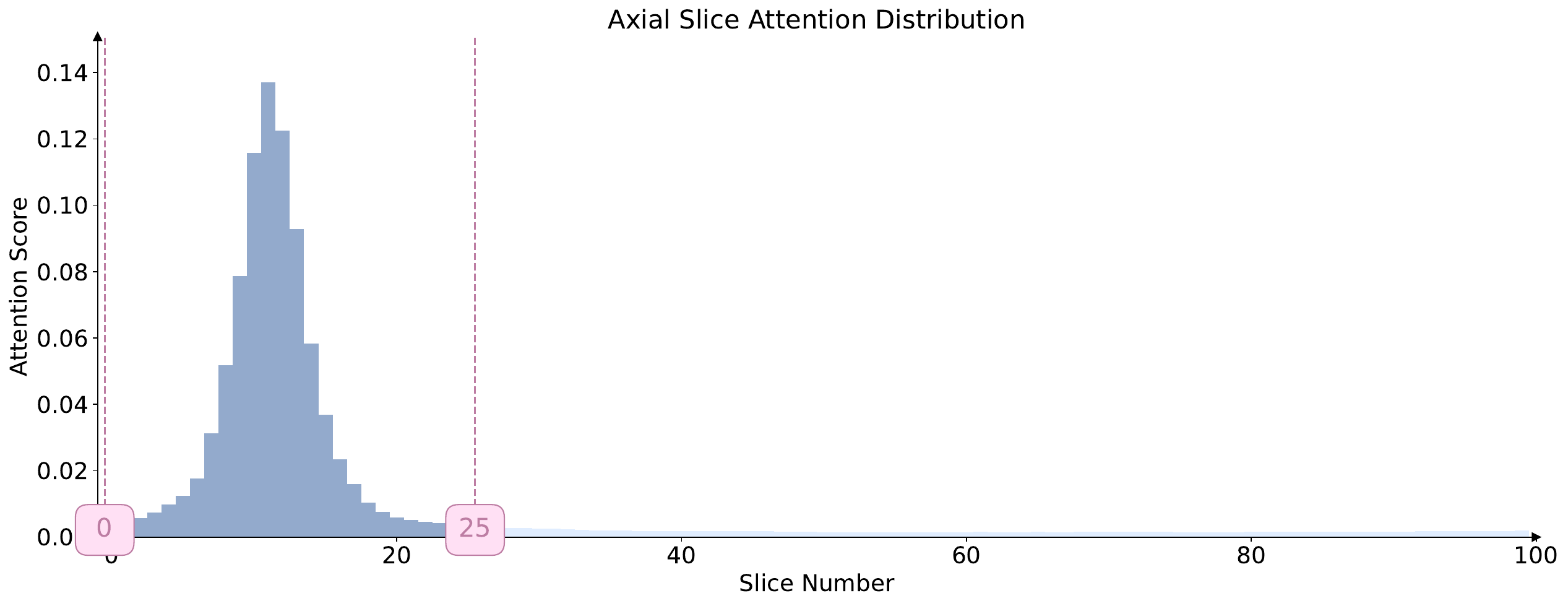}
        \end{subfigure}\\[1ex] 

        \begin{subfigure}{\linewidth}
            \includegraphics[width=\linewidth]{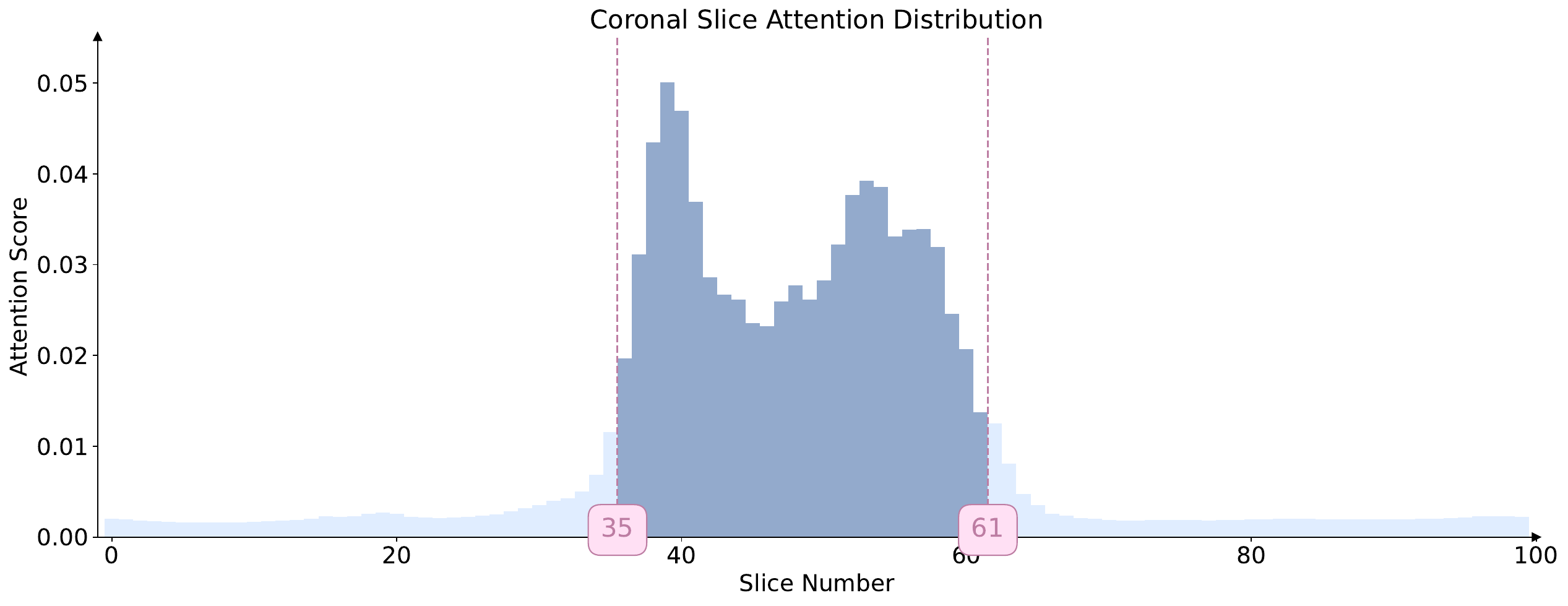}
        \end{subfigure}\\[1ex]

        \begin{subfigure}{\linewidth}
            \includegraphics[width=\linewidth]{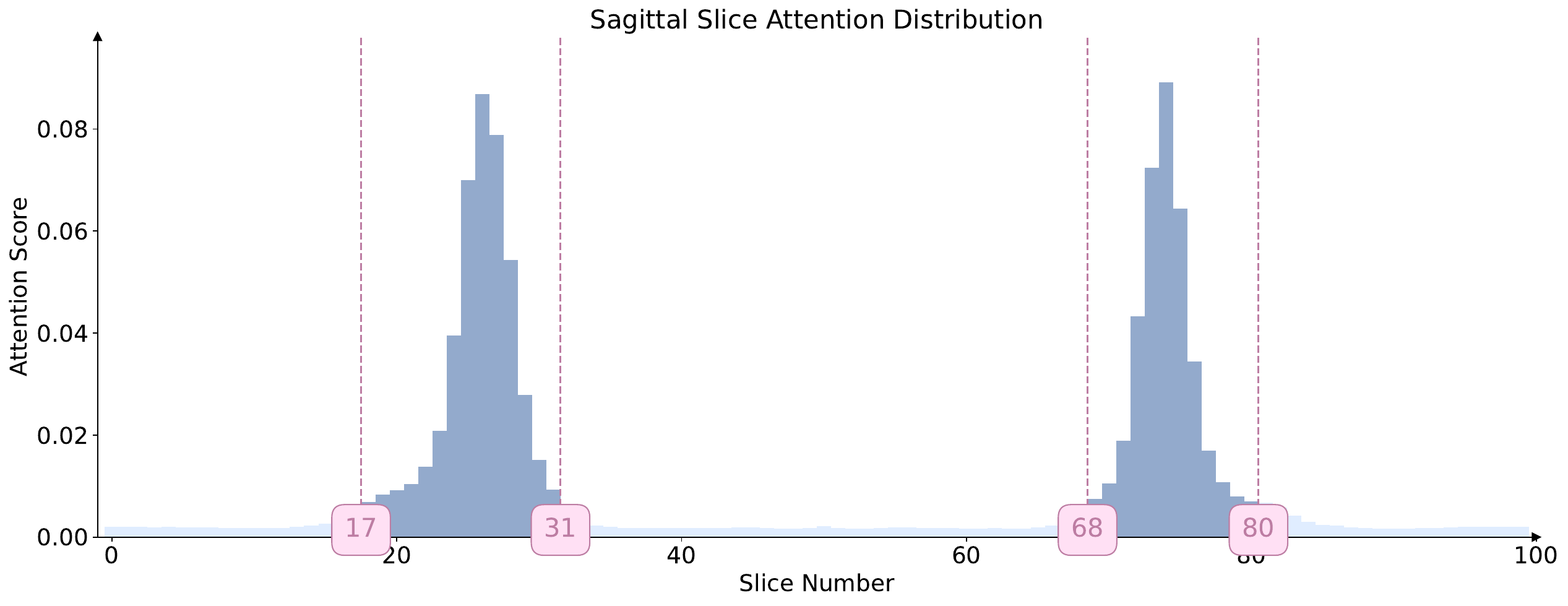}
        \end{subfigure}
    \end{minipage}\hfill
    \begin{minipage}{0.62\textwidth}
        \begin{subfigure}{0.49\linewidth}
            \includegraphics[width=\linewidth]{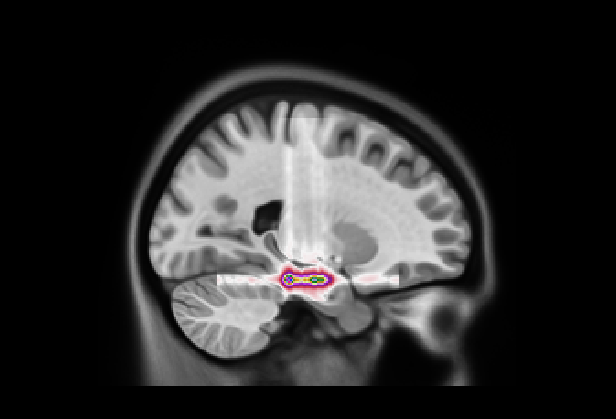}
        \end{subfigure}\hfill
        \begin{subfigure}{0.49\linewidth}
            \includegraphics[width=\linewidth]{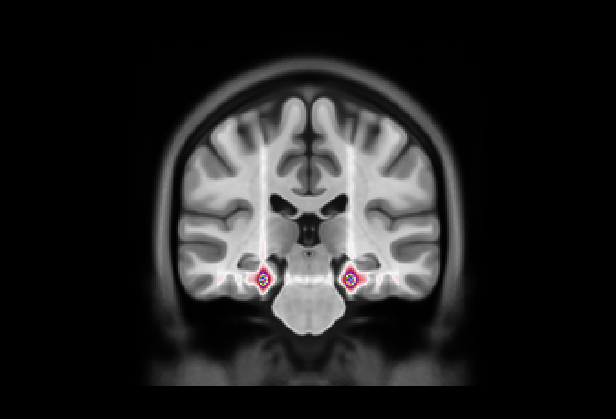}
        \end{subfigure}\\

        \begin{subfigure}{0.49\linewidth}
            \includegraphics[width=\linewidth]{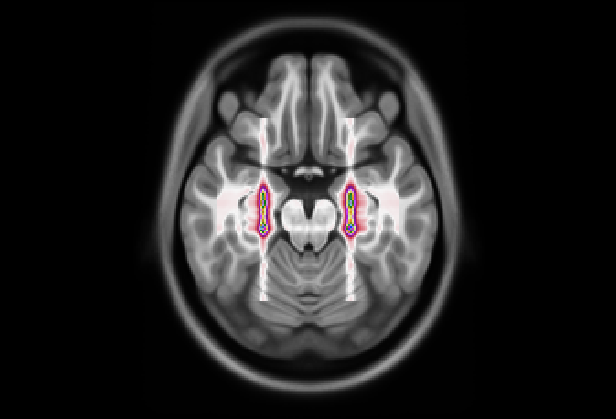}
        \end{subfigure}\hfill
        \begin{subfigure}{0.49\linewidth}
            \includegraphics[width=\linewidth]{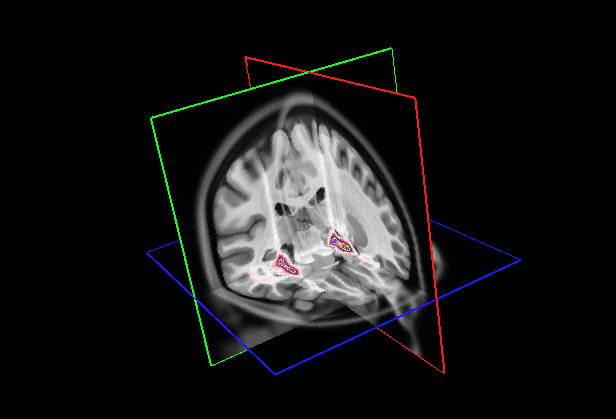}
        \end{subfigure}
    \end{minipage}
    \caption{(\textbf{Left}) Attention distribution for each plane averaged on all the five test to provide entire dataset distributions; (\textbf{Right}) Visualization of mean 3D attention map of entire dataset overlapped to MNI152 template}
    \label{fig:xai_visual_our}
\end{figure*}

\label{sec:xai_results}
This section analyzes the interpretability of our approach and those proposed in \citep{2d_slice_wang2024joint, 2d_slice_altay2021preclinical}.  
Our XAI method described in Section \ref{sec:xai_approach} allowed us to produce a 3D attentional map starting from the attentional weight distributions of the axial, coronal, and sagittal planes. 
The authors of AwareNet \citep{2d_slice_wang2024joint} designed the slice-aware module of this network to extract, as in our case, a distribution of attentional weights capable of summarizing the importance of each slice in the decision-making process. As a result, our approach can produce a 3D map also using the model proposed in \citep{2d_slice_wang2024joint}.

A recurring approach to interpreting decisions made by DL models is GradCAM \citep{selvaraju2017grad}. GradCAM is a technique for visualizing regions within an image that influence the classification decision of a convolutional neural network. It works by selecting a target convolutional layer, calculating the class score gradients relative to that layer's feature maps, averaging these gradients, and using them as weights to create a weighted sum of the feature maps. This sum is then passed through a ReLU function to produce a saliency map highlighting the areas that positively impact the class decision. 
However, GradCAM in this comparison scenario allows the production of 2D saliency maps, one for each slice, which deliver a qualitative but not quantitative result. To allow a fair comparison with other models, we have developed a variant of our XAI approach that allows us to generate 3D maps with GradCAM. The idea is to stack the produced 2D CAMs on top of each other to generate a 3D saliency map for each plane. Given \(A\), \(S\), and \(C\), the 3D salient maps generated for the respective axial, sagittal, and coronal planes, the final 3D salient map is obtained pointwise as:
\[ M[i, j, k] = {S}[i,j,k] \cdot {C}[i,j,k] \cdot {A}[i,j,k] \]
This 3D map can be used to quantitatively evaluate the impact of each brain area in the decision-making process with the metrics proposed in Section \ref{sec:xai_metrics}. This method also allows us to compare with Attention Transformer model proposed in \cite{2d_slice_altay2021preclinical}, which is not designed to yield attentional weights for each slice directly. In our implementation, we applied GradCAM to the last layer of the convolutional backbone. This choice is motivated by the fact that the last convolutional layer retains high-level spatial information that is crucial for localizing salient regions in the input image. 

In the following, we show the consistency of the attention weight distributions obtained from our model and AwareNet. Then, we present the results in qualitative and quantitative terms for each model using our XAI method both in the GradCAM case and with the attentional 3D maps.

\subsubsection{Attention consistency analysis}

Many studies focused on model interpretability by generating saliency maps for randomly chosen test subjects, enabling local qualitative analysis. However, this approach does not support the evaluation of interpretability consistency when the training and testing data vary. To examine this consistency in our approach, we considered the variability of attentional weight distributions across the five different folds used for cross-validation. Our evaluation leverages a clinical hypothesis specific to AD, which is believed to affect similar brain areas across different patients. 
For each fold, we calculated the weight distributions for each sample in the test set and derived an average distribution specific to that test set. By analyzing these average distributions across the five folds, we could assess whether our interpretability remained consistent, thereby supporting the generalizability of our findings across different subsets of data.

Fig. \ref{fig:consistency_all_views} presents our diagnosis network's average attentional weight distributions for each fold in the axial, coronal, and sagittal views. Upon examination of the histograms for each view, we observe a remarkable consistency in the distribution shapes across all five folds, indicating that our interpretability approach is stable despite the variation in the train/test set data. Specifically, the axial distributions reveal a consistent concentration of attentional weights around the initial slices. This trend suggests the model's recurrent focus on the brain's inferior regions, notably the areas where degenerative changes first manifest in AD, such as the \emph{hippocampus}. In the coronal view, attentional weights are notably centered, indicating that the model consistently identifies the central part of the brain as significant. This central focus might correspond to the \emph{medial temporal lobe}, including the \emph{hippocampus} and the surrounding regions, further substantiating the axial findings. The sagittal view is the only bimodal distribution, suggesting that the model pinpointed symmetrical areas along this plane. We hypothesize that the network was focusing on the \emph{hippocampus} since it adheres to all the constraints: situated in the inferior part of the brain, centrally located, and symmetrical. The consistency and specificity of these findings across multiple data folds strengthen the argument that our network could reliably identify specific brain regions as a critical biomarker for distinguishing between AD and CN subjects. 

We also analyzed the consistencies of AwareNet distributions \citep{2d_slice_wang2024joint} to compare the robustness and interpretability of different attention mechanisms. However, the average distributions produced by this model, as seen in Fig. \ref{fig:consistency_all_views_awarenet}, present sparsely distributed peaks and do not allow the identification of predominant slice ranges. Furthermore, the distributions between the folds are inconsistent. These results indicate that on this dataset, AwareNet could not produce consistent attentional weights that are useful for contributing to the interpretability of its decisions.

\subsection{Qualitative and quantitative results}

\begin{table*}[!p]
\centering
\caption{Importance measures of 20 largest brain regions computed with 3D Attention maps. \textbf{Our model} on the left (a); \textbf{AwareNet} on the right (b).}
\label{tab:metrics_awarenet_our_attention}

\begin{minipage}{.5\textwidth}
\centering
\setlength{\tabcolsep}{5pt}
\scriptsize
\begin{spacing}{1.5}
\begin{tabular}{lcccccc}
\toprule
Brain Region                        & \(V_r\) & \(\mu_r\) & \(\sigma_r\)  & \(A_{max,r}\) & \(A_{min,r}\) & \(P_r\) \\
\midrule
Hippocampus left                  &  1562  &   0.136   & 0.139         &  0.762        &  0.028        & 0.333                \\
Hippocampus right                 &  1426  &   0.126   & 0.133         &  0.783        &  0.028        & 0.304                \\
Parahippocampal left              &   688  &   0.129   & 0.137         &  0.884        &  0.028        & 0.254                \\
Parahippocampal right             &   534  &   0.129   & 0.148         &  1.000        &  0.028        & 0.197                \\
Amygdala left                     &   480  &   0.097   & 0.092         &  0.620        &  0.028        & 0.291                \\
Amygdala right                    &   427  &   0.095   & 0.087         &  0.569        &  0.028        & 0.259                \\
Inferior Lateral Ventricle right  &   232  &   0.113   & 0.129         &  0.677        &  0.028        & 0.219                \\
Inferior Lateral Ventricle left   &   212  &   0.106   & 0.105         &  0.589        &  0.028        & 0.200                \\
Cerebellum Gray Matter left       &   208  &   0.035   & 0.005         &  0.052        &  0.028        & 0.003                \\
Lateral Orbitofrontal left        &   194  &   0.033   & 0.004         &  0.045        &  0.028        & 0.013                \\
Fusiform right                    &   184  &   0.045   & 0.015         &  0.107        &  0.028        & 0.014                \\
Lateral Orbitofrontal right       &   140  &   0.034   & 0.004         &  0.046        &  0.028        & 0.009                \\
Cerebellum Gray Matter right      &   119  &   0.034   & 0.005         &  0.054        &  0.028        & 0.002                \\
Fusiform left                     &    88  &   0.040   & 0.010         &  0.070        &  0.028        & 0.007                \\
Entorhinal left                   &    16  &   0.034   & 0.003         &  0.041        &  0.030        & 0.005                \\
Ventral Diencephalon left         &     6  &   0.029   & 0.001         &  0.031        &  0.028        & 0.001                \\
Entorhinal right                  &     2  &   0.033   & 0.003         &  0.036        &  0.031        & 0.001                \\
-                                 &  -     &   -       & -             &  -            &  -            & -                    \\
-                                 &  -     &   -       & -             &  -            &  -            & -                    \\
-                                 &  -     &   -       & -             &  -            &  -            & -                    \\
\bottomrule
\end{tabular}
\end{spacing}
\caption*{(a)}
\end{minipage}%
\begin{minipage}{.5\textwidth}
\centering
\setlength{\tabcolsep}{5pt}
\scriptsize
\begin{spacing}{1.5}
\begin{tabular}{lcccccc}
\toprule
Brain Region                        & \(V_r\) & \(\mu_r\) & \(\sigma_r\)  & \(A_{max,r}\) & \(A_{min,r}\) & \(P_r\) \\
\midrule
Cerebellum Gray Matter - right      & 192 & 0.001 & 0.003 & 0.022 & 0.000 & 0.003 \\
Lateral Occipital - right           & 173 & 0.003 & 0.008 & 0.067 & 0.000 & 0.008 \\
Cerebellum Gray Matter - left       & 132 & 0.001 & 0.004 & 0.035 & 0.000 & 0.002 \\
Lateral Occipital - left            & 116 & 0.002 & 0.011 & 0.106 & 0.000 & 0.005 \\
Fusiform - left                     & 93 & 0.001 & 0.002 & 0.009 & 0.000 & 0.007 \\
Fusiform - right                    & 74 & 0.001 & 0.001 & 0.007 & 0.000 & 0.005 \\
Hippocampus - right                 & 74 & 0.004 & 0.013 & 0.076 & 0.000 & 0.016 \\
Lateral Orbitofrontal - right       & 70 & 0.000 & 0.001 & 0.003 & 0.000 & 0.005 \\
Entorhinal - right                  & 62 & 0.009 & 0.025 & 0.117 & 0.000 & 0.019 \\
Ventral Diencephalon - right        & 62 & 0.001 & 0.003 & 0.014 & 0.000 & 0.010 \\
Hippocampus - left                  & 55 & 0.003 & 0.012 & 0.088 & 0.000 & 0.012 \\
Brainstem - right                   & 53 & 0.002 & 0.010 & 0.076 & 0.000 & 0.003 \\
Lingual - right                     & 48 & 0.000 & 0.001 & 0.003 & 0.000 & 0.004 \\
Lateral Orbitofrontal - left        & 47 & 0.000 & 0.001 & 0.004 & 0.000 & 0.003 \\
Parahippocampal - right             & 47 & 0.000 & 0.000 & 0.002 & 0.000 & 0.017 \\
Cerebellum White Matter - right     & 36 & 0.000 & 0.000 & 0.002 & 0.000 & 0.003 \\
Ventral Diencephalon - left         & 35 & 0.002 & 0.005 & 0.022 & 0.000 & 0.006 \\
Entorhinal - left                   & 35 & 0.011 & 0.042 & 0.186 & 0.000 & 0.011 \\
Amygdala - right                    & 35 & 0.000 & 0.000 & 0.002 & 0.000 & 0.021 \\
Superior Frontal - right            & 27 & 0.000 & 0.000 & 0.000 & 0.000 & 0.001 \\
\bottomrule
\end{tabular}
\end{spacing}
\caption*{(b)}
\end{minipage}
\end{table*}

\begin{table*}[!p]
\centering
\caption{Importance measures of 20 largest brain regions computed with 3D GradCAM maps. \textbf{Our model} on the left (a); \textbf{Attention Transformer} on the right (b).}
\label{tab:metrics_transformer_our_attention}

\begin{minipage}{.5\textwidth}
\centering
\setlength{\tabcolsep}{5pt}
\scriptsize
\begin{spacing}{1.5}
\begin{tabular}{lcccccc}
\toprule
Brain Region                        & \(V_r\) & \(\mu_r\) & \(\sigma_r\)  & \(M_{max,r}\) & \(M_{min,r}\) & \(P_r\) \\
\midrule
Hippocampus left                    & 1722 & 0.874 & 0.057 & 0.992 & 0.773 & 0.367 \\
Hippocampus right                   & 965 & 0.835 & 0.041 & 0.963 & 0.773 & 0.206 \\
Inferior Lateral Ventricle left     & 648 & 0.861 & 0.054 & 1.000 & 0.773 & 0.612 \\
Inferior Lateral Ventricle right    & 435 & 0.830 & 0.038 & 0.944 & 0.773 & 0.411 \\
Superior Temporal left              & 164 & 0.810 & 0.022 & 0.860 & 0.774 & 0.006 \\
Amygdala right                      & 146 & 0.795 & 0.015 & 0.844 & 0.774 & 0.088 \\
Superior Temporal right             & 143 & 0.816 & 0.029 & 0.888 & 0.774 & 0.006 \\
Amygdala left                       & 118 & 0.795 & 0.021 & 0.867 & 0.773 & 0.072 \\
Middle Temporal left                & 96 & 0.825 & 0.028 & 0.873 & 0.774 & 0.003 \\
Middle Temporal right               & 95 & 0.823 & 0.028 & 0.887 & 0.774 & 0.003 \\
Parahippocampal left                & 91 & 0.829 & 0.038 & 0.928 & 0.774 & 0.034 \\
Entorhinal left                     & 77 & 0.825 & 0.032 & 0.891 & 0.774 & 0.024 \\
Fusiform left                       & 65 & 0.814 & 0.033 & 0.903 & 0.774 & 0.005 \\
Insula left                         & 8 & 0.787 & 0.009 & 0.806 & 0.775 & 0.001 \\
Insula right                        & 7 & 0.787 & 0.014 & 0.809 & 0.776 & 0.001 \\
Parahippocampal right               & 2 & 0.773 & 0.000 & 0.774 & 0.773 & 0.001 \\
Ventral Diencephalon left           & 2 & 0.779 & 0.005 & 0.784 & 0.774 & 0.000 \\
Inferior temporal left              & 1 & 0.775 & 0.000 & 0.775 & 0.775 & 0.000 \\
Putamen left                        & 1 & 0.776 & 0.000 & 0.776 & 0.776 & 0.000 \\
-                                   & - & -  & - & -  & - & - \\
\bottomrule
\end{tabular}
\end{spacing}
\caption*{(a)}
\end{minipage}%
\begin{minipage}{.5\textwidth}
\centering
\setlength{\tabcolsep}{5pt}
\scriptsize
\begin{spacing}{1.5}
\begin{tabular}{lcccccc}
\toprule
Brain Region                        & \(V_r\) & \(\mu_r\) & \(\sigma_r\)  & \(M_{max,r}\) & \(M_{min,r}\) & \(P_r\) \\
\midrule
Hippocampus - left                  & 1921 & 0.614 & 0.089 & 0.921 & 0.482 & 0.409 \\
Hippocampus - right                 & 1728 & 0.628 & 0.103 & 1.000 & 0.482 & 0.368 \\
Amygdala - right                    & 409 & 0.576 & 0.068 & 0.779 & 0.482 & 0.248 \\
Inferior Lateral Ventricle - right  & 322 & 0.625 & 0.090 & 0.847 & 0.482 & 0.304 \\
Inferior Lateral Ventricle - left   & 318 & 0.649 & 0.088 & 0.899 & 0.483 & 0.300 \\
Amygdala - left                     & 312 & 0.577 & 0.065 & 0.731 & 0.482 & 0.189 \\
Parahippocampal - right             & 176 & 0.524 & 0.029 & 0.599 & 0.484 & 0.065 \\
Parahippocampal - left              & 156 & 0.527 & 0.032 & 0.598 & 0.482 & 0.058 \\
Entorhinal - left                   & 132 & 0.545 & 0.041 & 0.651 & 0.482 & 0.041 \\
Ventral Diencephalon - left         & 124 & 0.574 & 0.066 & 0.783 & 0.483 & 0.020 \\
Entorhinal - right                  & 106 & 0.541 & 0.045 & 0.663 & 0.482 & 0.033 \\
Putamen - left                      & 46 & 0.549 & 0.040 & 0.628 & 0.484 & 0.007 \\
Ventral Diencephalon - right        & 38 & 0.537 & 0.050 & 0.687 & 0.483 & 0.006 \\
Pallidum - left                     & 23 & 0.519 & 0.025 & 0.583 & 0.485 & 0.014 \\
Fusiform - left                     & 5 & 0.502 & 0.013 & 0.522 & 0.487 & 0.000 \\
Fusiform - right                    & 2 & 0.484 & 0.000 & 0.484 & 0.483 & 0.000 \\
Putamen - right                     & 2 & 0.495 & 0.001 & 0.496 & 0.495 & 0.000 \\
-                                   & - & -  & - & -  & - & - \\
-                                   & - & -  & - & -  & - & - \\
-                                   & - & -  & - & -  & - & - \\
\bottomrule
\end{tabular}
\end{spacing}
\caption*{(b)}
\end{minipage}

\end{table*}

\begin{figure*}[t!]
    \centering
    \begin{subfigure}{\columnwidth}
        \includegraphics[width=\linewidth]{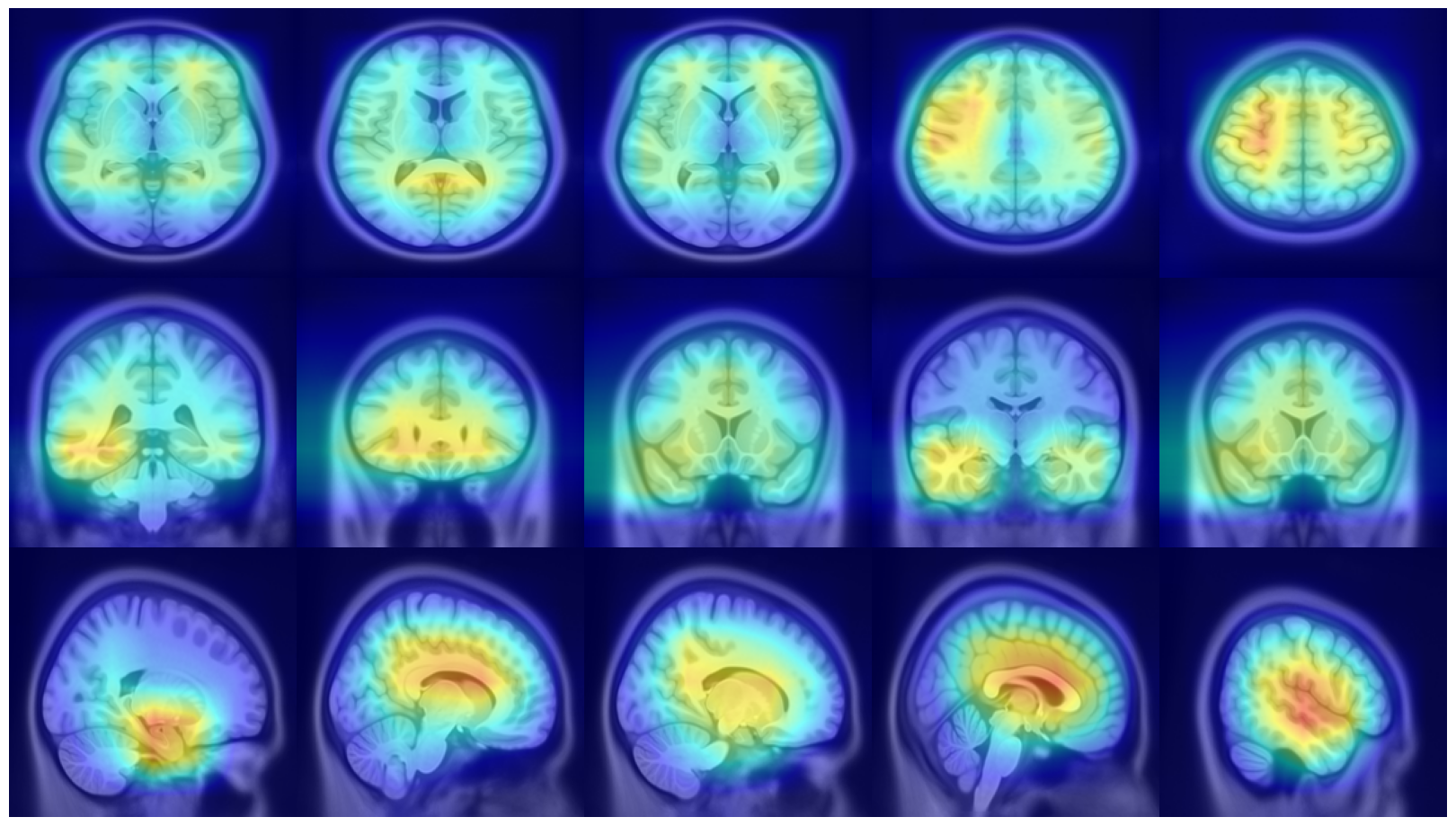}
        \caption{}
    \end{subfigure}
    \hfill
    \begin{subfigure}{\columnwidth}
        \includegraphics[width=\linewidth]{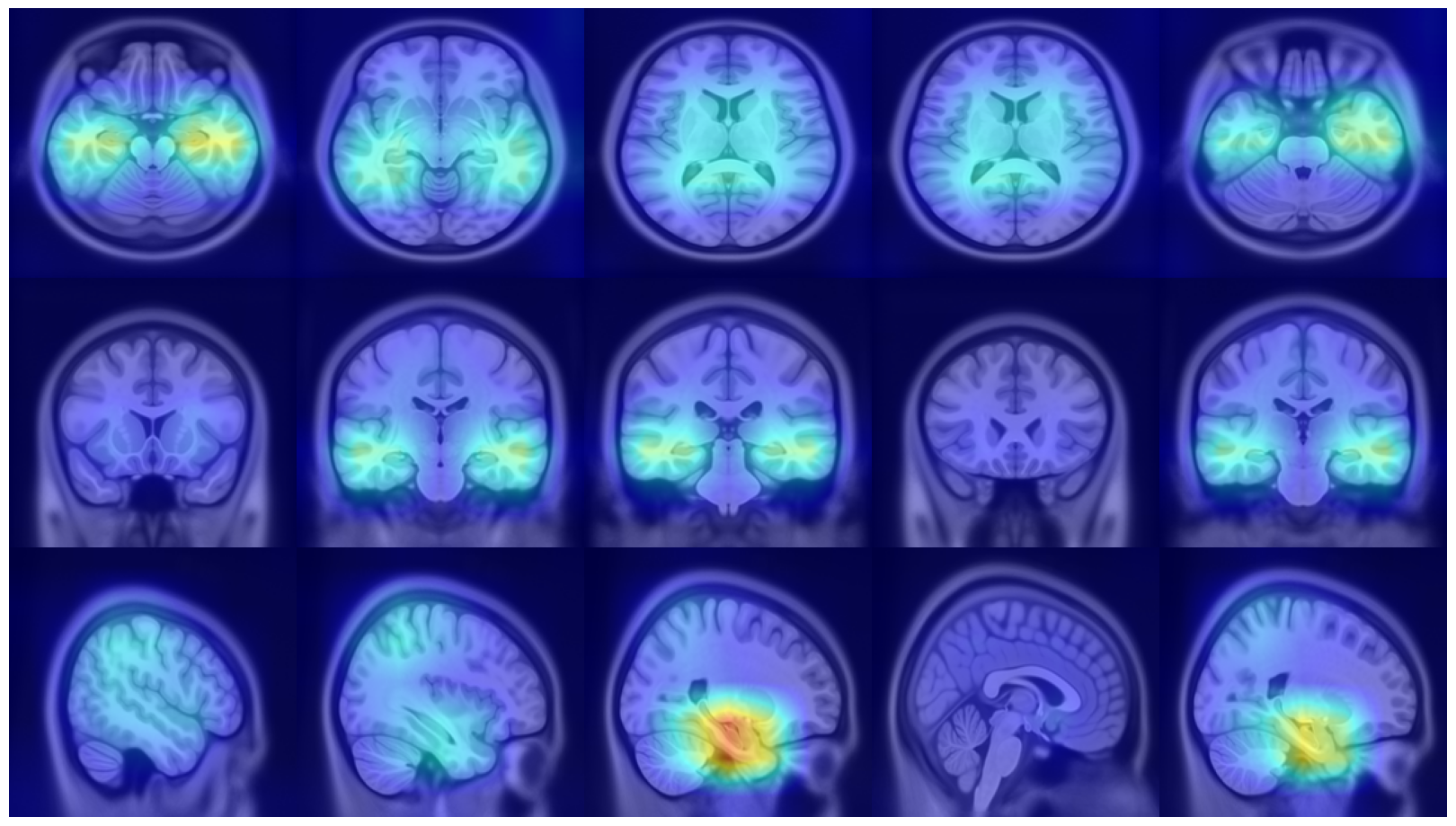}
        \caption{}
    \end{subfigure}
    \caption{\textbf{GradCAM++} visualization of five slices for each plane selected randomly for \textbf{our model} (a) and \textbf{Attention Transformer} (b)}
    \label{fig:gradcam_ours_transformer}
\end{figure*}

\begin{figure*}
    \centering
    \begin{minipage}{0.969\columnwidth}
        \begin{subfigure}{0.49\linewidth}
            \includegraphics[width=\linewidth]{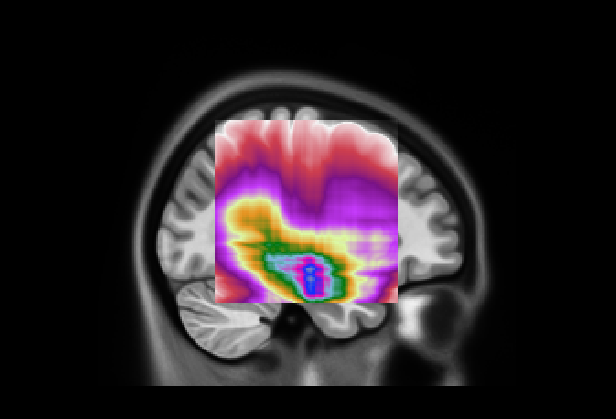}
        \end{subfigure}\hfill
        \begin{subfigure}{0.49\linewidth}
            \includegraphics[width=\linewidth]{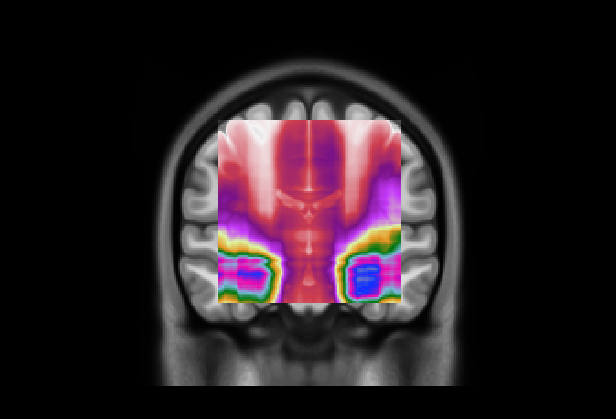}
        \end{subfigure}\\

        \begin{subfigure}{0.49\linewidth}
            \includegraphics[width=\linewidth]{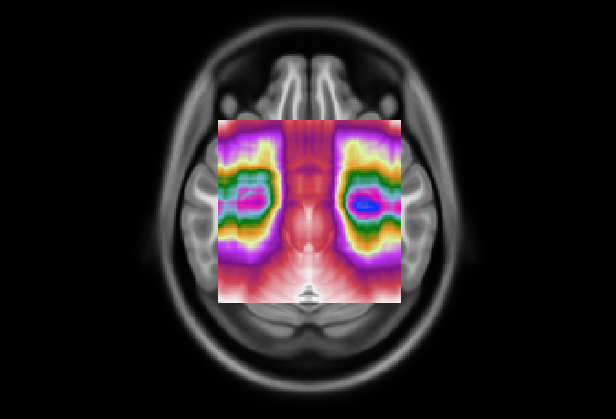}
        \end{subfigure}\hfill
        \begin{subfigure}{0.49\linewidth}
            \includegraphics[width=\linewidth]{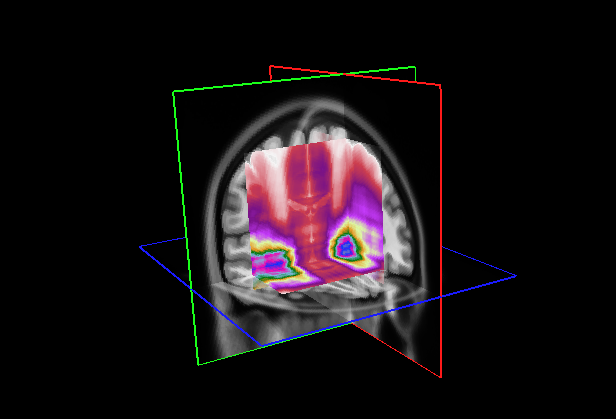}
        \end{subfigure}
    \end{minipage}\hfill
    \begin{minipage}{\columnwidth}
        \begin{subfigure}{0.49\linewidth}
            \includegraphics[width=\linewidth]{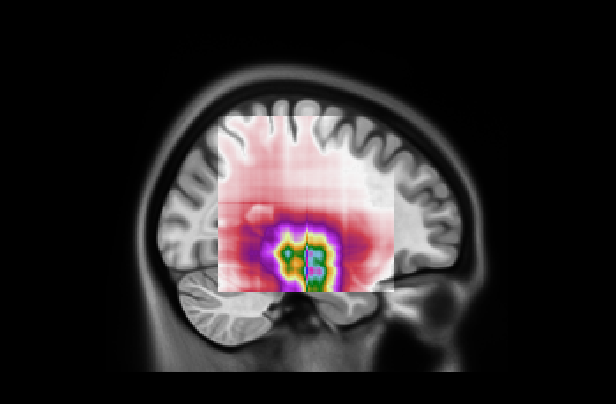}
        \end{subfigure}\hfill
        \begin{subfigure}{0.49\linewidth}
            \includegraphics[width=\linewidth]{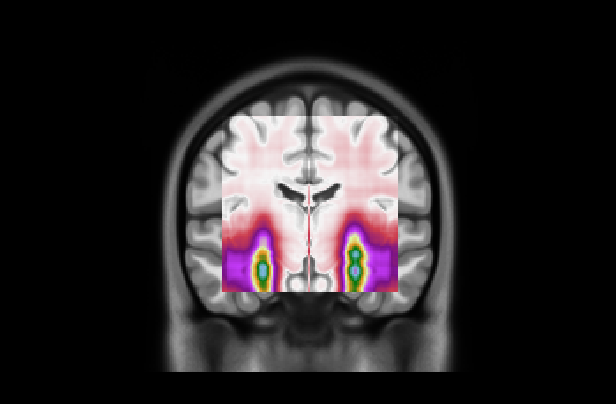}
        \end{subfigure}\\

        \begin{subfigure}{0.49\linewidth}
            \includegraphics[width=\linewidth]{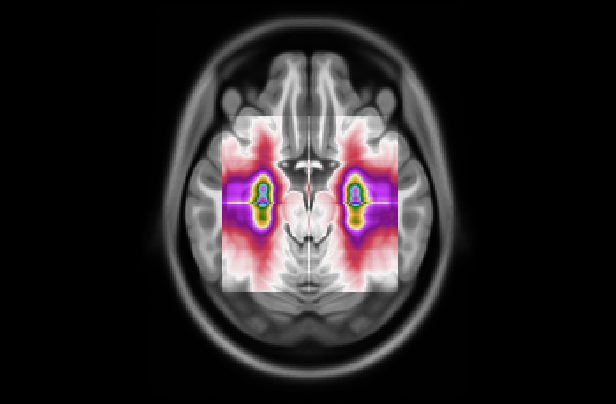}
        \end{subfigure}\hfill
        \begin{subfigure}{0.49\linewidth}
            \includegraphics[width=\linewidth]{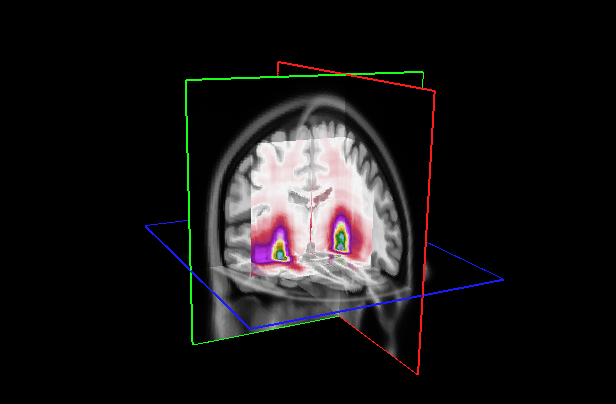}
        \end{subfigure}
    \end{minipage}
    \caption{(\textbf{Left}) Visualization of mean 3D GradCAM++ map of entire dataset overlapped to MNI152 template with \textbf{our model}; (\textbf{Right}) Visualization of mean 3D GradCAM++ map of entire dataset overlapped to MNI152 template with \textbf{Attention Transformer};}
    \label{fig:xai_visual_our_transformer}
\end{figure*}

This subsection examines the visual results and quantitative analysis concerning the brain areas emphasized by each model. Fig. \ref{fig:xai_visual_our}, on the left, displays the attentional weight distributions across the three planes, averaged over all five folds of the cross-validation. This averaging provides a comprehensive view of the data distribution across all images in the dataset. Starting from the entire dataset distributions, the 3D attentional map was created as detailed in Section \ref{sec:xai_approach}. The averaged 3D map was enhanced by a factor of 10 and overlaid on the MNI152 template, which is representative of a typical patient's brain. Combining this template with its corresponding atlas facilitates the identification of regions that, on average, received attention from the models. The right side of Fig. \ref{fig:xai_visual_our} shows the explainable MRI generated. The visual representation also indicates that the network targets the \emph{medial temporal lobe region}, as suggested by the distributions. This result is confirmed by the quantitative analysis shown in Table \ref{tab:metrics_awarenet_our_attention}, which reports the metrics for the 20 more extensive regions selected by our model and by AwareNet. As shown in Table \ref{tab:metrics_awarenet_our_attention}(a), the three largest regions focused by our diagnosis model are the \emph{hippocampus}, the \emph{parahippocampus}, and the \emph{amygdala}. In contrast, the 3D attentional map generated with AwareNet appears to focus on different regions. The first 3 regions highlighted are Cerebellum Gray Matter, Lateral Occipital, and Fusiform. The right part of the hippocampus appears only after them. From this result, it is also possible to note that with the same 99.9 percentile threshold for binarization, our model highlights a much more localized region. Specifically, our model selects 17 regions with a strong concentration in the top 14. On the contrary, AwareNet highlights 88 regions, 68 of which have been omitted in the Table \ref{tab:metrics_awarenet_our_attention}.

We also examined the interpretability of the Attention Transformer model proposed in \citep{2d_slice_altay2021preclinical}. For comparison, we created 2D saliency maps using the GradCAM++ algorithm \citep{Chattopadhay_2018_gradcam++} across all three views and all five test sets from the different folds. These maps were combined to create a unified average 3D saliency map as outlined in Section \ref{sec:xai_results}. This method was also applied to generate equivalent results from the saliency maps produced using our diagnostic model. As shown in Fig. \ref{fig:gradcam_ours_transformer}, the 2D maps produced by our method are generally sparser compared to those from the Attention Transformer. The method introduced in \citep{2d_slice_altay2021preclinical} employs a cross-attention mechanism via a Multi-Head. Therefore, it is plausible that the Multi-Head allows to generate 2D maps that align more meaningfully within the 3D context. This finding suggests that our approach may consider less contextual information from adjacent slices unless it is particularly relevant. In contrast, the cross-attention in the Attention Transformer might enable a more cohesive representation of the entire 3D space by considering both the local features within slices and their contextual interactions. This behavior is further clarified by creating 3D maps and overlaying them on the MNI152 template, similar to the attentional maps. As illustrated in Fig. \ref{fig:xai_visual_our_transformer} on the left, the 3D maps created using our model cover a broader and less concentrated area compared to those produced by the Attention Transformer, which are shown on the right. However, similar to the 3D attentional maps, both models predominantly focus on an area surrounding the hippocampus. As detailed in Table \ref{tab:metrics_transformer_our_attention}, both models identify key areas, such as the hippocampus and the amygdala, as significant. However, the emphasis on other regions varies markedly between the two. In the attention transformer model, there is a noticeable focus on the inferior lateral ventricles and the parahippocampal region, areas less emphasized by our model in this case. This result indicates that the Attention Transformer using cross-attention in combination with GradCAM can produce results similar to those obtained by our method with a 3D attentional map. As seen in Tables \ref{tab:metrics_awarenet_our_attention}(a) and \ref{tab:metrics_transformer_our_attention}(b), the first four areas on which our model focused with our approach are the same as those focused on by Attention Transformer with GradCAM. In contrast, our model with GradCAM shows broader involvement with regions such as the superior and middle temporal areas, which are not as prominent in the other cases.

\section{Discussion}
\label{sec:discussions}
This section discusses the main findings of our study, including the prevalence of VGG architectures, the efficacy of double transfer learning, the effectiveness of our method against state-of-the-art ones, and the implications of our XAI in the context of medical imaging and diagnostics. Additionally, we thoroughly discuss the limitations and potential areas for further investigation in our study.

\subsection{VGG for AD diagnosis}
Different studies on AD have suggested using models from the VGG architecture family \citep{2d_slice_hu2023conv, hu2023vgg, mehmood2021transfervgg}. The selection of this family is supported by their effective performance in medical imaging, mainly when using transfer learning techniques \citep{mehmood2021transfervgg}. While it is difficult to justify the clear superiority of this relatively simple architecture over more sophisticated ones such as EfficientNet and ResNet, its structure and pre-training stage may be the reason. The VGG models, for instance, VGG16, include a large number of parameters, approximately 138 million, with about 110 million dedicated to the classification layers and the remaining 28 million to convolutional layers. This structure suggests that most of the specific pattern recognition capabilities for class distinction learned from ImageNet likely reside in the classifier module. Therefore, the convolutional part may have acquired more general and transferable features than those in ResNet and EfficientNet, which rely on fully connected layers for classification.

\subsection{Transfer domain knowledge for AD prediction}
This study used a double transfer learning strategy to enhance model performance for the AD prediction task. The findings indicate that incorporating this additional domain-specific knowledge allows the model to more effectively differentiate subtle variations in brain morphology associated with early MCI progression. The utilization of such a strategy addresses the challenges posed by limited training data, which is a common issue in medical imaging tasks due to privacy concerns and the difficulty of obtaining large annotated datasets. However, it is crucial to note that the model's performance, while promising, needs further improvement. The modest MCC values indicate that the model, although capable of identifying relevant patterns, may still struggle with generalization across diverse patient profiles or imaging conditions. This limitation underscores the need to further refine the model's architecture and training regimen, possibly by integrating richer datasets or applying more sophisticated image augmentation techniques to enhance its robustness and clinical applicability. We did not perform an XAI evaluation on this task due to the evident model's current explanatory limitations, underscored by its modest correlation metric. As the model's predictive accuracy and reliability improve, integrating advanced XAI techniques to provide deeper insights into the decision-making process will become essential. This advancement will not only increase transparency but could also contribute to the discovery of new structural brain markers that will help predict the disease well in advance.

\subsection{Comparison with other attention-based methods}
The evaluation of our proposed model against state-of-the-art attention-based techniques reveals several noteworthy insights, particularly in employing a more straightforward, lightweight architecture in medical imaging tasks constrained by limited data availability. Our model's performance suggests that a meticulously designed, less complex network can rival or surpass more intricate systems in diagnosis and explainability. The successful application of 2D CNNs with attention mechanisms in tasks traditionally reserved for 3D CNNs is particularly intriguing. It suggests that simpler 2D networks can effectively extract and utilize the spatial information necessary for accurate diagnosis with the proper architectural considerations and training strategies. This finding could significantly impact the computational efficiency and accessibility of deep learning models for medical imaging, enabling their deployment in more diverse clinical environments with varying resource availability. 

\subsection{XAI Evaluation and Interpretability}
Our XAI methods analysis has showcased both the potential and the challenges of interpreting DL models. Our findings show that the Attention Transformer model with GradCAM and our model, both with and without GradCAM, consistently highlighted regions such as the hippocampus, parahippocampal gyrus, amygdala and ventricles which are well-documented in literature as being affected by AD \citep{rao2022hippocampus, van2000parahippocampal, poulin2011amygdala, ferrarini2006shape_ventricles}. 
However, our attention-based approach obtained the most relevant clinical results since the algorithm ordered the areas in the same order of importance that radiologists give to diagnose AD and is the only one that gives almost the same importance to the left and right portions of a brain part that is reasonable in clinical practice. 

A comprehensive assessment must incorporate clinical expertise to compare XAI methods and models. Neurologists and radiologists play a pivotal role in interpreting these XAI outputs, as their expertise in recognizing AD-specific biomarkers is crucial for validating the clinical relevance of the areas highlighted by the AI models. 

Even though we had the opportunity to validate the results with clinical experts, we know that involving clinical experts in the evaluation process presents its own challenges. It requires access to a panel of specialists willing to participate in such studies and a methodological framework that allows for averaging their insights to attain statistically significant conclusions. This process can be resource-intensive and difficult to implement across different studies, making it a less feasible option for consistent use. Metrics should be developed to simplify comparability across models and XAI methods in medical imaging. Such metrics could evaluate the relevance of highlighted regions within the context of a pathology. It should assess the localization and focus of maps and correlate these aspects with the known pathological features of the disease. By establishing these metrics, it could be easy to assess the comparison and validation of XAI approaches.

\subsection{Limitations}
In this section, we discuss several limitations identified in our study that warrant further exploration to enhance the robustness and applicability of our model. Firstly, the effectiveness of our method was assessed within a relatively limited dataset scenario, utilizing the ADNI1: Complete 1Yr 1.5T collection. While the model demonstrated promising results in this context, its performance against state-of-the-art models in larger datasets remains to be determined. Expanding the scope of testing to include a broader range of datasets with varying characteristics could help establish the scalability and general effectiveness of the proposed method. 

Secondly, our XAI approach requires training three separate models, each tailored to one of the principal anatomical planes: axial, coronal, and sagittal. This requirement increases the computational demand and complexity of the training process since each model must be individually optimized and evaluated. Lastly, our approach lacks of integration between the models corresponding to different anatomical planes. Currently, each model is trained independently, without considering the inherent correlations between these planes. This segmentation of the training process potentially overlooks critical spatial information that could be utilized to enhance diagnostic accuracy and explainability. Future improvements could involve the development of integrated multi-view learning strategies, which could simultaneously process and cross-validate information across different planes, offering a more comprehensive understanding of the imaging data.

To evaluate the consistency of interpretability in our approach, we examined the variability of attentional weight distributions across the five different folds used for cross-validation. Our evaluation leverages a clinical hypothesis specific to AD, which is believed to affect similar brain areas across different patients \citep{rao2022hippocampus}. However, we acknowledge that this assumption of spatial consistency may not hold for other diseases, e.g. cancer, where affected areas can vary significantly among patients. Therefore, this evaluation cannot be generalized a priori but is applicable in specific cases like AD, where the disease is supposed to impact similar regions in different patients.

\section{Conclusions}
\label{sec:conclusions}

In this study, we presented an innovative method for AD diagnosis using pre-trained 2D CNNs to classify 3D volumes. Our approach integrates an attention mechanism to enhance the interpretability and accuracy of diagnosing AD and differentiating stable from progressive mild cognitive impairment. Our method outperformed traditional baseline methods, achieving an MCC of 0.712 for distinguishing AD from CN subjects and 0.442 for sMCI from pMCI subjects. These results demonstrate the capability of our approach to classify these conditions effectively. A novel aspect of our work is the enhancement of model explainability. We successfully implemented voxel-level attention activation maps highlighting specific brain areas implicated in AD, such as the hippocampus, amygdala, parahippocampus, and inferior lateral ventricles. These regions are known to be crucial in AD pathology, making our model's outputs medically relevant. Our approach also includes a double transfer learning strategy that leverages pre-trained models to improve performance on limited datasets. This strategy utilizes knowledge transfer from the AD vs. CN task to enhance the model's sensitivity to subtle morphological changes associated with the progression of MCI, which are often hard to detect. In conclusion, our method advances diagnostic capabilities using 3D MRI scans for early AD detection and tries to address the critical need for explainability in medical imaging AI applications. By providing insights into the model's decision-making process, our approach helps bridge the gap between AI tools and clinical usability, making it a valuable asset for neurodegenerative disease research and potentially aiding in the clinical diagnosis and monitoring of AD progression.

\section*{Declaration of generative AI and AI-assisted technologies in the writing process}
During the preparation of this work the authors used ChatGPT in order to improve language and readability. After using this tool/service, the authors reviewed and edited the content as needed and take full responsibility for the content of the publication.

\section*{Acknowledgements}
Project ECS 0000024 “Ecosistema dell’innovazione - Rome Technopole” financed by EU in NextGenerationEU plan through MUR Decree n. 1051 23.06.2022 PNRR Missione 4 Componente 2 Investimento 1.5 - CUP H33C22000420001

\bibliographystyle{elsarticle-harv.bst} \biboptions{comma,sort,comma,authoryear}
\bibliography{refs}

\begin{thebibliography}{76}
\expandafter\ifx\csname natexlab\endcsname\relax\def\natexlab#1{#1}\fi
\providecommand{\url}[1]{\texttt{#1}}
\providecommand{\href}[2]{#2}
\providecommand{\path}[1]{#1}
\providecommand{\DOIprefix}{doi:}
\providecommand{\ArXivprefix}{arXiv:}
\providecommand{\URLprefix}{URL: }
\providecommand{\Pubmedprefix}{pmid:}
\providecommand{\doi}[1]{\href{http://dx.doi.org/#1}{\path{#1}}}
\providecommand{\Pubmed}[1]{\href{pmid:#1}{\path{#1}}}
\providecommand{\bibinfo}[2]{#2}
\ifx\xfnm\relax \def\xfnm[#1]{\unskip,\space#1}\fi
\bibitem[{Altay et~al.(2021)Altay, S{\'a}nchez, James, Faraone, Velipasalar and Salekin}]{2d_slice_altay2021preclinical}
\bibinfo{author}{Altay, F.}, \bibinfo{author}{S{\'a}nchez, G.R.}, \bibinfo{author}{James, Y.}, \bibinfo{author}{Faraone, S.V.}, \bibinfo{author}{Velipasalar, S.}, \bibinfo{author}{Salekin, A.}, \bibinfo{year}{2021}.
\newblock \bibinfo{title}{Preclinical stage alzheimer's disease detection using magnetic resonance image scans}, in: \bibinfo{booktitle}{Proceedings of the AAAI Conference on Artificial Intelligence}, pp. \bibinfo{pages}{15088--15097}.
\bibitem[{Avants et~al.(2008)Avants, Epstein, Grossman and Gee}]{avants2008symmetric}
\bibinfo{author}{Avants, B.B.}, \bibinfo{author}{Epstein, C.L.}, \bibinfo{author}{Grossman, M.}, \bibinfo{author}{Gee, J.C.}, \bibinfo{year}{2008}.
\newblock \bibinfo{title}{Symmetric diffeomorphic image registration with cross-correlation: evaluating automated labeling of elderly and neurodegenerative brain}.
\newblock \bibinfo{journal}{Medical image analysis} \bibinfo{volume}{12}, \bibinfo{pages}{26--41}.
\bibitem[{Avants et~al.(2014)Avants, Tustison, Stauffer, Song, Wu and Gee}]{avants2014insight}
\bibinfo{author}{Avants, B.B.}, \bibinfo{author}{Tustison, N.J.}, \bibinfo{author}{Stauffer, M.}, \bibinfo{author}{Song, G.}, \bibinfo{author}{Wu, B.}, \bibinfo{author}{Gee, J.C.}, \bibinfo{year}{2014}.
\newblock \bibinfo{title}{The insight toolkit image registration framework}.
\newblock \bibinfo{journal}{Frontiers in neuroinformatics} \bibinfo{volume}{8}, \bibinfo{pages}{44}.
\bibitem[{Basaia et~al.(2019)Basaia, Agosta, Wagner, Canu, Magnani, Santangelo, Filippi, Initiative et~al.}]{3d_basaia2019automated}
\bibinfo{author}{Basaia, S.}, \bibinfo{author}{Agosta, F.}, \bibinfo{author}{Wagner, L.}, \bibinfo{author}{Canu, E.}, \bibinfo{author}{Magnani, G.}, \bibinfo{author}{Santangelo, R.}, \bibinfo{author}{Filippi, M.}, \bibinfo{author}{Initiative, A.D.N.}, et~al., \bibinfo{year}{2019}.
\newblock \bibinfo{title}{Automated classification of alzheimer's disease and mild cognitive impairment using a single mri and deep neural networks}.
\newblock \bibinfo{journal}{NeuroImage: Clinical} \bibinfo{volume}{21}, \bibinfo{pages}{101645}.
\bibitem[{Brookmeyer et~al.(2007)Brookmeyer, Johnson, Ziegler-Graham and Arrighi}]{brookmeyer2007forecasting}
\bibinfo{author}{Brookmeyer, R.}, \bibinfo{author}{Johnson, E.}, \bibinfo{author}{Ziegler-Graham, K.}, \bibinfo{author}{Arrighi, H.M.}, \bibinfo{year}{2007}.
\newblock \bibinfo{title}{Forecasting the global burden of alzheimer’s disease}.
\newblock \bibinfo{journal}{Alzheimer's \& dementia} \bibinfo{volume}{3}, \bibinfo{pages}{186--191}.
\bibitem[{Carcagn{\`\i} et~al.(2023)Carcagn{\`\i}, Leo, Del~Coco, Distante and De~Salve}]{2d_slice_carcagni2023convolution}
\bibinfo{author}{Carcagn{\`\i}, P.}, \bibinfo{author}{Leo, M.}, \bibinfo{author}{Del~Coco, M.}, \bibinfo{author}{Distante, C.}, \bibinfo{author}{De~Salve, A.}, \bibinfo{year}{2023}.
\newblock \bibinfo{title}{Convolution neural networks and self-attention learners for alzheimer dementia diagnosis from brain mri}.
\newblock \bibinfo{journal}{Sensors} \bibinfo{volume}{23}, \bibinfo{pages}{1694}.
\bibitem[{Chattopadhay et~al.(2018)Chattopadhay, Sarkar, Howlader and Balasubramanian}]{Chattopadhay_2018_gradcam++}
\bibinfo{author}{Chattopadhay, A.}, \bibinfo{author}{Sarkar, A.}, \bibinfo{author}{Howlader, P.}, \bibinfo{author}{Balasubramanian, V.N.}, \bibinfo{year}{2018}.
\newblock \bibinfo{title}{Grad-cam++: Generalized gradient-based visual explanations for deep convolutional networks}, in: \bibinfo{booktitle}{2018 IEEE Winter Conference on Applications of Computer Vision (WACV)}, \bibinfo{publisher}{IEEE}.
\newblock \URLprefix \url{http://dx.doi.org/10.1109/WACV.2018.00097}, \DOIprefix\doi{10.1109/wacv.2018.00097}.
\bibitem[{Chicco and Jurman(2020)}]{mcc_chicco2020advantages}
\bibinfo{author}{Chicco, D.}, \bibinfo{author}{Jurman, G.}, \bibinfo{year}{2020}.
\newblock \bibinfo{title}{The advantages of the matthews correlation coefficient (mcc) over f1 score and accuracy in binary classification evaluation}.
\newblock \bibinfo{journal}{BMC genomics} \bibinfo{volume}{21}, \bibinfo{pages}{1--13}.
\bibitem[{Dosovitskiy et~al.(2020)Dosovitskiy, Beyer, Kolesnikov, Weissenborn, Zhai, Unterthiner, Dehghani, Minderer, Heigold, Gelly et~al.}]{vit_dosovitskiy2020image}
\bibinfo{author}{Dosovitskiy, A.}, \bibinfo{author}{Beyer, L.}, \bibinfo{author}{Kolesnikov, A.}, \bibinfo{author}{Weissenborn, D.}, \bibinfo{author}{Zhai, X.}, \bibinfo{author}{Unterthiner, T.}, \bibinfo{author}{Dehghani, M.}, \bibinfo{author}{Minderer, M.}, \bibinfo{author}{Heigold, G.}, \bibinfo{author}{Gelly, S.}, et~al., \bibinfo{year}{2020}.
\newblock \bibinfo{title}{An image is worth 16x16 words: Transformers for image recognition at scale}.
\newblock \bibinfo{journal}{arXiv preprint arXiv:2010.11929} .
\bibitem[{Duchesne et~al.(2008)Duchesne, Caroli, Geroldi, Barillot, Frisoni and Collins}]{duchesne2008mri}
\bibinfo{author}{Duchesne, S.}, \bibinfo{author}{Caroli, A.}, \bibinfo{author}{Geroldi, C.}, \bibinfo{author}{Barillot, C.}, \bibinfo{author}{Frisoni, G.B.}, \bibinfo{author}{Collins, D.L.}, \bibinfo{year}{2008}.
\newblock \bibinfo{title}{Mri-based automated computer classification of probable ad versus normal controls}.
\newblock \bibinfo{journal}{IEEE transactions on medical imaging} \bibinfo{volume}{27}, \bibinfo{pages}{509--520}.
\bibitem[{Ebrahimi et~al.(2021)Ebrahimi, Luo, Chiong, Initiative et~al.}]{2d_slice_ebrahimi2021deep}
\bibinfo{author}{Ebrahimi, A.}, \bibinfo{author}{Luo, S.}, \bibinfo{author}{Chiong, R.}, \bibinfo{author}{Initiative, A.D.N.}, et~al., \bibinfo{year}{2021}.
\newblock \bibinfo{title}{Deep sequence modelling for alzheimer's disease detection using mri}.
\newblock \bibinfo{journal}{Computers in Biology and Medicine} \bibinfo{volume}{134}, \bibinfo{pages}{104537}.
\bibitem[{Ewers et~al.(2011)Ewers, Sperling, Klunk, Weiner and Hampel}]{ewers2011neuroimaging}
\bibinfo{author}{Ewers, M.}, \bibinfo{author}{Sperling, R.A.}, \bibinfo{author}{Klunk, W.E.}, \bibinfo{author}{Weiner, M.W.}, \bibinfo{author}{Hampel, H.}, \bibinfo{year}{2011}.
\newblock \bibinfo{title}{Neuroimaging markers for the prediction and early diagnosis of alzheimer's disease dementia}.
\newblock \bibinfo{journal}{Trends in neurosciences} \bibinfo{volume}{34}, \bibinfo{pages}{430--442}.
\bibitem[{Falahati et~al.(2014)Falahati, Westman and Simmons}]{ML_falahati2014multivariate}
\bibinfo{author}{Falahati, F.}, \bibinfo{author}{Westman, E.}, \bibinfo{author}{Simmons, A.}, \bibinfo{year}{2014}.
\newblock \bibinfo{title}{Multivariate data analysis and machine learning in alzheimer's disease with a focus on structural magnetic resonance imaging}.
\newblock \bibinfo{journal}{Journal of Alzheimer's disease} \bibinfo{volume}{41}, \bibinfo{pages}{685--708}.
\bibitem[{Feng et~al.(2022)Feng, Provenzano, Small and Initiative}]{3d_feng2022deep}
\bibinfo{author}{Feng, X.}, \bibinfo{author}{Provenzano, F.A.}, \bibinfo{author}{Small, S.A.}, \bibinfo{author}{Initiative, A.D.N.}, \bibinfo{year}{2022}.
\newblock \bibinfo{title}{A deep learning mri approach outperforms other biomarkers of prodromal alzheimer’s disease}.
\newblock \bibinfo{journal}{Alzheimer's Research \& Therapy} \bibinfo{volume}{14}, \bibinfo{pages}{45}.
\bibitem[{Ferrarini et~al.(2006)Ferrarini, Palm, Olofsen, van Buchem, Reiber and Admiraal-Behloul}]{ferrarini2006shape_ventricles}
\bibinfo{author}{Ferrarini, L.}, \bibinfo{author}{Palm, W.M.}, \bibinfo{author}{Olofsen, H.}, \bibinfo{author}{van Buchem, M.A.}, \bibinfo{author}{Reiber, J.H.}, \bibinfo{author}{Admiraal-Behloul, F.}, \bibinfo{year}{2006}.
\newblock \bibinfo{title}{Shape differences of the brain ventricles in alzheimer's disease}.
\newblock \bibinfo{journal}{Neuroimage} \bibinfo{volume}{32}, \bibinfo{pages}{1060--1069}.
\bibitem[{Fonov et~al.(2011)Fonov, Evans, Botteron, Almli, McKinstry, Collins, Group et~al.}]{fonov2011unbiased}
\bibinfo{author}{Fonov, V.}, \bibinfo{author}{Evans, A.C.}, \bibinfo{author}{Botteron, K.}, \bibinfo{author}{Almli, C.R.}, \bibinfo{author}{McKinstry, R.C.}, \bibinfo{author}{Collins, D.L.}, \bibinfo{author}{Group, B.D.C.}, et~al., \bibinfo{year}{2011}.
\newblock \bibinfo{title}{Unbiased average age-appropriate atlases for pediatric studies}.
\newblock \bibinfo{journal}{Neuroimage} \bibinfo{volume}{54}, \bibinfo{pages}{313--327}.
\bibitem[{Fonov et~al.(2009)Fonov, Evans, McKinstry, Almli and Collins}]{fonov2009unbiased}
\bibinfo{author}{Fonov, V.S.}, \bibinfo{author}{Evans, A.C.}, \bibinfo{author}{McKinstry, R.C.}, \bibinfo{author}{Almli, C.R.}, \bibinfo{author}{Collins, D.}, \bibinfo{year}{2009}.
\newblock \bibinfo{title}{Unbiased nonlinear average age-appropriate brain templates from birth to adulthood}.
\newblock \bibinfo{journal}{NeuroImage} \bibinfo{volume}{47}, \bibinfo{pages}{S102}.
\bibitem[{Frisoni et~al.(2010)Frisoni, Fox, Jack~Jr, Scheltens and Thompson}]{frisoni2010clinical}
\bibinfo{author}{Frisoni, G.B.}, \bibinfo{author}{Fox, N.C.}, \bibinfo{author}{Jack~Jr, C.R.}, \bibinfo{author}{Scheltens, P.}, \bibinfo{author}{Thompson, P.M.}, \bibinfo{year}{2010}.
\newblock \bibinfo{title}{The clinical use of structural mri in alzheimer disease}.
\newblock \bibinfo{journal}{Nature Reviews Neurology} \bibinfo{volume}{6}, \bibinfo{pages}{67--77}.
\bibitem[{Goenka et~al.(2022)Goenka, Goenka and Tiwari}]{3d_patch_goenka2022patch}
\bibinfo{author}{Goenka, N.}, \bibinfo{author}{Goenka, A.}, \bibinfo{author}{Tiwari, S.}, \bibinfo{year}{2022}.
\newblock \bibinfo{title}{Patch-based classification for alzheimer disease using smri}, in: \bibinfo{booktitle}{2022 International Conference on Emerging Smart Computing and Informatics (ESCI)}, \bibinfo{organization}{IEEE}. pp. \bibinfo{pages}{1--5}.
\bibitem[{Gorgolewski et~al.(2016)Gorgolewski, Auer, Calhoun, Craddock, Das, Duff, Flandin, Ghosh, Glatard, Halchenko et~al.}]{bids_gorgolewski2016brain}
\bibinfo{author}{Gorgolewski, K.J.}, \bibinfo{author}{Auer, T.}, \bibinfo{author}{Calhoun, V.D.}, \bibinfo{author}{Craddock, R.C.}, \bibinfo{author}{Das, S.}, \bibinfo{author}{Duff, E.P.}, \bibinfo{author}{Flandin, G.}, \bibinfo{author}{Ghosh, S.S.}, \bibinfo{author}{Glatard, T.}, \bibinfo{author}{Halchenko, Y.O.}, et~al., \bibinfo{year}{2016}.
\newblock \bibinfo{title}{The brain imaging data structure, a format for organizing and describing outputs of neuroimaging experiments}.
\newblock \bibinfo{journal}{Scientific data} \bibinfo{volume}{3}, \bibinfo{pages}{1--9}.
\bibitem[{Haller et~al.(2011)Haller, Lovblad and Giannakopoulos}]{ML_haller2011principles}
\bibinfo{author}{Haller, S.}, \bibinfo{author}{Lovblad, K.O.}, \bibinfo{author}{Giannakopoulos, P.}, \bibinfo{year}{2011}.
\newblock \bibinfo{title}{Principles of classification analyses in mild cognitive impairment (mci) and alzheimer disease}.
\newblock \bibinfo{journal}{Journal of Alzheimer's Disease} \bibinfo{volume}{26}, \bibinfo{pages}{389--394}.
\bibitem[{He et~al.(2015)He, Zhang, Ren and Sun}]{kaiming_he2015delving}
\bibinfo{author}{He, K.}, \bibinfo{author}{Zhang, X.}, \bibinfo{author}{Ren, S.}, \bibinfo{author}{Sun, J.}, \bibinfo{year}{2015}.
\newblock \bibinfo{title}{Delving deep into rectifiers: Surpassing human-level performance on imagenet classification}, in: \bibinfo{booktitle}{Proceedings of the IEEE international conference on computer vision}, pp. \bibinfo{pages}{1026--1034}.
\bibitem[{He et~al.(2016)He, Zhang, Ren and Sun}]{resnet_He2016}
\bibinfo{author}{He, K.}, \bibinfo{author}{Zhang, X.}, \bibinfo{author}{Ren, S.}, \bibinfo{author}{Sun, J.}, \bibinfo{year}{2016}.
\newblock \bibinfo{title}{Deep residual learning for image recognition}, in: \bibinfo{booktitle}{2016 {IEEE} Conference on Computer Vision and Pattern Recognition, {CVPR} 2016, Las Vegas, NV, USA, June 27-30, 2016}, \bibinfo{publisher}{{IEEE} Computer Society}. pp. \bibinfo{pages}{770--778}.
\newblock \DOIprefix\doi{10.1109/CVPR.2016.90}.
\bibitem[{Hon and Khan(2017)}]{2d_slice_hon2017towards}
\bibinfo{author}{Hon, M.}, \bibinfo{author}{Khan, N.M.}, \bibinfo{year}{2017}.
\newblock \bibinfo{title}{Towards alzheimer's disease classification through transfer learning}, in: \bibinfo{booktitle}{2017 IEEE International conference on bioinformatics and biomedicine (BIBM)}, \bibinfo{organization}{IEEE}. pp. \bibinfo{pages}{1166--1169}.
\bibitem[{Hu et~al.(2023a)Hu, Li, Wang, Zhang, Hou, Initiative et~al.}]{2d_slice_hu2023conv}
\bibinfo{author}{Hu, Z.}, \bibinfo{author}{Li, Y.}, \bibinfo{author}{Wang, Z.}, \bibinfo{author}{Zhang, S.}, \bibinfo{author}{Hou, W.}, \bibinfo{author}{Initiative, A.D.N.}, et~al., \bibinfo{year}{2023}a.
\newblock \bibinfo{title}{Conv-swinformer: Integration of cnn and shift window attention for alzheimer’s disease classification}.
\newblock \bibinfo{journal}{Computers in Biology and Medicine} \bibinfo{volume}{164}, \bibinfo{pages}{107304}.
\bibitem[{Hu et~al.(2023b)Hu, Wang, Jin and Hou}]{hu2023vgg}
\bibinfo{author}{Hu, Z.}, \bibinfo{author}{Wang, Z.}, \bibinfo{author}{Jin, Y.}, \bibinfo{author}{Hou, W.}, \bibinfo{year}{2023}b.
\newblock \bibinfo{title}{Vgg-tswinformer: Transformer-based deep learning model for early alzheimer’s disease prediction}.
\newblock \bibinfo{journal}{Computer Methods and Programs in Biomedicine} \bibinfo{volume}{229}, \bibinfo{pages}{107291}.
\bibitem[{Huang et~al.(2017)Huang, Liu, Van Der~Maaten and Weinberger}]{huang2017densely}
\bibinfo{author}{Huang, G.}, \bibinfo{author}{Liu, Z.}, \bibinfo{author}{Van Der~Maaten, L.}, \bibinfo{author}{Weinberger, K.Q.}, \bibinfo{year}{2017}.
\newblock \bibinfo{title}{Densely connected convolutional networks}, in: \bibinfo{booktitle}{Proceedings of the IEEE conference on computer vision and pattern recognition}, pp. \bibinfo{pages}{4700--4708}.
\bibitem[{Jack et~al.(2003)Jack, Slomkowski, Gracon, Hoover, Felmlee, Stewart, Xu, Shiung, O’brien, Cha et~al.}]{jack2003mri}
\bibinfo{author}{Jack, C.}, \bibinfo{author}{Slomkowski, M.}, \bibinfo{author}{Gracon, S.}, \bibinfo{author}{Hoover, T.}, \bibinfo{author}{Felmlee, J.}, \bibinfo{author}{Stewart, K.}, \bibinfo{author}{Xu, Y.}, \bibinfo{author}{Shiung, M.}, \bibinfo{author}{O’brien, P.}, \bibinfo{author}{Cha, R.}, et~al., \bibinfo{year}{2003}.
\newblock \bibinfo{title}{Mri as a biomarker of disease progression in a therapeutic trial of milameline for ad}.
\newblock \bibinfo{journal}{Neurology} \bibinfo{volume}{60}, \bibinfo{pages}{253--260}.
\bibitem[{Jenkinson et~al.(2012)Jenkinson, Beckmann, Behrens, Woolrich and Smith}]{jenkinson2012fsl}
\bibinfo{author}{Jenkinson, M.}, \bibinfo{author}{Beckmann, C.F.}, \bibinfo{author}{Behrens, T.E.}, \bibinfo{author}{Woolrich, M.W.}, \bibinfo{author}{Smith, S.M.}, \bibinfo{year}{2012}.
\newblock \bibinfo{title}{Fsl}.
\newblock \bibinfo{journal}{Neuroimage} \bibinfo{volume}{62}, \bibinfo{pages}{782--790}.
\bibitem[{Jetley et~al.(2018)Jetley, Lord, Lee and Torr}]{attention_exp_jetley2018learn}
\bibinfo{author}{Jetley, S.}, \bibinfo{author}{Lord, N.A.}, \bibinfo{author}{Lee, N.}, \bibinfo{author}{Torr, P.H.}, \bibinfo{year}{2018}.
\newblock \bibinfo{title}{Learn to pay attention}.
\newblock \bibinfo{journal}{arXiv preprint arXiv:1804.02391} .
\bibitem[{Jin et~al.(2019)Jin, Xu, Zhao, Hu, Yang, Liu, Jiang and Liu}]{3d_jin2019attention}
\bibinfo{author}{Jin, D.}, \bibinfo{author}{Xu, J.}, \bibinfo{author}{Zhao, K.}, \bibinfo{author}{Hu, F.}, \bibinfo{author}{Yang, Z.}, \bibinfo{author}{Liu, B.}, \bibinfo{author}{Jiang, T.}, \bibinfo{author}{Liu, Y.}, \bibinfo{year}{2019}.
\newblock \bibinfo{title}{Attention-based 3d convolutional network for alzheimer’s disease diagnosis and biomarkers exploration}, in: \bibinfo{booktitle}{2019 IEEE 16Th international symposium on biomedical imaging (ISBI 2019)}, \bibinfo{organization}{IEEE}. pp. \bibinfo{pages}{1047--1051}.
\bibitem[{Kang et~al.(2021)Kang, Lin, Zhang, Shen, Wu, Initiative et~al.}]{2d_slice_kang2021multi}
\bibinfo{author}{Kang, W.}, \bibinfo{author}{Lin, L.}, \bibinfo{author}{Zhang, B.}, \bibinfo{author}{Shen, X.}, \bibinfo{author}{Wu, S.}, \bibinfo{author}{Initiative, A.D.N.}, et~al., \bibinfo{year}{2021}.
\newblock \bibinfo{title}{Multi-model and multi-slice ensemble learning architecture based on 2d convolutional neural networks for alzheimer's disease diagnosis}.
\newblock \bibinfo{journal}{Computers in Biology and Medicine} \bibinfo{volume}{136}, \bibinfo{pages}{104678}.
\bibitem[{Klaiber et~al.(2021)Klaiber, Sauter, Baumgartl and Buettner}]{3d_review_klaiber2021systematic}
\bibinfo{author}{Klaiber, M.}, \bibinfo{author}{Sauter, D.}, \bibinfo{author}{Baumgartl, H.}, \bibinfo{author}{Buettner, R.}, \bibinfo{year}{2021}.
\newblock \bibinfo{title}{A systematic literature review on transfer learning for 3d-cnns}, in: \bibinfo{booktitle}{2021 international joint conference on neural networks (IJCNN)}, \bibinfo{organization}{IEEE}. pp. \bibinfo{pages}{1--10}.
\bibitem[{Kl{\"o}ppel et~al.(2008)Kl{\"o}ppel, Stonnington, Chu, Draganski, Scahill, Rohrer, Fox, Jack~Jr, Ashburner and Frackowiak}]{kloppel2008automatic}
\bibinfo{author}{Kl{\"o}ppel, S.}, \bibinfo{author}{Stonnington, C.M.}, \bibinfo{author}{Chu, C.}, \bibinfo{author}{Draganski, B.}, \bibinfo{author}{Scahill, R.I.}, \bibinfo{author}{Rohrer, J.D.}, \bibinfo{author}{Fox, N.C.}, \bibinfo{author}{Jack~Jr, C.R.}, \bibinfo{author}{Ashburner, J.}, \bibinfo{author}{Frackowiak, R.S.}, \bibinfo{year}{2008}.
\newblock \bibinfo{title}{Automatic classification of mr scans in alzheimer's disease}.
\newblock \bibinfo{journal}{Brain} \bibinfo{volume}{131}, \bibinfo{pages}{681--689}.
\bibitem[{Kushol et~al.(2022)Kushol, Masoumzadeh, Huo, Kalra and Yang}]{2d_slice_kushol2022addformer}
\bibinfo{author}{Kushol, R.}, \bibinfo{author}{Masoumzadeh, A.}, \bibinfo{author}{Huo, D.}, \bibinfo{author}{Kalra, S.}, \bibinfo{author}{Yang, Y.H.}, \bibinfo{year}{2022}.
\newblock \bibinfo{title}{Addformer: Alzheimer’s disease detection from structural mri using fusion transformer}, in: \bibinfo{booktitle}{2022 IEEE 19th International Symposium On Biomedical Imaging (ISBI)}, \bibinfo{organization}{IEEE}. pp. \bibinfo{pages}{1--5}.
\bibitem[{Lam et~al.(2013)Lam, Masellis, Freedman, Stuss and Black}]{lam2013clinical}
\bibinfo{author}{Lam, B.}, \bibinfo{author}{Masellis, M.}, \bibinfo{author}{Freedman, M.}, \bibinfo{author}{Stuss, D.T.}, \bibinfo{author}{Black, S.E.}, \bibinfo{year}{2013}.
\newblock \bibinfo{title}{Clinical, imaging, and pathological heterogeneity of the alzheimer's disease syndrome}.
\newblock \bibinfo{journal}{Alzheimer's research \& therapy} \bibinfo{volume}{5}, \bibinfo{pages}{1--14}.
\bibitem[{Liu et~al.(2023)Liu, Yuan, Li, Xu and Sheng}]{3d_patch_liu2023patch}
\bibinfo{author}{Liu, F.}, \bibinfo{author}{Yuan, S.}, \bibinfo{author}{Li, W.}, \bibinfo{author}{Xu, Q.}, \bibinfo{author}{Sheng, B.}, \bibinfo{year}{2023}.
\newblock \bibinfo{title}{Patch-based deep multi-modal learning framework for alzheimer’s disease diagnosis using multi-view neuroimaging}.
\newblock \bibinfo{journal}{Biomedical Signal Processing and Control} \bibinfo{volume}{80}, \bibinfo{pages}{104400}.
\bibitem[{Loewenstein et~al.(2006)Loewenstein, Acevedo, Agron, Issacson, Strauman, Crocco, Barker and Duara}]{loewenstein2006cognitive}
\bibinfo{author}{Loewenstein, D.A.}, \bibinfo{author}{Acevedo, A.}, \bibinfo{author}{Agron, J.}, \bibinfo{author}{Issacson, R.}, \bibinfo{author}{Strauman, S.}, \bibinfo{author}{Crocco, E.}, \bibinfo{author}{Barker, W.W.}, \bibinfo{author}{Duara, R.}, \bibinfo{year}{2006}.
\newblock \bibinfo{title}{Cognitive profiles in alzheimer’s disease and in mild cognitive impairment of different etiologies}.
\newblock \bibinfo{journal}{Dementia and geriatric cognitive disorders} \bibinfo{volume}{21}, \bibinfo{pages}{309--315}.
\bibitem[{Mehmood et~al.(2021)Mehmood, Yang, Feng, Wang, Ahmad, Khan, Maqsood and Yaqub}]{mehmood2021transfervgg}
\bibinfo{author}{Mehmood, A.}, \bibinfo{author}{Yang, S.}, \bibinfo{author}{Feng, Z.}, \bibinfo{author}{Wang, M.}, \bibinfo{author}{Ahmad, A.S.}, \bibinfo{author}{Khan, R.}, \bibinfo{author}{Maqsood, M.}, \bibinfo{author}{Yaqub, M.}, \bibinfo{year}{2021}.
\newblock \bibinfo{title}{A transfer learning approach for early diagnosis of alzheimer’s disease on mri images}.
\newblock \bibinfo{journal}{Neuroscience} \bibinfo{volume}{460}, \bibinfo{pages}{43--52}.
\bibitem[{Nicoll et~al.(2019)Nicoll, Buckland, Harrison, Page, Harris, Love, Neal, Holmes and Boche}]{nicoll2019persistent}
\bibinfo{author}{Nicoll, J.A.}, \bibinfo{author}{Buckland, G.R.}, \bibinfo{author}{Harrison, C.H.}, \bibinfo{author}{Page, A.}, \bibinfo{author}{Harris, S.}, \bibinfo{author}{Love, S.}, \bibinfo{author}{Neal, J.W.}, \bibinfo{author}{Holmes, C.}, \bibinfo{author}{Boche, D.}, \bibinfo{year}{2019}.
\newblock \bibinfo{title}{Persistent neuropathological effects 14 years following amyloid-$\beta$ immunization in alzheimer’s disease}.
\newblock \bibinfo{journal}{Brain} \bibinfo{volume}{142}, \bibinfo{pages}{2113--2126}.
\bibitem[{Pan et~al.(2020)Pan, Zeng, Jia, Huang, Frizzell and Song}]{2d_slice_pan2020early}
\bibinfo{author}{Pan, D.}, \bibinfo{author}{Zeng, A.}, \bibinfo{author}{Jia, L.}, \bibinfo{author}{Huang, Y.}, \bibinfo{author}{Frizzell, T.}, \bibinfo{author}{Song, X.}, \bibinfo{year}{2020}.
\newblock \bibinfo{title}{Early detection of alzheimer’s disease using magnetic resonance imaging: a novel approach combining convolutional neural networks and ensemble learning}.
\newblock \bibinfo{journal}{Frontiers in neuroscience} \bibinfo{volume}{14}, \bibinfo{pages}{259}.
\bibitem[{Park et~al.(2023)Park, Jung and Suk}]{3d_patch_park2023deep}
\bibinfo{author}{Park, C.}, \bibinfo{author}{Jung, W.}, \bibinfo{author}{Suk, H.I.}, \bibinfo{year}{2023}.
\newblock \bibinfo{title}{Deep joint learning of pathological region localization and alzheimer’s disease diagnosis}.
\newblock \bibinfo{journal}{Scientific reports} \bibinfo{volume}{13}, \bibinfo{pages}{11664}.
\bibitem[{Paszke et~al.(2019)Paszke, Gross, Massa, Lerer, Bradbury, Chanan, Killeen, Lin, Gimelshein, Antiga et~al.}]{paszke2019pytorch}
\bibinfo{author}{Paszke, A.}, \bibinfo{author}{Gross, S.}, \bibinfo{author}{Massa, F.}, \bibinfo{author}{Lerer, A.}, \bibinfo{author}{Bradbury, J.}, \bibinfo{author}{Chanan, G.}, \bibinfo{author}{Killeen, T.}, \bibinfo{author}{Lin, Z.}, \bibinfo{author}{Gimelshein, N.}, \bibinfo{author}{Antiga, L.}, et~al., \bibinfo{year}{2019}.
\newblock \bibinfo{title}{Pytorch: An imperative style, high-performance deep learning library}.
\newblock \bibinfo{journal}{Advances in neural information processing systems} \bibinfo{volume}{32}.
\bibitem[{Poulin et~al.(2011)Poulin, Dautoff, Morris, Barrett, Dickerson, Initiative et~al.}]{poulin2011amygdala}
\bibinfo{author}{Poulin, S.P.}, \bibinfo{author}{Dautoff, R.}, \bibinfo{author}{Morris, J.C.}, \bibinfo{author}{Barrett, L.F.}, \bibinfo{author}{Dickerson, B.C.}, \bibinfo{author}{Initiative, A.D.N.}, et~al., \bibinfo{year}{2011}.
\newblock \bibinfo{title}{Amygdala atrophy is prominent in early alzheimer's disease and relates to symptom severity}.
\newblock \bibinfo{journal}{Psychiatry Research: Neuroimaging} \bibinfo{volume}{194}, \bibinfo{pages}{7--13}.
\bibitem[{Qiu et~al.(2020)Qiu, Joshi, Miller, Xue, Zhou, Karjadi, Chang, Joshi, Dwyer, Zhu et~al.}]{3d_patch_qiu2020development}
\bibinfo{author}{Qiu, S.}, \bibinfo{author}{Joshi, P.S.}, \bibinfo{author}{Miller, M.I.}, \bibinfo{author}{Xue, C.}, \bibinfo{author}{Zhou, X.}, \bibinfo{author}{Karjadi, C.}, \bibinfo{author}{Chang, G.H.}, \bibinfo{author}{Joshi, A.S.}, \bibinfo{author}{Dwyer, B.}, \bibinfo{author}{Zhu, S.}, et~al., \bibinfo{year}{2020}.
\newblock \bibinfo{title}{Development and validation of an interpretable deep learning framework for alzheimer’s disease classification}.
\newblock \bibinfo{journal}{Brain} \bibinfo{volume}{143}, \bibinfo{pages}{1920--1933}.
\bibitem[{Rao et~al.(2021)Rao, Zhao, Zhu, Lu and Zhou}]{gfnet_rao2021global}
\bibinfo{author}{Rao, Y.}, \bibinfo{author}{Zhao, W.}, \bibinfo{author}{Zhu, Z.}, \bibinfo{author}{Lu, J.}, \bibinfo{author}{Zhou, J.}, \bibinfo{year}{2021}.
\newblock \bibinfo{title}{Global filter networks for image classification}.
\newblock \bibinfo{journal}{Advances in neural information processing systems} \bibinfo{volume}{34}, \bibinfo{pages}{980--993}.
\bibitem[{Rao et~al.(2022)Rao, Ganaraja, Murlimanju, Joy, Krishnamurthy and Agrawal}]{rao2022hippocampus}
\bibinfo{author}{Rao, Y.L.}, \bibinfo{author}{Ganaraja, B.}, \bibinfo{author}{Murlimanju, B.}, \bibinfo{author}{Joy, T.}, \bibinfo{author}{Krishnamurthy, A.}, \bibinfo{author}{Agrawal, A.}, \bibinfo{year}{2022}.
\newblock \bibinfo{title}{Hippocampus and its involvement in alzheimer’s disease: a review}.
\newblock \bibinfo{journal}{3 Biotech} \bibinfo{volume}{12}, \bibinfo{pages}{55}.
\bibitem[{Rathore et~al.(2017)Rathore, Habes, Iftikhar, Shacklett and Davatzikos}]{ML_rathore2017review}
\bibinfo{author}{Rathore, S.}, \bibinfo{author}{Habes, M.}, \bibinfo{author}{Iftikhar, M.A.}, \bibinfo{author}{Shacklett, A.}, \bibinfo{author}{Davatzikos, C.}, \bibinfo{year}{2017}.
\newblock \bibinfo{title}{A review on neuroimaging-based classification studies and associated feature extraction methods for alzheimer's disease and its prodromal stages}.
\newblock \bibinfo{journal}{NeuroImage} \bibinfo{volume}{155}, \bibinfo{pages}{530--548}.
\bibitem[{Routier et~al.(2021)Routier, Burgos, D{\'\i}az, Bacci, Bottani, El-Rifai, Fontanella, Gori, Guillon, Guyot et~al.}]{clinica_routier2021}
\bibinfo{author}{Routier, A.}, \bibinfo{author}{Burgos, N.}, \bibinfo{author}{D{\'\i}az, M.}, \bibinfo{author}{Bacci, M.}, \bibinfo{author}{Bottani, S.}, \bibinfo{author}{El-Rifai, O.}, \bibinfo{author}{Fontanella, S.}, \bibinfo{author}{Gori, P.}, \bibinfo{author}{Guillon, J.}, \bibinfo{author}{Guyot, A.}, et~al., \bibinfo{year}{2021}.
\newblock \bibinfo{title}{Clinica: An open-source software platform for reproducible clinical neuroscience studies}.
\newblock \bibinfo{journal}{Frontiers in Neuroinformatics} \bibinfo{volume}{15}, \bibinfo{pages}{689675}.
\bibitem[{Russakovsky et~al.(2015)Russakovsky, Deng, Su, Krause, Satheesh, Ma, Huang, Karpathy, Khosla, Bernstein et~al.}]{russakovsky2015imagenet}
\bibinfo{author}{Russakovsky, O.}, \bibinfo{author}{Deng, J.}, \bibinfo{author}{Su, H.}, \bibinfo{author}{Krause, J.}, \bibinfo{author}{Satheesh, S.}, \bibinfo{author}{Ma, S.}, \bibinfo{author}{Huang, Z.}, \bibinfo{author}{Karpathy, A.}, \bibinfo{author}{Khosla, A.}, \bibinfo{author}{Bernstein, M.}, et~al., \bibinfo{year}{2015}.
\newblock \bibinfo{title}{Imagenet large scale visual recognition challenge}.
\newblock \bibinfo{journal}{International journal of computer vision} \bibinfo{volume}{115}, \bibinfo{pages}{211--252}.
\bibitem[{Samper-Gonz{\'a}lez et~al.(2018)Samper-Gonz{\'a}lez, Burgos, Bottani, Fontanella, Lu, Marcoux, Routier, Guillon, Bacci, Wen et~al.}]{clinica_samper2018reproducible}
\bibinfo{author}{Samper-Gonz{\'a}lez, J.}, \bibinfo{author}{Burgos, N.}, \bibinfo{author}{Bottani, S.}, \bibinfo{author}{Fontanella, S.}, \bibinfo{author}{Lu, P.}, \bibinfo{author}{Marcoux, A.}, \bibinfo{author}{Routier, A.}, \bibinfo{author}{Guillon, J.}, \bibinfo{author}{Bacci, M.}, \bibinfo{author}{Wen, J.}, et~al., \bibinfo{year}{2018}.
\newblock \bibinfo{title}{Reproducible evaluation of classification methods in alzheimer's disease: Framework and application to mri and pet data}.
\newblock \bibinfo{journal}{NeuroImage} \bibinfo{volume}{183}, \bibinfo{pages}{504--521}.
\bibitem[{Schlemper et~al.(2019)Schlemper, Oktay, Schaap, Heinrich, Kainz, Glocker and Rueckert}]{attention_exp_schlemper2019attention}
\bibinfo{author}{Schlemper, J.}, \bibinfo{author}{Oktay, O.}, \bibinfo{author}{Schaap, M.}, \bibinfo{author}{Heinrich, M.}, \bibinfo{author}{Kainz, B.}, \bibinfo{author}{Glocker, B.}, \bibinfo{author}{Rueckert, D.}, \bibinfo{year}{2019}.
\newblock \bibinfo{title}{Attention gated networks: Learning to leverage salient regions in medical images}.
\newblock \bibinfo{journal}{Medical image analysis} \bibinfo{volume}{53}, \bibinfo{pages}{197--207}.
\bibitem[{Selvaraju et~al.(2017)Selvaraju, Cogswell, Das, Vedantam, Parikh and Batra}]{selvaraju2017grad}
\bibinfo{author}{Selvaraju, R.R.}, \bibinfo{author}{Cogswell, M.}, \bibinfo{author}{Das, A.}, \bibinfo{author}{Vedantam, R.}, \bibinfo{author}{Parikh, D.}, \bibinfo{author}{Batra, D.}, \bibinfo{year}{2017}.
\newblock \bibinfo{title}{Grad-cam: Visual explanations from deep networks via gradient-based localization}, in: \bibinfo{booktitle}{Proceedings of the IEEE international conference on computer vision}, pp. \bibinfo{pages}{618--626}.
\bibitem[{Simonyan and Zisserman(2014)}]{vgg_SimonyanZ14a}
\bibinfo{author}{Simonyan, K.}, \bibinfo{author}{Zisserman, A.}, \bibinfo{year}{2014}.
\newblock \bibinfo{title}{Very deep convolutional networks for large-scale image recognition}.
\newblock \bibinfo{journal}{CoRR} \bibinfo{volume}{abs/1409.1556}.
\newblock \URLprefix \url{http://arxiv.org/abs/1409.1556}, \href{http://arxiv.org/abs/1409.1556}{{\tt arXiv:1409.1556}}.
\bibitem[{Singh et~al.(2020)Singh, Sengupta and Lakshminarayanan}]{EXP_singh2020explainable}
\bibinfo{author}{Singh, A.}, \bibinfo{author}{Sengupta, S.}, \bibinfo{author}{Lakshminarayanan, V.}, \bibinfo{year}{2020}.
\newblock \bibinfo{title}{Explainable deep learning models in medical image analysis}.
\newblock \bibinfo{journal}{Journal of imaging} \bibinfo{volume}{6}, \bibinfo{pages}{52}.
\bibitem[{Smith(2002)}]{bet_smith2002fast}
\bibinfo{author}{Smith, S.M.}, \bibinfo{year}{2002}.
\newblock \bibinfo{title}{Fast robust automated brain extraction}.
\newblock \bibinfo{journal}{Human brain mapping} \bibinfo{volume}{17}, \bibinfo{pages}{143--155}.
\bibitem[{Tan and Le(2021)}]{effnet_tan2021efficientnetv2}
\bibinfo{author}{Tan, M.}, \bibinfo{author}{Le, Q.V.}, \bibinfo{year}{2021}.
\newblock \bibinfo{title}{Efficientnetv2: Smaller models and faster training}.
\newblock \bibinfo{journal}{arXiv preprint arXiv:2104.00298} .
\bibitem[{Tanveer et~al.(2021)Tanveer, Rashid, Ganaie, Reza, Razzak and Hua}]{2d_slice_tanveer2021classification}
\bibinfo{author}{Tanveer, M.}, \bibinfo{author}{Rashid, A.H.}, \bibinfo{author}{Ganaie, M.}, \bibinfo{author}{Reza, M.}, \bibinfo{author}{Razzak, I.}, \bibinfo{author}{Hua, K.L.}, \bibinfo{year}{2021}.
\newblock \bibinfo{title}{Classification of alzheimer’s disease using ensemble of deep neural networks trained through transfer learning}.
\newblock \bibinfo{journal}{IEEE Journal of Biomedical and Health Informatics} \bibinfo{volume}{26}, \bibinfo{pages}{1453--1463}.
\bibitem[{Tustison et~al.(2010)Tustison, Avants, Cook, Zheng, Egan, Yushkevich and Gee}]{n4itk_tustinson_2010}
\bibinfo{author}{Tustison, N.J.}, \bibinfo{author}{Avants, B.B.}, \bibinfo{author}{Cook, P.A.}, \bibinfo{author}{Zheng, Y.}, \bibinfo{author}{Egan, A.}, \bibinfo{author}{Yushkevich, P.A.}, \bibinfo{author}{Gee, J.C.}, \bibinfo{year}{2010}.
\newblock \bibinfo{title}{N4itk: Improved n3 bias correction}.
\newblock \bibinfo{journal}{IEEE Transactions on Medical Imaging} \bibinfo{volume}{29}, \bibinfo{pages}{1310--1320}.
\newblock \DOIprefix\doi{10.1109/TMI.2010.2046908}.
\bibitem[{Van~Hoesen et~al.(2000)Van~Hoesen, Augustinack, Dierking, Redman and Thangavel}]{van2000parahippocampal}
\bibinfo{author}{Van~Hoesen, G.W.}, \bibinfo{author}{Augustinack, J.C.}, \bibinfo{author}{Dierking, J.}, \bibinfo{author}{Redman, S.J.}, \bibinfo{author}{Thangavel, R.}, \bibinfo{year}{2000}.
\newblock \bibinfo{title}{The parahippocampal gyrus in alzheimer's disease: clinical and preclinical neuroanatomical correlates}.
\newblock \bibinfo{journal}{Annals of the New York Academy of Sciences} \bibinfo{volume}{911}, \bibinfo{pages}{254--274}.
\bibitem[{Van~der Velden et~al.(2022)Van~der Velden, Kuijf, Gilhuijs and Viergever}]{EXP_van2022explainable}
\bibinfo{author}{Van~der Velden, B.H.}, \bibinfo{author}{Kuijf, H.J.}, \bibinfo{author}{Gilhuijs, K.G.}, \bibinfo{author}{Viergever, M.A.}, \bibinfo{year}{2022}.
\newblock \bibinfo{title}{Explainable artificial intelligence (xai) in deep learning-based medical image analysis}.
\newblock \bibinfo{journal}{Medical Image Analysis} \bibinfo{volume}{79}, \bibinfo{pages}{102470}.
\bibitem[{Vemuri et~al.(2009)Vemuri, Wiste, Weigand, Shaw, Trojanowski, Weiner, Knopman, Petersen, Jack et~al.}]{vemuri2009mri}
\bibinfo{author}{Vemuri, P.}, \bibinfo{author}{Wiste, H.}, \bibinfo{author}{Weigand, S.}, \bibinfo{author}{Shaw, L.}, \bibinfo{author}{Trojanowski, J.}, \bibinfo{author}{Weiner, M.}, \bibinfo{author}{Knopman, D.}, \bibinfo{author}{Petersen, R.}, \bibinfo{author}{Jack, C.}, et~al., \bibinfo{year}{2009}.
\newblock \bibinfo{title}{Mri and csf biomarkers in normal, mci, and ad subjects: diagnostic discrimination and cognitive correlations}.
\newblock \bibinfo{journal}{Neurology} \bibinfo{volume}{73}, \bibinfo{pages}{287--293}.
\bibitem[{Venugopalan et~al.(2021)Venugopalan, Tong, Hassanzadeh and Wang}]{3d_venugopalan2021multimodal}
\bibinfo{author}{Venugopalan, J.}, \bibinfo{author}{Tong, L.}, \bibinfo{author}{Hassanzadeh, H.R.}, \bibinfo{author}{Wang, M.D.}, \bibinfo{year}{2021}.
\newblock \bibinfo{title}{Multimodal deep learning models for early detection of alzheimer’s disease stage}.
\newblock \bibinfo{journal}{Scientific reports} \bibinfo{volume}{11}, \bibinfo{pages}{3254}.
\bibitem[{Viswan et~al.(2024)Viswan, Shaffi, Mahmud, Subramanian and Hajamohideen}]{EXP_no_gradcam_viswan2024explainable}
\bibinfo{author}{Viswan, V.}, \bibinfo{author}{Shaffi, N.}, \bibinfo{author}{Mahmud, M.}, \bibinfo{author}{Subramanian, K.}, \bibinfo{author}{Hajamohideen, F.}, \bibinfo{year}{2024}.
\newblock \bibinfo{title}{Explainable artificial intelligence in alzheimer’s disease classification: A systematic review}.
\newblock \bibinfo{journal}{Cognitive Computation} \bibinfo{volume}{16}, \bibinfo{pages}{1--44}.
\bibitem[{Wang et~al.(2024)Wang, Piao, Huang, Gao, Zhang, Li, Shan, Initiative et~al.}]{2d_slice_wang2024joint}
\bibinfo{author}{Wang, C.}, \bibinfo{author}{Piao, S.}, \bibinfo{author}{Huang, Z.}, \bibinfo{author}{Gao, Q.}, \bibinfo{author}{Zhang, J.}, \bibinfo{author}{Li, Y.}, \bibinfo{author}{Shan, H.}, \bibinfo{author}{Initiative, A.D.N.}, et~al., \bibinfo{year}{2024}.
\newblock \bibinfo{title}{Joint learning framework of cross-modal synthesis and diagnosis for alzheimer’s disease by mining underlying shared modality information}.
\newblock \bibinfo{journal}{Medical Image Analysis} \bibinfo{volume}{91}, \bibinfo{pages}{103032}.
\bibitem[{Wang et~al.(2018)Wang, Phillips, Sui, Liu, Yang and Cheng}]{2d_slice_wang2018classification}
\bibinfo{author}{Wang, S.H.}, \bibinfo{author}{Phillips, P.}, \bibinfo{author}{Sui, Y.}, \bibinfo{author}{Liu, B.}, \bibinfo{author}{Yang, M.}, \bibinfo{author}{Cheng, H.}, \bibinfo{year}{2018}.
\newblock \bibinfo{title}{Classification of alzheimer’s disease based on eight-layer convolutional neural network with leaky rectified linear unit and max pooling}.
\newblock \bibinfo{journal}{Journal of medical systems} \bibinfo{volume}{42}, \bibinfo{pages}{1--11}.
\bibitem[{Wen et~al.(2020)Wen, Thibeau-Sutre, Diaz-Melo, Samper-Gonz{\'a}lez, Routier, Bottani, Dormont, Durrleman, Burgos, Colliot et~al.}]{review_wen2020convolutional}
\bibinfo{author}{Wen, J.}, \bibinfo{author}{Thibeau-Sutre, E.}, \bibinfo{author}{Diaz-Melo, M.}, \bibinfo{author}{Samper-Gonz{\'a}lez, J.}, \bibinfo{author}{Routier, A.}, \bibinfo{author}{Bottani, S.}, \bibinfo{author}{Dormont, D.}, \bibinfo{author}{Durrleman, S.}, \bibinfo{author}{Burgos, N.}, \bibinfo{author}{Colliot, O.}, et~al., \bibinfo{year}{2020}.
\newblock \bibinfo{title}{Convolutional neural networks for classification of alzheimer's disease: Overview and reproducible evaluation}.
\newblock \bibinfo{journal}{Medical image analysis} \bibinfo{volume}{63}, \bibinfo{pages}{101694}.
\bibitem[{Winblad et~al.(2016)Winblad, Amouyel, Andrieu, Ballard, Brayne, Brodaty, Cedazo-Minguez, Dubois, Edvardsson, Feldman et~al.}]{winblad2016defeating}
\bibinfo{author}{Winblad, B.}, \bibinfo{author}{Amouyel, P.}, \bibinfo{author}{Andrieu, S.}, \bibinfo{author}{Ballard, C.}, \bibinfo{author}{Brayne, C.}, \bibinfo{author}{Brodaty, H.}, \bibinfo{author}{Cedazo-Minguez, A.}, \bibinfo{author}{Dubois, B.}, \bibinfo{author}{Edvardsson, D.}, \bibinfo{author}{Feldman, H.}, et~al., \bibinfo{year}{2016}.
\newblock \bibinfo{title}{Defeating alzheimer's disease and other dementias: a priority for european science and society}.
\newblock \bibinfo{journal}{The Lancet Neurology} \bibinfo{volume}{15}, \bibinfo{pages}{455--532}.
\bibitem[{Wong(2020)}]{wong2020economic}
\bibinfo{author}{Wong, W.}, \bibinfo{year}{2020}.
\newblock \bibinfo{title}{Economic burden of alzheimer disease and managed care considerations.}
\newblock \bibinfo{journal}{The American journal of managed care} \bibinfo{volume}{26}, \bibinfo{pages}{S177--S183}.
\bibitem[{Wu et~al.(2022)Wu, Zhou, Zeng, Qian and Song}]{3d_wu2022attention}
\bibinfo{author}{Wu, Y.}, \bibinfo{author}{Zhou, Y.}, \bibinfo{author}{Zeng, W.}, \bibinfo{author}{Qian, Q.}, \bibinfo{author}{Song, M.}, \bibinfo{year}{2022}.
\newblock \bibinfo{title}{An attention-based 3d cnn with multi-scale integration block for alzheimer's disease classification}.
\newblock \bibinfo{journal}{IEEE Journal of Biomedical and Health Informatics} \bibinfo{volume}{26}, \bibinfo{pages}{5665--5673}.
\bibitem[{Wyman et~al.(2013)Wyman, Harvey, Crawford, Bernstein, Carmichael, Cole, Crane, DeCarli, Fox, Gunter et~al.}]{adni_wyman2013standardization}
\bibinfo{author}{Wyman, B.T.}, \bibinfo{author}{Harvey, D.J.}, \bibinfo{author}{Crawford, K.}, \bibinfo{author}{Bernstein, M.A.}, \bibinfo{author}{Carmichael, O.}, \bibinfo{author}{Cole, P.E.}, \bibinfo{author}{Crane, P.K.}, \bibinfo{author}{DeCarli, C.}, \bibinfo{author}{Fox, N.C.}, \bibinfo{author}{Gunter, J.L.}, et~al., \bibinfo{year}{2013}.
\newblock \bibinfo{title}{Standardization of analysis sets for reporting results from adni mri data}.
\newblock \bibinfo{journal}{Alzheimer's \& Dementia} \bibinfo{volume}{9}, \bibinfo{pages}{332--337}.
\bibitem[{Yarkoni et~al.(2019)Yarkoni, Markiewicz, de~la Vega, Gorgolewski, Salo, Halchenko, McNamara, DeStasio, Poline, Petrov, Hayot-Sasson, Nielson, Carlin, Kiar, Whitaker, DuPre, Wagner, Tirrell, Jas, Hanke, Poldrack, Esteban, Appelhoff, Holdgraf, Staden, Thirion, Kleinschmidt, Lee, di~Castello, Notter and Blair}]{pybids_Yarkoni2019}
\bibinfo{author}{Yarkoni, T.}, \bibinfo{author}{Markiewicz, C.J.}, \bibinfo{author}{de~la Vega, A.}, \bibinfo{author}{Gorgolewski, K.J.}, \bibinfo{author}{Salo, T.}, \bibinfo{author}{Halchenko, Y.O.}, \bibinfo{author}{McNamara, Q.}, \bibinfo{author}{DeStasio, K.}, \bibinfo{author}{Poline, J.B.}, \bibinfo{author}{Petrov, D.}, \bibinfo{author}{Hayot-Sasson, V.}, \bibinfo{author}{Nielson, D.M.}, \bibinfo{author}{Carlin, J.}, \bibinfo{author}{Kiar, G.}, \bibinfo{author}{Whitaker, K.}, \bibinfo{author}{DuPre, E.}, \bibinfo{author}{Wagner, A.}, \bibinfo{author}{Tirrell, L.S.}, \bibinfo{author}{Jas, M.}, \bibinfo{author}{Hanke, M.}, \bibinfo{author}{Poldrack, R.A.}, \bibinfo{author}{Esteban, O.}, \bibinfo{author}{Appelhoff, S.}, \bibinfo{author}{Holdgraf, C.}, \bibinfo{author}{Staden, I.}, \bibinfo{author}{Thirion, B.}, \bibinfo{author}{Kleinschmidt, D.F.}, \bibinfo{author}{Lee, J.A.}, \bibinfo{author}{di~Castello, M.V.O.}, \bibinfo{author}{Notter, M.P.}, \bibinfo{author}{Blair, R.}, \bibinfo{year}{2019}.
\newblock \bibinfo{title}{Pybids: Python tools for bids datasets}.
\newblock \bibinfo{journal}{Journal of Open Source Software} \bibinfo{volume}{4}, \bibinfo{pages}{1294}.
\newblock \URLprefix \url{https://doi.org/10.21105/joss.01294}, \DOIprefix\doi{10.21105/joss.01294}.
\bibitem[{Yarkoni et~al.(2023)Yarkoni, Markiewicz, de~la Vega, Gorgolewski, Salo, Halchenko, Papadopoulos~Orfanos, Esteban, Gau, McNamara, DeStasio, Poline, Johnson, Kalenkovich, Petrov, Nielson, Kent, Kent, Appelhoff, Van~Dyken, Goncalves, Bansal, Hayot-Sasson, Carlin, Kiar, Whitaker, Ghosh, Wagner, DuPre, Janke, Ivanov, Gillman, Wennberg, Tirrell, Tilley~II, Li, Legarreta, Waller, Jas, Hanke, Guenther, Poldrack, Rokem, Boulay, Mumford, Thual, Holdgraf, Staden, Staph, Drew, Sinha, Rovai, Adebimpe, Thirion, Kleinschmidt, Dickie, Ben-Zvi, Lee, Kruper, Visconti~di Oleggio~Castello, Notter, Roca, Blair, Pati and Sundaravadivelu}]{pybids_yarkoni_2023_8253830}
\bibinfo{author}{Yarkoni, T.}, \bibinfo{author}{Markiewicz, C.J.}, \bibinfo{author}{de~la Vega, A.}, \bibinfo{author}{Gorgolewski, K.J.}, \bibinfo{author}{Salo, T.}, \bibinfo{author}{Halchenko, Y.O.}, \bibinfo{author}{Papadopoulos~Orfanos, D.}, \bibinfo{author}{Esteban, O.}, \bibinfo{author}{Gau, R.}, \bibinfo{author}{McNamara, Q.}, \bibinfo{author}{DeStasio, K.}, \bibinfo{author}{Poline, J.B.}, \bibinfo{author}{Johnson, H.}, \bibinfo{author}{Kalenkovich, E.}, \bibinfo{author}{Petrov, D.}, \bibinfo{author}{Nielson, D.M.}, \bibinfo{author}{Kent, J.}, \bibinfo{author}{Kent, J.D.}, \bibinfo{author}{Appelhoff, S.}, \bibinfo{author}{Van~Dyken, P.}, \bibinfo{author}{Goncalves, M.}, \bibinfo{author}{Bansal, S.}, \bibinfo{author}{Hayot-Sasson, V.}, \bibinfo{author}{Carlin, J.}, \bibinfo{author}{Kiar, G.}, \bibinfo{author}{Whitaker, K.}, \bibinfo{author}{Ghosh, S.}, \bibinfo{author}{Wagner, A.}, \bibinfo{author}{DuPre, E.}, \bibinfo{author}{Janke, A.}, \bibinfo{author}{Ivanov, A.}, \bibinfo{author}{Gillman, A.},
  \bibinfo{author}{Wennberg, J.}, \bibinfo{author}{Tirrell, L.S.}, \bibinfo{author}{Tilley~II, S.}, \bibinfo{author}{Li, A.}, \bibinfo{author}{Legarreta, J.H.}, \bibinfo{author}{Waller, L.}, \bibinfo{author}{Jas, M.}, \bibinfo{author}{Hanke, M.}, \bibinfo{author}{Guenther, N.}, \bibinfo{author}{Poldrack, R.}, \bibinfo{author}{Rokem, A.}, \bibinfo{author}{Boulay, C.}, \bibinfo{author}{Mumford, J.}, \bibinfo{author}{Thual, A.}, \bibinfo{author}{Holdgraf, C.}, \bibinfo{author}{Staden, I.}, \bibinfo{author}{Staph, J.A.}, \bibinfo{author}{Drew, W.}, \bibinfo{author}{Sinha, A.}, \bibinfo{author}{Rovai, A.}, \bibinfo{author}{Adebimpe, A.}, \bibinfo{author}{Thirion, B.}, \bibinfo{author}{Kleinschmidt, D.F.}, \bibinfo{author}{Dickie, E.W.}, \bibinfo{author}{Ben-Zvi, G.}, \bibinfo{author}{Lee, J.A.}, \bibinfo{author}{Kruper, J.}, \bibinfo{author}{Visconti~di Oleggio~Castello, M.}, \bibinfo{author}{Notter, M.P.}, \bibinfo{author}{Roca, P.}, \bibinfo{author}{Blair, R.}, \bibinfo{author}{Pati, S.},
  \bibinfo{author}{Sundaravadivelu, S.}, \bibinfo{year}{2023}.
\newblock \bibinfo{title}{Pybids: Python tools for bids datasets}.
\newblock \URLprefix \url{https://doi.org/10.5281/zenodo.8253830}, \DOIprefix\doi{10.5281/zenodo.8253830}.
\bibitem[{Yosinski et~al.(2014)Yosinski, Clune, Bengio and Lipson}]{yosinski2014transferable}
\bibinfo{author}{Yosinski, J.}, \bibinfo{author}{Clune, J.}, \bibinfo{author}{Bengio, Y.}, \bibinfo{author}{Lipson, H.}, \bibinfo{year}{2014}.
\newblock \bibinfo{title}{How transferable are features in deep neural networks?}
\newblock \bibinfo{journal}{Advances in neural information processing systems} \bibinfo{volume}{27}.
\bibitem[{Zhang et~al.(2022)Zhang, Teng, Liu, Liu and He}]{2d_slice_zhang2022diagnosis}
\bibinfo{author}{Zhang, Y.}, \bibinfo{author}{Teng, Q.}, \bibinfo{author}{Liu, Y.}, \bibinfo{author}{Liu, Y.}, \bibinfo{author}{He, X.}, \bibinfo{year}{2022}.
\newblock \bibinfo{title}{Diagnosis of alzheimer's disease based on regional attention with smri gray matter slices}.
\newblock \bibinfo{journal}{Journal of neuroscience methods} \bibinfo{volume}{365}, \bibinfo{pages}{109376}.
\bibitem[{Zhou et~al.(2023)Zhou, Wang, Yu, Wang and Zhang}]{review_zhou2023survey}
\bibinfo{author}{Zhou, Q.}, \bibinfo{author}{Wang, J.}, \bibinfo{author}{Yu, X.}, \bibinfo{author}{Wang, S.}, \bibinfo{author}{Zhang, Y.}, \bibinfo{year}{2023}.
\newblock \bibinfo{title}{A survey of deep learning for alzheimer’s disease}.
\newblock \bibinfo{journal}{Machine Learning and Knowledge Extraction} \bibinfo{volume}{5}, \bibinfo{pages}{611--668}.

\end{thebibliography}




\end{document}